\documentclass[longauth]{aa}

\usepackage[varg]{txfonts}
\usepackage{graphicx}
\usepackage{url}
\usepackage{hyperref}
\usepackage{natbib}
\usepackage{color}
\usepackage{multirow}

\bibpunct{(}{)}{;}{a}{}{,}

\begin{document}


\title{Stellar populations of galaxies in the ALHAMBRA survey up to $z \sim 1$\thanks{Based on observations collected at the Centro Astron\'omico Hispano Alem\'an (CAHA) at Calar Alto, operated jointly by the Max-Planck Institut für Astronomie and the Instituto de Astrof\'isica de Andaluc\'ia (CSIC). Table~\ref{tab:ptable} is fully available in electronic form at the CDS via anonymous ftp to \texttt{cdsarc.u-strasbg.fr} (130.79.128.5) or via \texttt{http://cdsweb.u-strasbg.fr/cgi-bin/qcat?J/A+A/}}}
\subtitle{II. Stellar content of quiescent galaxies within the dust-corrected stellar mass--colour and the $UVJ$ colour--colour diagrams}

%
\author{L.~A.~D\'iaz-Garc\'ia\inst{\ref{a1},\ref{a0}}
\and A.~J.~Cenarro\inst{\ref{a01}}
\and C.~L\'opez-Sanjuan\inst{\ref{a01}}
\and I.~Ferreras\inst{\ref{a2}}
\and M.~Cervi\~no\inst{\ref{a3},\ref{a4}}
\and A.~Fern\'andez-Soto\inst{\ref{a5},\ref{a6}}
\and R.~M.~Gonz\'alez~Delgado\inst{\ref{a3}}
\and I.~M\'arquez\inst{\ref{a3}}
\and M.~Povi\'c\inst{\ref{a3}}
\and I.~San Roman\inst{\ref{a1}}
\and K.~Viironen\inst{\ref{a01}}
\and M.~Moles\inst{\ref{a1}}
\and D.~Crist\'obal-Hornillos\inst{\ref{a01}}
\and A.~L{\'o}pez-Comazzi\inst{\ref{a1}}
\and E.~Alfaro\inst{\ref{a3}}
\and T.~Aparicio-Villegas\inst{\ref{a7}}
\and N.~Ben\'itez\inst{\ref{a3}}
\and T.~Broadhurst\inst{\ref{a8},\ref{a9}}
\and J.~Cabrera-Ca\~no\inst{\ref{a10}}
\and F.~J.~Castander\inst{\ref{a11}}
\and J.~Cepa\inst{\ref{a4},\ref{a12}}
\and C.~Husillos\inst{\ref{a3}}
\and L.~Infante\inst{\ref{a13},\ref{a14}}
\and J.~A.~L.~Aguerri\inst{\ref{a4},\ref{a12}}
\and V.~J.~Mart\'inez\inst{\ref{a16},\ref{a17},\ref{a6}}
\and J.~Masegosa\inst{\ref{a3}}
\and A.~Molino\inst{\ref{a15}}
\and A.~del~Olmo\inst{\ref{a3}}
\and J.~Perea\inst{\ref{a3}}
\and F.~Prada\inst{\ref{a3}}
\and J.~M.~Quintana\inst{\ref{a3}}
}
%
\institute{
Centro de Estudios de F\'isica del Cosmos de Arag\'on (CEFCA), Plaza San Juan 1, Floor 2, E--44001 Teruel, Spain\label{a1}\\  \email{ladiaz@asiaa.sinica.edu.tw}
\and Academia Sinica Institute of Astronomy \& Astrophysics (ASIAA), 11F of Astronomy-Mathematics Building, AS/NTU, No.~1, Section 4, Roosevelt Road, Taipei 10617, Taiwan\label{a0}
\and Centro de Estudios de F\'isica del Cosmos de Arag\'on (CEFCA) - Unidad Asociada al CSIC, Plaza San Juan 1, Floor 2, E--44001 Teruel, Spain\label{a01}
\and Mullard Space Science Laboratory, University College London, Holmbury St Mary, Dorking, Surrey RH5 6NT, United Kingdom\label{a2}
\and IAA-CSIC, Glorieta de la Astronom\'ia s/n, 18008 Granada, Spain\label{a3}
\and Instituto de Astrof\'isica de Canarias, V\'ia L\'actea s/n, 38200 La Laguna, Tenerife, Spain\label{a4}
\and Instituto de F\'isica de Cantabria (CSIC-UC), E-39005 Santander, Spain\label{a5}
\and Unidad Asociada Observatorio Astron\'omico (IFCA-UV), E-46980, Paterna, Spain\label{a6}
\and Observat\'orio Nacional-MCT, Rua Jos\'e Cristino, 77. CEP 20921-400, Rio de Janeiro-RJ, Brazil\label{a7}
\and Department of Theoretical Physics, University of the Basque Country UPV/EHU, 48080 Bilbao, Spain\label{a8}
\and IKERBASQUE, Basque Foundation for Science, Bilbao, Spain\label{a9}
\and Departamento de F\'isica At\'omica, Molecular y Nuclear, Facultad de F\'isica, Universidad de Sevilla, 41012 Sevilla, Spain\label{a10}
\and Institut de Ci\`encies de l'Espai (IEEC-CSIC), Facultat de Ci\`encies, Campus UAB, 08193 Bellaterra, Spain\label{a11}
\and Departamento de Astrof\'isica, Facultad de F\'isica, Universidad de La Laguna, 38206 La Laguna, Spain\label{a12}
\and Instituto de Astrof\'{\i}sica, Universidad Cat\'olica de Chile, Av. Vicuna Mackenna 4860, 782-0436 Macul, Santiago, Chile\label{a13}
\and Centro de Astro-Ingenier\'{\i}a, Universidad Cat\'olica de Chile, Av. Vicuna Mackenna 4860, 782-0436 Macul, Santiago, Chile\label{a14}
\and Observatori Astron\`omic, Universitat de Val\`encia, C/ Catedr\`atic Jos\'e Beltr\'an 2, E-46980, Paterna, Spain\label{a16}
\and Departament d'Astronomia i Astrof\'isica, Universitat de Val\`encia, E-46100, Burjassot, Spain\label{a17}
\and Instituto de Astronom{\'{\i}}a, Geof{\'{\i}}sica e Ci\'encias Atmosf\'ericas, Universidade de S{\~{a}}o Paulo, S{\~{a}}o Paulo, Brazil\label{a15}
}

\date{Received ? / Accepted ?}

\abstract{}  
         {Our aim is to determine the distribution of stellar population parameters (extinction, age, metallicity, and star formation rates) of quiescent galaxies within the rest-frame stellar mass--colour diagrams and $UVJ$ colour--colour diagrams corrected for extinction up to $z\sim1$. These novel diagrams reduce the contamination in samples of quiescent galaxies owing to dust-reddened galaxies, and they provide useful constraints on stellar population parameters only using rest-frame colours and/or stellar mass.}            
         {We set constraints on the stellar population parameters of quiescent galaxies combining the ALHAMBRA multi-filter photo-spectra with our fitting code for spectral energy distribution,  MUlti-Filter FITting (MUFFIT), making use of composite stellar population models based on two independent sets of simple stellar population (SSP) models. The extinction obtained by MUFFIT allowed us to remove dusty star-forming (DSF) galaxies from the sample of red $UVJ$ galaxies. The distributions of stellar population parameters across these rest-frame diagrams are revealed after the dust correction and are fitted by LOESS, a bi-dimensional and locally weighted regression method, to reduce uncertainty effects.}     
         {Quiescent galaxy samples defined via classical $UVJ$ diagrams are typically contaminated by a $\sim20$~\% fraction of DSF galaxies. A significant part of the galaxies in the green valley are actually obscured star-forming galaxies ($\sim30$--$ 65$~\%). Consequently, the transition of galaxies from the blue cloud to the red sequence, and hence the related mechanisms for quenching, seems to be much more efficient and faster than previously reported. The rest-frame stellar mass--colour  and $UVJ$ colour--colour diagrams are  useful  for constraining the age, metallicity, extinction, and star formation rate of quiescent galaxies by only their redshift, rest-frame colours, and/or stellar mass. Dust correction plays an important role in understanding how quiescent galaxies are distributed in these diagrams and is key to performing a pure selection of quiescent galaxies via intrinsic colours.}    
         {}  


\titlerunning{Stellar populations of galaxies within diagrams involving colours}

\authorrunning{L.~A.~D\'iaz-Garc\'ia et al.}

\maketitle



\section{Introduction}\label{sec:introduction}

Galaxies exhibit a well-known bimodal distribution of colours; they are  usually referred to as red and blue galaxies, with well-differentiated properties \citep[e.g.][]{Bell2004,Baldry2004,Williams2009,Ilbert2010,Peng2010,Arnouts2013,Moresco2013,Povic2013,Fritz2014,Schawinski2014}. Red galaxies present more evolved stellar populations with lower star formation levels, and so are usually called passive or quiescent. In contrast,  blue galaxies exhibit younger stellar populations, typically with signatures of strong star formation processes. Understanding the formation and evolution of the so-called quiescent or passive galaxies is still a challenge today;  these galaxies started to form stars at very early epochs, to later shut down their star formation by mechanisms that are   a matter of debate even today \citep[see references in][]{Peng2015}.

One of the most extended diagrams for disentangling different types of galaxies involves rest-frame colours and absolute magnitudes \citep[e.g.][]{Bell2004,Baldry2004,Brown2007}. These diagrams, usually referred to  as colour--magnitude diagrams (CMDs), clearly show the bimodal distribution of galaxies previously mentioned. Within this diagram, galaxies are usually referred to as red-sequence galaxies (predominantly old and including very massive galaxies in the nearby Universe) and blue cloud galaxies (numerous and with a prominent on-going star formation). In addition, some authors have  proposed that there is a third group of galaxies lying in the colour region between these two dominant groups of galaxies, which we call the green valley. Interestingly, the presence of active galactic nuclei (AGNs) in green valley galaxies is remarkable \citep[][]{Nandra2007,Bundy2008,Georgakakis2008,Silverman2008,Hickox2009,Schawinski2009,Povic2012}. Owing to this overdensity of galaxies hosting an AGN in the green valley, some authors have proposed that the presence of AGNs may be related to a quenching mechanism that produces a transition of galaxies from the blue cloud to the red sequence \citep[e.g.][]{Faber2007,Schawinski2007}. However, there is also evidence to show that galaxies belonging to the green valley are mainly dust-reddened star-forming (hereafter DSF) galaxies \citep{Bell2005,Cowie2008,Brammer2009,Cardamone2010}. 

During the last decade, the rest-frame colour--colour diagrams,  especially the $UVJ$ diagram, have become one of the most extended methods used to select quiescent and star-forming galaxies \citep[e.g.][]{Daddi2004,Williams2009,Arnouts2013}. The advantage of these diagrams is that they use two colours,  one in the blue part of the spectral energy distribution (SED), $U-V$ or $NUV-r$,  the other  in the red part, $V-J$ or $r-K$; the two colours  reduce the contamination of DSF galaxies in the quiescent sample with predominant red colours and low star formation rates \cite[e.g.][]{Williams2009,Arnouts2013,Ilbert2013,Moresco2013}. These colour--colour diagrams are efficient because the galaxies empirically present more separated loci between the red and blue populations, whereas in CMDs the two galaxy populations appear largely overlapped \citep{Wyder2007,Cowie2008,Brammer2009}. Even though $UVJ$ diagrams have been demonstrated to be a reliable method for splitting quiescent from star-forming galaxies, these diagrams present a level of contamination in the selection of quiescent galaxies that depends on redshift and stellar mass \citep{Williams2009,Moresco2013}. In some cases the number of star-forming interlopers may reach  $30$~\%  of galaxies at certain redshift and mass regimes, or at least  $15$~\%  after imposing a more restrictive $U-V$ colour limit than the one defined originally \citep{Moresco2013}.

Alternative methods for selecting galaxies by spectral type comprise colours and stellar masses \citep{Peng2010}, but these are almost equivalent to CMDs because the absolute magnitudes or luminosities are tightly linked to stellar mass. Therefore, colour--stellar mass diagrams present similar and non-negligible contamination of DSF galaxies in the red part since they are not able to disentangle the reddening by dust. Other works \citep[e.g.][]{Noeske2007,Whitaker2012,Moustakas2013,Fumagalli2014} suggest introducing star formation rates (SFRs) along with stellar masses as a criterion to separate quiescent and star-forming galaxies. The disadvantage of this diagram is that we need a previous estimation of the mentioned stellar population parameters, which in many cases is not available or is hard to calculate. In some cases, SFRs and specific star formation rates (sSFRs, defined as the star formation rate per unit of stellar mass, i.e.~$SFR/M_\star$) can   also be used as an additional (or the unique) criterion to select quiescent galaxies below a pre-defined threshold \citep[e.g.][]{Ilbert2010,Pozzetti2010,DominguezSanchez2011,DominguezSanchez2016}.

Recent studies show that some stellar population parameters are related to the range of colours that galaxies occupy in $UVJ$ colour--colour  diagrams \citep{Belli2015,DominguezSanchez2016,Martis2016,Pacifici2016,Yano2016,Fang2017}. There is a general agreement amongst the results, which indicates  that certain ranges of SFR and sSFR values are well constrained in precise colour ranges within $UVJ$ diagrams. This means that galaxies in the typical colour ranges enclosing quiescent galaxies exhibit low sSFR values, whereas the higher sSFR values lie on the lower parts of this kind of diagrams. There is evidence to support a continuous sSFR gradient perpendicular to the empirical colour limit separating quiescent from star-forming galaxies in $UVJ$ diagrams \citep[see e.g.][]{Belli2015,Fang2017}. In addition, and as expected from extinction laws, the most dust-reddened galaxies  populate the reddest parts of these diagrams \citep[see e.g.][]{Martis2016,Fang2017}. This suggests that rest-frame $UVJ$ diagrams can be used as stellar population estimators to predict, or at least constrain, the stellar population properties of galaxies. However, this potential idea, especially for quiescent galaxies, has not been extensively exploited in detail yet. Usually stellar populations are only plotted on these diagrams with the unique purpose of a sanity check. Large-scale multi-filter surveys can provide a large number of galaxies to largely populate the full range of colours for any kind of rest-frame colour--colour diagrams at wide redshift ranges and stellar masses. 

In this series of papers, our aim is to improve our knowledge of the evolution of quiescent galaxies since $z\sim1$ through the use of spectrophotometric data from the Advanced Large Homogeneous Area Medium Band Redshift Astronomical survey \citep[ALHAMBRA;][]{Moles2008}. In this second paper in the series we focus on the distribution of the stellar population parameters of quiescent galaxies within the $UVJ$ diagram. For the first time, we dissected the loci of age, metallicity, extinction, and sSFR simultaneously for a complete and numerous sample of quiescent galaxies from the ALHAMBRA survey. In addition, we extended this study to the stellar mass--colour diagram, which illustrates the influence of stellar mass more clearly and complements the reliability around the selection of quiescent galaxies. 

This paper is organised as follows.  In Sect.~\ref{sec:data_sp} we briefly present the main features of the ALHAMBRA survey. In Sect.~\ref{sec:methods}, we summarise the SED fitting methodology embedded in MUFFIT\footnote{MUlti-Filter FITting for stellar population diagnostics} and the set of simple stellar population (SSP) models used to constrain the stellar populations of galaxies. Section~\ref{sec:sample} details the process to build a reliable sample of quiescent galaxies, and to determine the stellar mass completeness of the sample. The techniques used to obtain SFRs, and therefore sSFRs, from the SED fitting results provided by MUFFIT are shown in Sect.~\ref{sec:dusty}. The main result of this work, which is the distribution of the stellar population parameters of quiescent galaxies within stellar mass--colour and colour--colour rest-frame diagrams, is given in Sect.~\ref{sec:uvj}. Implications from our results are discussed in Sect.~\ref{sec:discussion}. The conclusions of this work are summarised in Sect.~\ref{sec:conclusions}.

Throughout this paper we adopted a Lambda cold dark matter ($\Lambda$CDM) cosmology with $H_0 = 71$~km~s$^{-1}$, $\Omega_\mathrm{M}=0.27$, and $\Omega_\mathrm{\Lambda}=0.73$. All magnitudes are in the AB-system \citep{Oke1983}. Stellar masses are quoted in solar mass units $[M_\sun]$. In this work, we assumed a \citet{Chabrier2003} and Kroupa Universal \citep[][]{Kroupa2001} initial stellar mass functions (IMF, more details in Sect.~\ref{sec:methods}).


\section{Photometric data from the ALHAMBRA survey}\label{sec:data_sp}

The ALHAMBRA survey\footnote{\url{http://www.alhambrasurvey.com}} provides flux in $23$ photometric bands\footnote{\url{http://svo2.cab.inta-csic.es/theory/fps3/}} in AB-magnitudes \citep[corrected for point spread function,  PSF;][]{Coe2006}, $20$ in the optical range $\lambda\lambda\ 3500$--$9700$~\AA\ \citep[][top hat medium-band filters,  $\sim300$~\AA\ width, overlapping close to zero between contiguous bands]{Aparicio2010} and $3$ in the near-infrared (NIR) spectral window $\lambda\lambda\ 1.0$--$2.3$~{$\mu$m} ($J$, $H$, and $K_\mathrm{s}$ bands), for each source in seven non-contiguous fields along the northern hemisphere. The current effective area is $\sim2.8$~deg$^2$ and the total on-target exposure time is $\sim700$~h ($\sim608$~h for bands in the optical range, and $\sim92$~h for the NIR). Magnitude limits for optical bands ($5\sigma$ level and $3\arcsec$ aperture) range from $m_\mathrm{AB}\sim24$ for the $14$ bluer filters, decreasing towards the red, reaching $m_\mathrm{AB}\sim22$ in the reddest one \citep{Molino2014}. For the NIR range, magnitude limits are equal to $J\sim22.4$, $H\sim21.3$, and $K_\mathrm{s}\sim20.0$ \citep[][$50$\% of recovery efficiency depth and point-like sources]{Cristobal2009}. The observations in the optical range were carried out with the wide-field camera LAICA\footnote{\url{http://www.caha.es/CAHA/Instruments/LAICA}} (four\ CCDs of $4096 \times 4096$ pixels and pixel scale $0.225\arcsec~\mathrm{pixel^{-1}}$), whereas in the NIR regime they were performed with the Omega-$2000$\footnote{\url{http://www.caha.es/CAHA/Instruments/O2000}} (one CCD of $2048 \times 2048$ pixels and plate scale $0.45\arcsec~ \mathrm{pixel^{-1}}$), both at the $3.5$~m telescope in the Calar Alto Observatory\footnote{\url{http://www.caha.es}} (CAHA). 

We adopted the ALHAMBRA Gold catalogue\footnote{\url{http://cosmo.iaa.es/content/alhambra-gold-catalog}} \citep{Molino2014} as the reference photometric catalogue. This catalogue provides accurate photometry (non-fixed apertures) for developing stellar population studies of galaxies \citep[][]{DiazGarcia2015} for $\sim95\,000$ galaxies. This catalogue also contains precise photo-$z$ predictions ($\sigma_z \sim 0.012$) obtained from the Bayesian Photometric Redshift \citep[BPZ2.0, ][]{Benitez2000,Molino2014}. Our galaxy set comprises all the sources in the Gold catalogue classified as galaxies (STAR/GALAXY discriminator parameter lower than \texttt{Stellar\_flag} $\le 0.5$) and with a minimum photometric weight of $70$\% on the detection image (\texttt{PercW} $\ge 0.7$). This constraint removes sources that are close to the image edges. This catalogue also provides the synthetic AB-magnitude $m_{F814W}$, which is mainly employed for detection purposes. The synthetic band $m_{F814W}$ was not used during the SED fitting process and it sets the Gold catalogue selection as $m_{F814W} \le 23$ \citep[see][]{Molino2014}.


\section{Stellar population parameters in ALHAMBRA}\label{sec:methods}

Throughout this work we focus on the distribution of the stellar population parameters of quiescent galaxies within the stellar mass--colour and colour--colour diagrams. To explore this topic, we employed the photometric data of each galaxy in ALHAMBRA and the code MUFFIT \citep[MUlti-Filter FITting for stellar population diagnostics;][]{DiazGarcia2015}. Briefly, MUFFIT is a SED fitting code carefully developed to set constraints on stellar population parameters of galaxies and it is specifically optimised to deal with multi-band photometric data. MUFFIT is based on an error-weighted $\chi^2$-test using composite models of stellar populations (mixtures of two SSPs) and it has proved to be  a powerful tool with high capabilities for constraining stellar population properties of galaxies from ALHAMBRA-like surveys \citep[see e.g.][]{DiazGarcia2015}.

As our analysis is based on galaxy colours, we included the iterative process to remove bands affected by strong emission lines, meaning all the stellar population properties reported in this work were obtained from the stellar continuum even when strong nebular or AGN emission lines are present. Moreover, MUFFIT performs Monte Carlo simulations to explore errors in the stellar population properties owing to degeneracies and photon-noise uncertainties. As a result of the SED fitting analysis  performed by MUFFIT, we obtained age and metallicity (both luminosity- and mass-weighted), photo $z$ (treated as another free parameter within the $1\sigma$ confidence level reported in the Gold catalogue), stellar mass, rest-frame luminosities, and extinction for all the galaxies in ALHAMBRA.

For the analysis, we separately built two sets of composite stellar population models from two sets of SSP models to assess potential systematics from the use of a given population synthesis model. Firstly, we selected the \citet{Bruzual2003} SSP models (hereafter BC03; Padova $1994$ tracks, ages from $0.06$ to $14$~Gyr, and metallicities $[\mathrm{M/H}]=-1.65$, $-0.64$, $-0.33$, $0.09$, $0.55$) with a \citet{Chabrier2003} IMF. The BC03 spectral coverage, $\lambda\lambda~91$~\AA--$160$~$\mu\mathrm{m}$, allowed us to perform our analysis in an extensive redshift range. This is relevant for this  SED fitting analysis because the population of galaxies in ALHAMBRA easily extends up to redshift $z\sim1.5$. Secondly, the other set of models comprises the \citet[][EMILES\footnote{\url{http://miles.iac.es}}] {Vazdekis2016} SSP models, which includes empirical stellar spectra and is an extension of the \citet[][MIUSCAT]{Vazdekis2003,Vazdekis2010,Vazdekis2012} SSP models, where the spectral coverage extends up to $\lambda\lambda~1\,680$~\AA--$5$~{$\mu\mathrm{m}$}.  It is noteworthy that this set includes the optical MIUSCAT SSP predictions \citep{Vazdekis2012,Ricciardelli2012}, the NIR predictions of MIUSCATIR \citep{Rock2015}, and the UV extension from the New Generation Stellar Library \citep[NGSL][]{Koleva2012}. This updated version of the models also includes two sets of theoretical isochrones: the scaled-solar isochrones of \citet[][hereafter Padova00]{Girardi2000} and \citet[][BaSTI in the following]{Pietrinferni2004}, which were taken into account for the analysis. In particular, for EMILES we took $22$ ages in the range $0.05$--$14$~Gyr, and metallicities $[\mathrm{M/H}]=-1.31$, $-0.71$, $-0.40$, $0.00$, $0.22$ for Padova00 and $[\mathrm{M/H}]=-1.26$, $-0.96$, $-0.66$, $-0.35$, $0.06$, $0.26$, $0.40$ for BaSTI, both with a \citet{Kroupa2001} Universal IMF.

Previously to the SED fitting analysis, the photometry of the ALHAMBRA galaxies was also corrected for Milky Way dust using MUFFIT, the \citet{Schlegel1998} dust maps, and the extinction law of \citet{Fitzpatrick1999}. This extinction law covers a wider spectral range than that observed in ALHAMBRA since $z\sim2$, and it is suitable for dereddening any photospectroscopic data. In addition, \citet[][]{Fitzpatrick1999} provides robust estimations of uncertainties for dereddening methodologies, which were also included in the error budget during the photometric analysis. Along this work, we only use SSP models with ages that are cosmologically relevant to build the sample of composite models of stellar populations; in other words, they cannot be much older than the age of the Universe at any redshift assuming a $\Lambda$CDM cosmology. However, this constraint on age did not alter our results significantly. For the whole set of SSP models, extinctions were added as a foreground screen\footnote{Throughout this work there is no distinction between extinction and attenuation law.} with values in the range $A_V=0.0$--$3.1$ (assuming a constant $R_V=3.1$ and the same dust attenuation in each of the composite model components), also following the extinction law of \citet{Fitzpatrick1999} for consistency.

As there are discrepancies among the CCDs of the LAICA camera, each galaxy was analysed according to the precise photometric system or CCD in which it was imaged. For the Omega-2000 camera, this process is not necessary because it only contains a unique CCD. Throughout this work, the mass-weighted age and metallicity ($\mathrm{Age_M}$ and $\mathrm{[M/H]_M}$, respectively) were preferred to the luminosity values. The mass-weighted parameters are more representative of the total stellar content of the galaxy and they are not linked to a definition of luminosity weight, although luminosity-weighted parameters were also estimated. A young population may dominate the luminosity of a galaxy, and consequently its luminosity-weighted age, even when its contribution in mass is very low \citep{Trager2000,Conroy2013,Vazdekis2016}.

\begin{sidewaystable*}
\caption{Stellar population parameters of the sample of quiescent galaxies using BC03 SSP models. From left to right: ALHAMBRA source ID, right ascension, declination, magnitude in the detection band, photometric redshift, stellar mass, extinction, mass-weighted age and metallicity, $m_{F365}-m_{F551}$ and $m_{F551}-J$ rest-frame colours and their values after dust correction (intrinsic colours), and average signal-to-noise ratio. The parameter uncertainties are attached below each quantity.} \label{tab:ptable}
\centering
\begin{tabular}{cccccccccccccc}
\hline\hline
$\mathrm{Source ID}$ & $\mathrm{RA}$ & $\mathrm{DEC}$ & $m_{F814W}$ & $z$ & $\log_{10} M_\star$ & $A_V$ & $\mathrm{Age_M}$ & $\mathrm{[M/H]_M}$ & {\small$m_{F365}-m_{F551}$} & {\small$m_{F551}-J$} & {\small$(m_{F365}-m_{F551})_\mathrm{int}$} & {\small$(m_{F551}-J)_\mathrm{int}$} & $<S/N>$ \\ 
$814-$ & $\mathrm{[deg]}$ & $\mathrm{[deg]}$ & $\mathrm{[AB]}$ &  & $\mathrm{[M_\odot]}$ & $\mathrm{[AB]}$ & $\mathrm{[Gyr]}$ & $\mathrm{[dex]}$ & $\mathrm{[AB]}$ & $\mathrm{[AB]}$ & $\mathrm{[AB]}$ & $\mathrm{[AB]}$ &  \\ 
\hline
$62402442$ & $214.4301$ & $52.3397$ & $16.71$ & $0.119$ & $10.66$ & $0.27$ & $4.5$ & $-0.24$ & $1.88$ & $1.24$ & $1.73$ & $1.09$ & $40.64$ \\ 
 & & & $ \pm 0.02$ & $ \pm 0.003$ & $ \pm 0.05$ & $ \pm 0.07$ & $ \pm 1.3$ & $ \pm 0.15$ & $ \pm 0.04$ & $ \pm 0.02$ & $ \pm 0.04$ & $ \pm 0.05$ & \\ 
$62300829$ & $213.5037$ & $52.3945$ & $16.53$ & $0.100$ & $10.64$ & $0.08$ & $6.0$ & $-0.05$ & $1.91$ & $1.20$ & $1.87$ & $1.16$ & $39.73$ \\ 
 & & & $ \pm 0.02$ & $ <0.001$ & $ \pm 0.08$ & $ \pm 0.08$ & $ \pm 2.4$ & $ \pm 0.17$ & $ \pm 0.02$ & $ \pm 0.02$ & $ \pm 0.05$ & $ \pm 0.05$ & \\ 
$74405663$ & $243.0770$ & $54.1213$ & $17.10$ & $0.119$ & $10.55$ & $0.15$ & $4.1$ & $-0.20$ & $1.78$ & $1.19$ & $1.68$ & $1.10$ & $39.52$ \\ 
 & & & $ \pm 0.02$ & $ \pm 0.003$ & $ \pm 0.04$ & $ \pm 0.16$ & $ \pm 0.5$ & $ \pm 0.10$ & $ \pm 0.04$ & $ \pm 0.02$ & $ \pm 0.20$ & $ \pm 0.09$ & \\ 
$22202797$ & $36.8434$ & $1.1945$ & $17.31$ & $0.104$ & $10.33$ & $0.10$ & $5.2$ & $-0.15$ & $1.86$ & $1.14$ & $1.80$ & $1.09$ & $38.77$ \\ 
 & & & $ \pm 0.02$ & $ \pm 0.007$ & $ \pm 0.10$ & $ \pm 0.08$ & $ \pm 1.9$ & $ \pm 0.20$ & $ \pm 0.06$ & $ \pm 0.03$ & $ \pm 0.06$ & $ \pm 0.05$ & \\ 
$74103404$ & $243.0550$ & $54.6608$ & $17.14$ & $0.230$ & $11.31$ & $0.02$ & $5.4$ & $0.12$ & $1.89$ & $1.24$ & $1.87$ & $1.22$ & $38.69$ \\ 
 & & & $ \pm 0.02$ & $ <0.001$ & $ \pm 0.04$ & $ \pm 0.05$ & $ \pm 0.9$ & $ \pm 0.16$ & $ \pm 0.02$ & $ \pm 0.02$ & $ \pm 0.04$ & $ \pm 0.03$ & \\ 
$21205253$ & $37.0285$ & $1.1497$ & $17.37$ & $0.136$ & $10.66$ & $0.07$ & $7.8$ & $-0.05$ & $1.96$ & $1.18$ & $1.93$ & $1.14$ & $38.66$ \\ 
 & & & $ \pm 0.02$ & $ \pm 0.005$ & $ \pm 0.08$ & $ \pm 0.09$ & $ \pm 2.0$ & $ \pm 0.16$ & $ \pm 0.04$ & $ \pm 0.03$ & $ \pm 0.05$ & $ \pm 0.06$ & \\ 
$74401062$ & $243.1835$ & $54.2408$ & $18.02$ & $0.140$ & $10.33$ & $0.13$ & $4.8$ & $-0.12$ & $1.86$ & $1.12$ & $1.79$ & $1.05$ & $38.64$ \\ 
 & & & $ \pm 0.02$ & $ \pm 0.001$ & $ \pm 0.06$ & $ \pm 0.09$ & $ \pm 1.5$ & $ \pm 0.13$ & $ \pm 0.02$ & $ \pm 0.02$ & $ \pm 0.05$ & $ \pm 0.06$ & \\ 
$74405753$ & $243.1381$ & $54.1192$ & $17.97$ & $0.151$ & $10.47$ & $0.11$ & $5.0$ & $0.03$ & $1.94$ & $1.23$ & $1.88$ & $1.17$ & $38.61$ \\ 
 & & & $ \pm 0.02$ & $ \pm 0.004$ & $ \pm 0.06$ & $ \pm 0.10$ & $ \pm 1.6$ & $ \pm 0.13$ & $ \pm 0.03$ & $ \pm 0.02$ & $ \pm 0.06$ & $ \pm 0.06$ & \\ 
$74406006$ & $243.2873$ & $54.1121$ & $17.14$ & $0.144$ & $10.80$ & $0.13$ & $7.1$ & $0.02$ & $1.99$ & $1.23$ & $1.92$ & $1.16$ & $38.59$ \\ 
 & & & $ \pm 0.02$ & $ \pm 0.006$ & $ \pm 0.10$ & $ \pm 0.11$ & $ \pm 2.2$ & $ \pm 0.15$ & $ \pm 0.05$ & $ \pm 0.03$ & $ \pm 0.07$ & $ \pm 0.07$ & \\ 
$51406997$ & $189.5499$ & $61.7978$ & $16.77$ & $0.108$ & $10.59$ & $0.15$ & $6.7$ & $-0.04$ & $1.97$ & $1.20$ & $1.88$ & $1.11$ & $38.57$ \\ 
 & & & $ \pm 0.02$ & $ \pm 0.007$ & $ \pm 0.09$ & $ \pm 0.10$ & $ \pm 2.4$ & $ \pm 0.14$ & $ \pm 0.06$ & $ \pm 0.03$ & $ \pm 0.07$ & $ \pm 0.07$ & \\ 
$74101458$ & $243.4018$ & $54.7131$ & $17.64$ & $0.140$ & $10.58$ & $0.02$ & $5.3$ & $-0.22$ & $1.74$ & $1.18$ & $1.73$ & $1.17$ & $38.41$ \\ 
 & & & $ \pm 0.02$ & $ <0.001$ & $ \pm 0.03$ & $ \pm 0.04$ & $ \pm 0.8$ & $ \pm 0.14$ & $ \pm 0.02$ & $ \pm 0.02$ & $ \pm 0.02$ & $ \pm 0.03$ & \\ 
$51210550$ & $188.7010$ & $62.2081$ & $17.53$ & $0.110$ & $10.26$ & $0.02$ & $5.3$ & $-0.19$ & $1.79$ & $1.14$ & $1.78$ & $1.13$ & $38.31$ \\ 
 & & & $ \pm 0.02$ & $ \pm 0.002$ & $ \pm 0.05$ & $ \pm 0.04$ & $ \pm 1.1$ & $ \pm 0.23$ & $ \pm 0.04$ & $ \pm 0.02$ & $ \pm 0.04$ & $ \pm 0.03$ & \\ 
$74104138$ & $243.0991$ & $54.6405$ & $17.40$ & $0.206$ & $11.16$ & $0.11$ & $7.2$ & $0.07$ & $2.04$ & $1.21$ & $1.98$ & $1.15$ & $38.30$ \\ 
 & & & $ \pm 0.02$ & $ \pm 0.006$ & $ \pm 0.07$ & $ \pm 0.11$ & $ \pm 2.2$ & $ \pm 0.10$ & $ \pm 0.05$ & $ \pm 0.03$ & $ \pm 0.07$ & $ \pm 0.06$ & \\ 
$41401013$ & $150.4250$ & $2.0662$ & $15.98$ & $0.133$ & $11.02$ & $0.17$ & $4.1$ & $0.13$ & $1.88$ & $1.26$ & $1.79$ & $1.16$ & $38.26$ \\ 
 & & & $ \pm 0.02$ & $ \pm 0.006$ & $ \pm 0.11$ & $ \pm 0.07$ & $ \pm 2.4$ & $ \pm 0.19$ & $ \pm 0.04$ & $ \pm 0.02$ & $ \pm 0.05$ & $ \pm 0.05$ & \\ 
$74406197$ & $243.3334$ & $54.1071$ & $17.96$ & $0.156$ & $10.48$ & $0.06$ & $4.6$ & $0.10$ & $1.88$ & $1.19$ & $1.85$ & $1.16$ & $38.25$ \\ 
 & & & $ \pm 0.02$ & $ \pm 0.005$ & $ \pm 0.07$ & $ \pm 0.08$ & $ \pm 1.1$ & $ \pm 0.16$ & $ \pm 0.05$ & $ \pm 0.03$ & $ \pm 0.04$ & $ \pm 0.05$ & \\ 
$73400609$ & $243.4698$ & $54.2539$ & $17.54$ & $0.142$ & $10.64$ & $0.04$ & $7.0$ & $-0.11$ & $1.92$ & $1.18$ & $1.91$ & $1.16$ & $38.12$ \\ 
 & & & $ \pm 0.02$ & $ \pm 0.005$ & $ \pm 0.06$ & $ \pm 0.07$ & $ \pm 1.8$ & $ \pm 0.12$ & $ \pm 0.05$ & $ \pm 0.04$ & $ \pm 0.04$ & $ \pm 0.04$ & \\ 
$74408909$ & $243.4100$ & $54.0274$ & $16.64$ & $0.145$ & $11.03$ & $0.06$ & $7.6$ & $-0.18$ & $1.92$ & $1.17$ & $1.88$ & $1.14$ & $38.06$ \\ 
 & & & $ \pm 0.02$ & $ \pm 0.006$ & $ \pm 0.07$ & $ \pm 0.09$ & $ \pm 1.9$ & $ \pm 0.13$ & $ \pm 0.06$ & $ \pm 0.04$ & $ \pm 0.06$ & $ \pm 0.07$ & \\ 
$74104674$ & $243.3939$ & $54.6296$ & $18.24$ & $0.220$ & $10.73$ & $0.05$ & $5.1$ & $0.08$ & $1.93$ & $1.12$ & $1.90$ & $1.10$ & $38.04$ \\ 
 & & & $ \pm 0.02$ & $ \pm 0.001$ & $ \pm 0.10$ & $ \pm 0.08$ & $ \pm 2.0$ & $ \pm 0.12$ & $ \pm 0.02$ & $ \pm 0.02$ & $ \pm 0.06$ & $ \pm 0.05$ & \\ 
$22102185$ & $37.3262$ & $1.2085$ & $18.41$ & $0.250$ & $10.74$ & $0.12$ & $4.0$ & $0.16$ & $1.96$ & $1.22$ & $1.89$ & $1.15$ & $38.00$ \\ 
 & & & $ \pm 0.02$ & $ \pm 0.004$ & $ \pm 0.08$ & $ \pm 0.10$ & $ \pm 1.5$ & $ \pm 0.13$ & $ \pm 0.04$ & $ \pm 0.03$ & $ \pm 0.06$ & $ \pm 0.06$ & \\ 
$74108616$ & $243.0808$ & $54.5163$ & $18.12$ & $0.120$ & $10.23$ & $0.02$ & $6.1$ & $-0.43$ & $1.71$ & $1.12$ & $1.70$ & $1.10$ & $37.98$ \\ 
 & & & $ \pm 0.02$ & $ \pm 0.001$ & $ \pm 0.03$ & $ \pm 0.04$ & $ \pm 0.8$ & $ \pm 0.11$ & $ \pm 0.02$ & $ \pm 0.02$ & $ \pm 0.03$ & $ \pm 0.04$ & \\ 
\end{tabular}
\tablefoot{The full version of this table is available in electronic form at the CDS via anonymous ftp to \url{cdsarc.u-strasbg.fr} (130.79.128.5) or via \url{http://cdsweb.u-strasbg.fr/cgi-bin/qcat?J/A+A/}. Only a portion of the table is shown in the printed version for illustrating purposes.}
\end{sidewaystable*}

Table~\ref{tab:ptable} illustrates part of the stellar population parameters and uncertainties derived by MUFFIT using BC03 for a subset of quiescent galaxies (see Sect.~\ref{sec:sample} below) used throughout this research. The complete catalogue\footnote{See also \url{http://archive.cefca.es/catalogues/alhambra_quiescent_galaxies-dr1}} with additional data will be available at the CDS\footnote{\url{http://cdsweb.u-strasbg.fr/cgi-bin/qcat?J/A+A/}}.


\section{Definition of the quiescent sample}\label{sec:sample}

The definition of a reliable sample of quiescent galaxies with low levels of contamination is a sensitive and tricky process because both DSF galaxies and cool stars in ground-based surveys present colours similar to those of the quiescent galaxy population. The contamination of these sources can represent a substantial part of the sample at certain redshift and mass ranges and their effects should be removed or minimised (see Sects.~\ref{sec:sample_uvj}--\ref{sec:stars}). Even in multi-filter photometric surveys, which are not biased by selection effects other than the photometric depth, the definition of a complete sample in stellar mass (Sect.~\ref{sec:completeness}) with accurate enough photo-$z$ predictions (Sect.~\ref{sec:photoz}) is key to reliably driving this study because the stellar mass is tightly related to the star formation history of each galaxy \citep[e.g.][]{Cowie1996,Trager2000,Tremonti2004,Gallazzi2005,Gallazzi2006,Jimenez2007,Panter2008,GonzalezDelgado2014a,GonzalezDelgado2014b}. 


\subsection{Dust-corrected $UVJ$ diagram}\label{sec:sample_uvj}

In this section  we present how  to diminish the contamination of DSF galaxies in our sample of quiescent galaxies using a $UVJ$ diagram. Otherwise, the sample would contain a subset of younger and obscured galaxies that differs from the largely evolved quiescent galaxies. Instead of using the rest-frame $U-V$ and $V-J$ colours  to build the $UVJ$-diagram, we took the bands $F365W$, $F551W$, and $J$ from ALHAMBRA to compute the rest-frame colours $m_{F365}-m_{F551}$ and $m_{F551}-J$. We note that the effective wavelengths of these ALHAMBRA bands are the closest ones to $U$, $V$, and $J$. To make a reliable sample selection, we took advantage of the stellar population results provided by MUFFIT. We used both the $k$-corrected magnitudes, along with the extinction values provided by MUFFIT to build a new and improved version of the $UVJ$ diagram free of dust effects, which allowed us to clean the quiescent sample of obscured star-forming galaxies. In brief, the $k$-corrected magnitudes are computed from the SED fitting analysis by convolving the ALHAMBRA photometric system with the same combination of rest-frame spectroscopic SSP models (in a general case age, metallicity, IMF, overabundances, stellar mass, and the weight of each SSP component in the mixture) that reproduces the SED of each galaxy. The uncertainties in the $k$-corrected magnitudes are also included in the analysis and they were obtained using the results from the Monte Carlo approach \citep[for further details, see][]{DiazGarcia2015}. From these mixtures, it is also straightforward to rebuild the same combination of rest-frame models with null extinction ($A_V = 0.0$), and therefore corrected for extinction.

\begin{figure*}
\centering
\includegraphics[trim=0 3mm 0 1mm,width=17cm,clip=True]{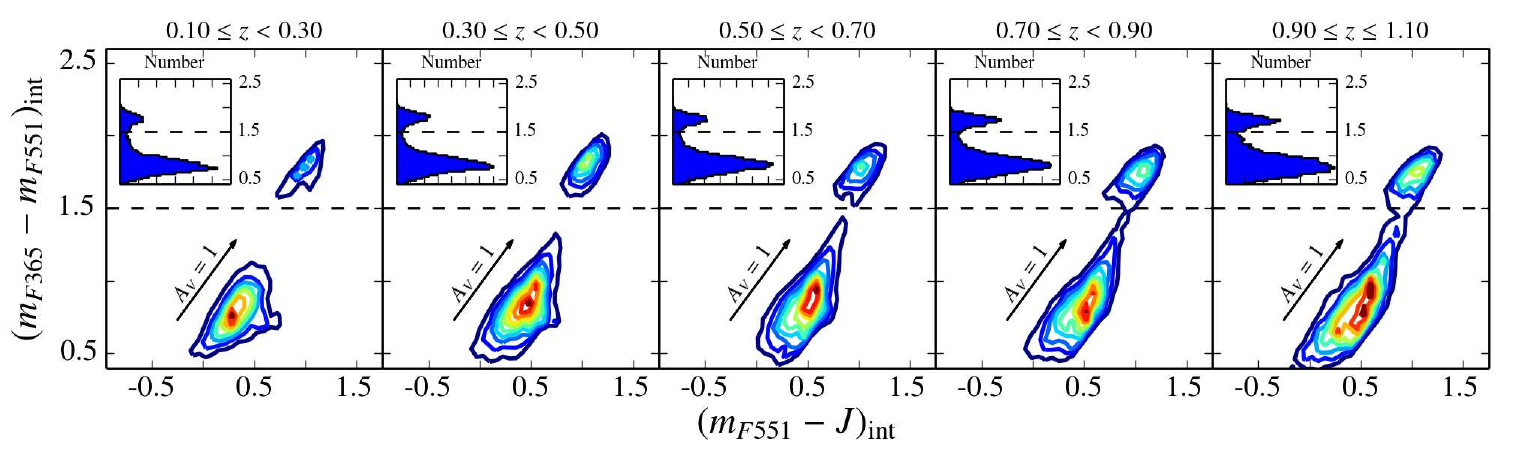}
\includegraphics[trim=0 3mm 0 1mm,width=17cm,clip=True]{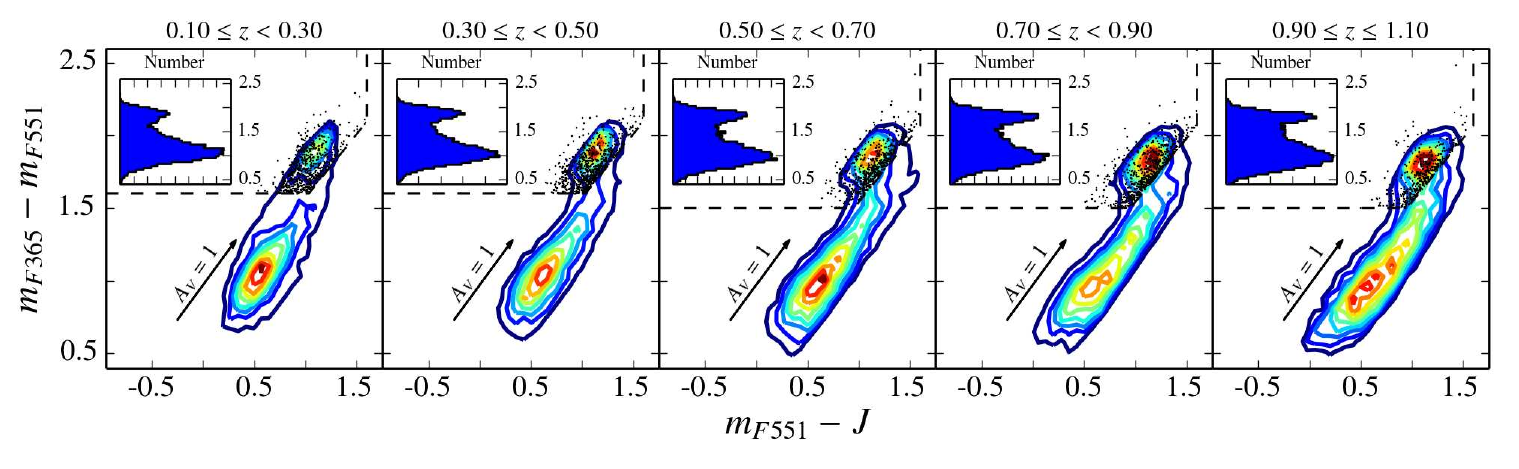}
\caption{Density surface and distribution of rest-frame colours for galaxies from the ALHAMBRA survey obtained using BC03 SSP models. Top: Rest-frame intrinsic colours $(m_{F551}-J)_\mathrm{int}$ ($X$-axis) and $(m_{F365}-m_{F551})_\mathrm{int}$ ($Y$-axis) after correcting for extinction at different redshifts. Bottom: Rest-frame colours without removing dust effects. Redder (bluer) density-curve colours are related to higher (lower) densities. Inner panels, histograms of the intrinsic (top) and observed (bottom) rest-frame colour $m_{F365}-m_{F551}$. Dashed lines in the top panels illustrate our limiting value $(m_{F365}-m_{F551})_\mathrm{int}=1.5$ for quiescent galaxies, and in the bottom panels the quiescent $UVJ$ sample defined by \citet[][Eq.~(\ref{eq:quiescent})]{Moresco2013}. Black dots are galaxies labelled as quiescent by the $UVJ$ criteria of \citet{Moresco2013} that lie in the star-forming region after removing extinction effects. We illustrate the colour variations owing to a reddening of $A_V=1$ (black arrow), assuming the extinction law of \citet{Fitzpatrick1999}.}
\label{fig:ccd_noav}
\end{figure*}

In Fig.~\ref{fig:ccd_noav} we present the $UVJ$ diagram corrected for reddening obtained for ALHAMBRA with the bands previously mentioned, which includes all the galaxies at $0.1 \le z \le 1.1$ and down to $m_{F814W}=23$. By looking at the distribution of the rest-frame intrinsic colour $(m_{F365}-m_{F551})_\mathrm{int}$ (see insets in Fig.~\ref{fig:ccd_noav}), we can easily set the limit for quiescent galaxies as $(m_{F365}-m_{F551})_\mathrm{int} \ge 1.5$, which is roughly constant with redshift up to $z\sim1.1$. Although this colour limit is not strictly located in the minimum between the red and blue peaks (corresponding to the quiescent and star-forming populations, respectively), its value was defined to agree with the limit established in \citet[][see Eq.~(\ref{eq:quiescent}) below]{Moresco2013} and to be slightly higher  to avoid the now poorly populated green valley, whose limits are difficult to define due to the small number of sources. After the definition of a quiescent sample complete in stellar mass (Sect.~\ref{sec:completeness}), the sample remains almost unaltered. For comparison reasons, we present the same $UVJ$ diagram without the extinction correction  in Fig.~\ref{fig:ccd_noav}. The range of colours defined by \citet{Moresco2013} to select quiescent galaxies (see Eq.~(\ref{eq:quiescent})) is also shown in this diagram, which is less contaminated by obscured star-forming galaxies than the range originally proposed by \citet{Williams2009}. Formally, 

\begin{equation}\label{eq:quiescent}
\left\lbrace
\begin{array}{ll}
(m_{F365}-m_{F551}) >& 0.88 \times (m_{F551}-J)+0.69,\ \mathrm{for}\ z \le 0.5 \\
(m_{F365}-m_{F551}) >& 0.88 \times (m_{F551}-J)+0.66,\ \mathrm{for}\ z > 0.5 \\
\end{array}
\right.
,\end{equation}
where $m_{F365}-m_{F551} > 1.6\ (m_{F365}-m_{F551}>1.5)$ at $z \le 0.5\ (z>0.5)$ and $m_{F551}-J < 1.6$, all quantities at  rest frame and in AB-magnitudes.

\begin{figure*}
\centering
\includegraphics[trim= 0 17.1mm 0 0,width=17cm,clip=True]{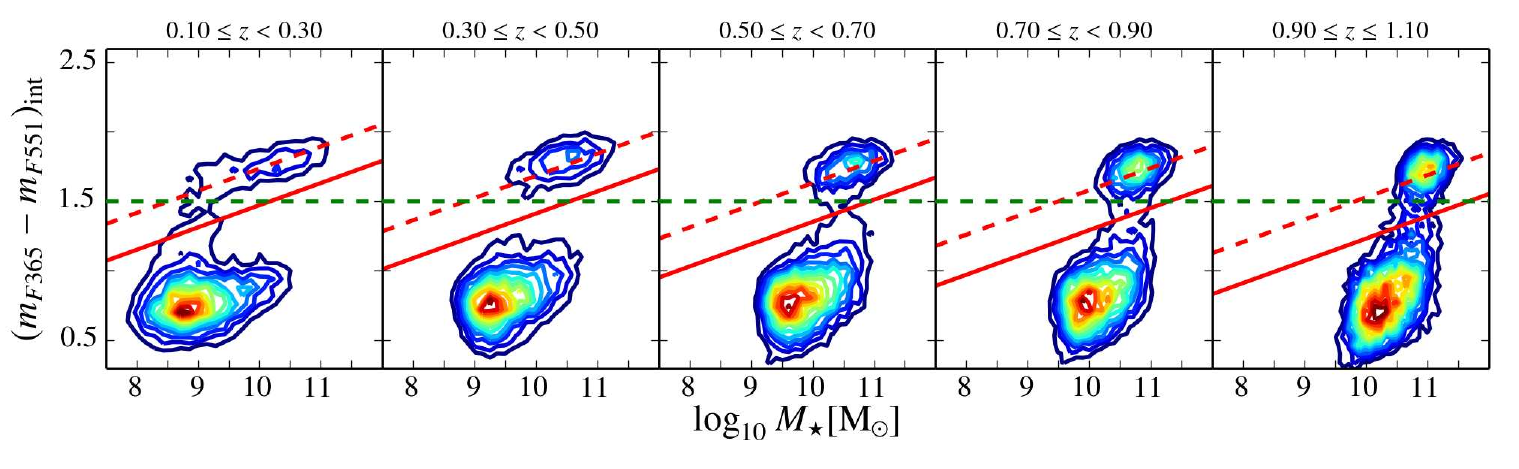}
\includegraphics[trim= 0 0 0 8mm,width=17cm,clip=True]{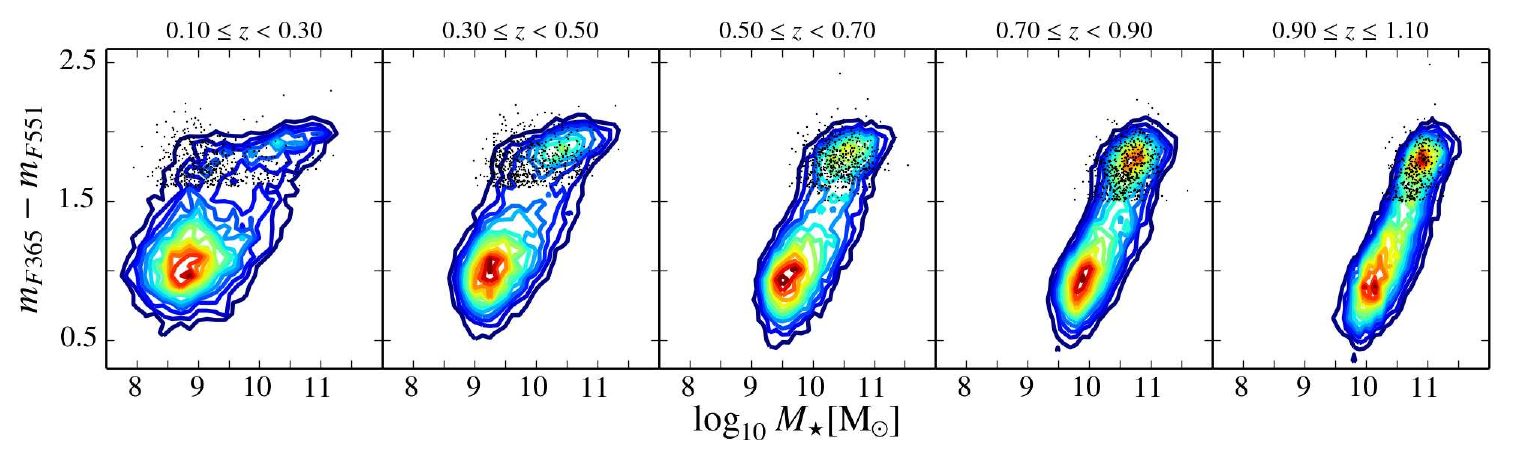}
\caption{Density surface and distribution of stellar mass vs. rest-frame colour $(m_{F365}-m_{F551})$ for galaxies from the ALHAMBRA survey using BC03 SSP models. Top: Rest-frame intrinsic colour $(m_{F365}-m_{F551})_\mathrm{int}$ ($Y$-axis) after correcting for extinction at different redshift. Bottom: Rest-frame colours without removing dust effects. Redder (bluer) density-curve colours are related to high (low) densities. Dashed green lines in the top panels illustrate the limiting value $(m_{F365}-m_{F551})_\mathrm{int}=1.5$ used for selecting quiescent galaxies in Sect.~\ref{sec:sample_uvj}. Dashed and solid red lines show the fit to the quiescent sequence, $(m_{F365}-m_{F551})_\mathrm{int}^\mathrm{Q}$, and the limiting intrinsic colours, $(m_{F365}-m_{F551})_\mathrm{int}^\mathrm{lim}$, values of quiescent galaxies respectively (see details in Sect.~\ref{sec:mcd}). Black dots in the bottom panels are galaxies labelled as quiescent with the $UVJ$ criteria of \citet{Moresco2013} that lie in the star-forming region after removing extinction effects.}
\label{fig:CMDE}
\end{figure*}

Independently of the SSP model set used, BC03 and EMILES, the extinction correction applied on the $UVJ$ colours yielded striking results. Firstly, and as expected, removing dust effects makes  the colour bimodality of galaxies clearer in the $UVJ$ diagram (see Fig.~\ref{fig:ccd_noav}). This occurs because a great part of the galaxies in the green valley (the bridge between the red and blue galaxies) are obscured star-forming galaxies. If we define the green valley as those galaxies whose colours are close to the limiting values expressed by Eq.~(\ref{eq:quiescent}), more precisely $\pm 0.1$~mag, we find that a $\sim30$~\% fraction of the galaxies are actually obscured star-forming galaxies. For other definitions of the green valley, this amount can be even higher. Their intrinsic colour $(m_{F365}-m_{F551})_\mathrm{int}$ reveals that these galaxies really lie within the star-forming population, although they also present a large dust content that reddens their observed colours \citep[][obtained a similar result for $U-V$]{Brammer2009,Whitaker2010}. This means that the green valley is largely depopulated after accounting for extinction \citep[Fig.~\ref{fig:ccd_noav}; see also][]{Bell2005,Cowie2008,Brammer2009,Cardamone2010}.

Secondly, and for a sample complete down to $m_{F814W}=23$, the histogram of the $(m_{F365}-m_{F551})_\mathrm{int}$ colour exhibits a local minimum at $\sim 1.45$, which can be imposed as the bluest colour limit to fairly select quiescent galaxies in ALHAMBRA. This limit to separate quiescent from star-forming galaxies also remains roughly constant since $z \le 1.1$, and it does not present any remarkable evolution. In spite of the presence of a bridge between the bulk of intrinsic red and blue galaxies at $(m_{F365}-m_{F551})_\mathrm{int} \sim 1.45$, this colour range is not populated by many galaxies.

Thirdly, those galaxies labelled as quiescent by Eq.~(\ref{eq:quiescent}) and that belong to the star-forming sample after the dust correction (intrinsic $m_{F365}-m_{F551}<1.5$, see Fig.~\ref{fig:ccd_noav}) are typically concentrated close to the edges of Eq.~(\ref{eq:quiescent}), supporting the reliability of the extinction values provided by MUFFIT. Otherwise, the distribution of DSF galaxies would uniformly populate the red part of the $UVJ$ diagram as a consequence of degeneracies, where the effects of age, metallicity, and extinction on the stellar continuum would not be properly differentiated. 

Finally, DSF galaxies comprise  $\sim20$~\%  of the quiescent sample defined through Eq.~(\ref{eq:quiescent}) (see Fig.~\ref{fig:ccd_noav}). Our results clearly establish that more massive quiescent galaxies are less biased by DSF galaxies than the less massive ones (see Fig.~\ref{fig:CMDE} as well). More precisely, the most massive part of the sample is weakly contaminated by DSF galaxies when they are defined by Eq.~(\ref{eq:quiescent}) ($\log_{10}M_\star \ge 11$, $4$--$10$~\% from $z\sim 0.1$ to $z \sim 1.1$), while the low-mass ones may be significantly biased by them (e.g.~$\sim 40$~\% for $9.2 \le \log_{10}M_\star \le 9.6$ at $0.1 \le z \le 0.3$). In Table~\ref{tab:uvj_cont}, we illustrate the number of galaxies labelled as DSF galaxies at different redshift and stellar mass bins, meaning the level of contamination that we would expect when we build samples of quiescent galaxies via red $UVJ$ colour--colour diagrams such as Eq.~(\ref{eq:quiescent}) \citep[see also][]{Moresco2013}.

\begin{table*}
\centering
\caption{Percentage of red $UVJ$ galaxies (defined by Eq.~(\ref{eq:quiescent})) in ALHAMBRA with a bluer intrinsic colour than $(m_{F365}-m_{F551})_\mathrm{int} < 1.5$ after a dust correction at different redshift and stellar mass bins.}
\label{tab:uvj_cont}
\centering
\begin{tabular}{rccccc}
\hline\hline
 & \multirow{2}{*}{$0.1 \le z < 0.3$} & \multirow{2}{*}{$0.3 \le z < 0.5$} & \multirow{2}{*}{$0.5 \le z < 0.7$} & \multirow{2}{*}{$0.7 \le z < 0.9$} & \multirow{2}{*}{$0.9 \le z \le 1.1$} \\
&&&&& \\
\hline
&&&&& \\
$\ 9.6 \le \log_{10}M_\star < 10.0$ & $24.8$  & -- & -- & -- & -- \\
$10.0 \le \log_{10}M_\star < 10.4$  & $15.3$  & $ 14.8$ & -- & -- & -- \\
$10.4 \le \log_{10}M_\star < 10.8$  & $ 5.9$ & $ 8.1$ & $ 19.2$ & -- & -- \\
$10.8 \le \log_{10}M_\star < 11.2$  & $ 5.1$ & $ 2.8$ & $ 11.2$ & $ 11.6$ & -- \\
$\log_{10}M_\star \ge 11.2$         & $ 7.3$ & $ 6.0$ & $  5.4$ & $  4.0$ & $  6.2$ \\
&&&&& \\
\hline
\end{tabular}
\end{table*}

In addition to the colour cut $(m_{F365}-m_{F551})_\mathrm{int} \ge 1.5$, we restrict this study to the redshift interval $0.1\le z \le 1.1$ because  the stellar mass completeness constraint largely reduces the number of quiescent galaxies further than $z > 1.1$ and because at $z<0.1$ the number of quiescent galaxies in ALHAMBRA is also very low, especially for the most massive ones, which are substantially less frequent. The total number of quiescent galaxies at this point is $\sim15500$, where discrepancies in number owing to the use of BC03 and EMILES are lower than  $\sim2$~\%.


\subsection{Stellar mass--colour diagram corrected for extinction}\label{sec:mcd}

One of the main goals of $UVJ$-like diagrams is to reduce the number of DSF galaxies in quiescent galaxy samples \citep[see e.g.][]{Moresco2013}. We show  above that removing dust effects makes  the colour bimodality of galaxies clearer in the $UVJ$ diagram (Fig.~\ref{fig:ccd_noav}). However, we did not know whether the selection in this diagram is creating a stellar-mass bias in the quiescent sample. For this reason, we explored the rest-frame stellar mass--colour diagram corrected for extinction (MCDE). This diagram was conceived to both unmask the DSF galaxies and include the stellar mass of galaxies (physical parameter) avoiding any bias in the sample selection. The MCDE diagram is actually a step forward or improvement with respect to classical CMD diagrams \citep[e.g.][]{Bell2004,Baldry2004,Brown2007} or the stellar mass--colour diagram \citep[e.g.][see Fig.~\ref{fig:CMDE}]{Peng2010,Schawinski2014}, which are typically contaminated by DSF galaxies. 

As expected, the MCDE makes   the colour bimodality of galaxies clearer (see Fig.~\ref{fig:CMDE}), where the dust-corrected colour $(m_{F365}-m_{F551})_\mathrm{int}$ is the key to disentangling quiescent galaxies from the red sources that are actually DSF galaxies (see Fig.~\ref{fig:CMDE}). Otherwise, disentangling DSF galaxies via colour--colour diagrams or CMD is not completely reliable. Again, the presence of galaxies in the green valley is much rarer, pointing out the large presence of DSF galaxies in the green valley colour range when a dust correction is not performed. Figure~\ref{fig:CMDE} shows that the colour cut $(m_{F365}-m_{F551})_\mathrm{int} \ge 1.5$ defined in Sect.~\ref{sec:sample_uvj} properly selects quiescent galaxies at first order up to $z\sim1$ and $m_{F814} \le 23$, although only for this work. For a general case, we deal with this concern in more detail in Sect.~\ref{sec:discussion_diagrams}. However, to better define the colour limit of the quiescent galaxy population (see Fig.~\ref{fig:CMDE}) we follow a process of three steps:

\begin{itemize}
\item[i)]{From the distribution of ALHAMBRA galaxies on the MCDE, we defined a set of straight lines to separate the two populations by eye. As in Sect.~\ref{sec:sample_uvj}, we focus on the redshift range at $0.1\le z \le 1.1$;}
\item[ii)]{We stated the representative intrinsic colour $(m_{F365}-m_{F551})_\mathrm{int}$ of quiescent galaxies as a function of stellar mass and redshift, denoted as $(m_{F365}-m_{F551})_\mathrm{int}^\mathrm{Q}$. To do this, we performed a least-squares fitting  with all the quiescent galaxies above the subsample defined by eye in the previous step, that is, the red galaxies in the MCDE. For the fit we assumed a linear relation between the intrinsic colour $(m_{F365}-m_{F551})_\mathrm{int}$ and stellar mass;}
\item[iii)]{We set colour limits for quiescent galaxies, $(m_{F365}-m_{F551})_\mathrm{int}^\mathrm{lim}$ as a function of redshift. To set this limit, we assumed that it exhibits the same dependence on stellar mass as $(m_{F365}-m_{F551})_\mathrm{int}^\mathrm{Q}$. Nevertheless, the ordinate depends on the intrinsic colour spread, $(m_{F365}-m_{F551})_\mathrm{int}^\mathrm{rot}$, of quiescent galaxies, 
\begin{equation}
(m_{F365}-m_{F551})_\mathrm{int}^\mathrm{rot} = (m_{F365}-m_{F551})_\mathrm{int}-(m_{F365}-m_{F551})_\mathrm{int}^\mathrm{Q}\ ,
\label{eq:CMDE_rot}
\end{equation}
which is properly described by a log-normal distribution (see Appendix~\ref{sec:appendix_mle}). As this distribution is affected by uncertainties, which also vary with redshift, we adopted the maximum likelihood method (MLE) developed by \citet[][further details in Appendix~\ref{sec:appendix_mle}]{LopezSanjuan2014} to remove uncertainty effects and reveal the intrinsic distribution of values. We define the maximum $(m_{F365}-m_{F551})_\mathrm{int}$ value of quiescent galaxies as the $3$$\sigma$ limit of the log-normal distribution revealed from the MLE.}
\end{itemize}

As a result, we obtain that  the representative intrinsic colours $(m_{F365}-m_{F551})_\mathrm{int}^\mathrm{Q}$ and the limiting values $(m_{F365}-m_{F551})_\mathrm{int}^\mathrm{lim}$ of quiescent galaxies are both properly expressed by an equation of the form
\begin{equation}
(m_{F365}-m_{F551})_\mathrm{int}^\mathrm{Q,lim} = a \cdot \left(\log_{10} M_\star/M_\sun - 10 \right) + b \cdot (z-0.1) + c\ ,
\label{eq:CMDE_main}
\end{equation}
where $a$, $b$, and $c$ depend on the SSP model set used. In Table~\ref{tab:cmde_main}, we show the values $a$, $b$, and $c$ obtained for $(m_{F365}-m_{F551})_\mathrm{int}^\mathrm{Q}$ and $(m_{F365}-m_{F551})_\mathrm{int}^\mathrm{lim}$ using BC03, EMILES+BaSTI, and EMILES+Padova00 SSP models.

\begin{table}
\centering
\caption{Parameters $a$, $b$, and $c$ that best fit the representative intrinsic colours $(m_{F365}-m_{F551})_\mathrm{int}^\mathrm{Q}$ (top panel) and the limiting values $(m_{F365}-m_{F551})_\mathrm{int}^\mathrm{lim}$ (bottom panel) of quiescent galaxies using Eq.~(\ref{eq:CMDE_main}) and the SSP models of BC03 and EMILES (BaSTI and Padova00 isochrones).}
\label{tab:cmde_main}
\centering
\begin{tabular}{lccc}
\hline\hline
\multirow{2}{*}{$(m_{F365}-m_{F551})_\mathrm{int}^\mathrm{Q}$} & \multirow{2}{*}{$a$} & \multirow{2}{*}{$b$} & \multirow{2}{*}{$c$} \\
&&& \\
\hline
&&& \\
BC03            & $0.16\pm0.01$ & $-0.26\pm0.12$ & $1.76\pm0.15$ \\
BaSTI    & $0.15\pm0.01$ & $-0.24\pm0.12$ & $1.73\pm0.15$ \\
Padova00 & $0.15\pm0.01$ & $-0.19\pm0.11$ & $1.75\pm0.15$ \\
&&& \\
\hline
\multirow{2}{*}{$(m_{F365}-m_{F551})_\mathrm{int}^\mathrm{lim}$} & \multirow{2}{*}{$a$} & \multirow{2}{*}{$b$} & \multirow{2}{*}{$c$} \\
&&& \\
\hline
&&& \\
BC03            & $0.16\pm0.01$ & $-0.30\pm0.26$ & $1.50\pm0.18$ \\
BaSTI    & $0.15\pm0.01$ & $-0.24\pm0.17$ & $1.46\pm0.15$ \\
Padova00 & $0.15\pm0.01$ & $-0.13\pm0.06$ & $1.47\pm0.13$ \\
&&& \\
\hline
\end{tabular}
\end{table}

From this analysis and the distribution of colours of quiescent galaxies in the MCDE, we find that (i) the slope of the quiescent sequence in this diagram is compatible with a constant value of $\sim0.16$ since $z\sim1$; (ii) the colour spread of values $(m_{F365}-m_{F551})_\mathrm{int}$ has not changed significantly since $z\sim1$, which exhibits a value in the range $\sim0.26$--$0.28$ (AB magnitudes, $3\sigma$ limit; see Table~\ref{tab:cmde_main}); (iii) the selection of green valley galaxies is subject to a higher contamination of DSF galaxies when making use of the CMD; and (iv) this contamination of DSF galaxies depends on the SSP model set used and amounts to  $50$--$75$~\%. For further results concerning green valley galaxies, see  Sect.~\ref{sec:discussion_green}.


\subsection{Visual inspection}\label{sec:visual}
To increase the purity of the sample, we also carried out an individual inspection of the $\sim 15000$ galaxies with intrinsic red colours obtained in Sects.~\ref{sec:sample_uvj} and \ref{sec:mcd} at $0.1\le z \le 1.1$. We removed from the sample spurious detections and those galaxies that were compromised by very nearby sources and/or were imaged in bad CCD regions. To develop this process, we simultaneously checked one-by-one all the galaxy stamps, the adequacy of the photometric aperture (mainly the efficiency on the detection-deblending of sources and surroundings), and that the photo-spectrum did not present strong irregularities such  as   time-variable sources or sources close to stellar spikes, which cannot be reproduced by stellar population models. In the end we removed  $\sim5$~\% of sources from the quiescent sample after the visual inspection.


\subsection{Removal of cool stars}\label{sec:stars}
The ALHAMBRA Gold catalogue provides a statistical star/galaxy classification to understand whether each source in the catalogue is  a galaxy or a star \citep[\texttt{Stellar\_flag}, see details in][]{Molino2014}. This parameter was originally used to define our sample of galaxies (see Sect.~\ref{sec:data_sp}) and it simultaneously accounts for the morphology, apparent magnitude, and two colours to provide a statistical approach. Unfortunately, this parameter is less effective at decreasing the signal-to-noise-ratio, and consequently this classification is uncertain for sources fainter than $m_{F814W}=22.5$,  and a value \texttt{Stellar\_flag}$=0.5$ is assigned. Although we do not expect a large degree of contamination from stars at $22.5 \le m_{F814W} < 23$ for the full sample, the ALHAMBRA fields may contain stars from the Milky Way halo, which are mainly composed of cool red stars. Our quiescent sample, composed of red sources, may   therefore be partly contaminated by stars that were treated as galaxies.

This problem  was faced through a new MUFFIT module devoted to analysing stars. This performs a SED fitting process that is similar to the version for galaxies, but using the \citet{Coelho2005} star models instead. This methodology   reduced the contamination of cool stars in ALHAMBRA from $24$~\% to $4$~\% after comparing our faint star detection predictions with the star/galaxy classification provided by the Cosmological Evolution Survey \citep[COSMOS, classified as point-like sources thanks to its tiny PSF;][]{Leauthaud2007}  in a shared subsample of red sources at $22.5 \le m_{F814W} < 23$. It is worth mentioning that owing to the spectral coverage of the \citet{Coelho2005} stellar models ($300$~nm -- $1.8\mu\mathrm{m}$), the flux from the ALHAMBRA $K_\mathrm{s}$ band is not used for the stellar SED fitting analysis. A more extended explanation of the whole process, as well as the comparison with COSMOS to check the reliability of the methodology, is detailed in Appendix~\ref{sec:appendix_faint}. In the end, we removed $439$ star candidates from the sample at $22.5 \le m_{F814W} < 23$.


\subsection{Stellar mass completeness}\label{sec:completeness}

We modelled the stellar mass completeness, $\mathcal{C}$, through a Fermi-Dirac distribution (see Appendix~\ref{sec:appendix_smc}) that is parametrised by two redshift-dependent parameters, $M_\mathrm{F}$ and $\Delta_\mathrm{F}$. The parameter $M_\mathrm{F}$ reflects the stellar mass value (in dex units) for which the completeness is equal to $50$~\% ($\mathcal{C}=0.5$), while $\Delta_\mathrm{F}$ is related to the rate of decrease in the number of galaxies. We found that this kind of distribution properly reproduces the decay on the less massive galaxies of our flux-limited sample ($m_{F814W} \le 23$) at any redshift. The process used to derive these parameters takes advantage of stellar mass functions from deeper surveys (for further details, see Appendix~\ref{sec:appendix_smc}), in particular from the COSMOS survey, which specifically provides them for quiescent galaxies \citep[][computed using BC03 SSP models]{Ilbert2010} and partly overlaps with ALHAMBRA. In this process, we assumed that any discrepancy between the low-mass end of the ALHAMBRA stellar mass function and the COSMOS value was caused by the mass incompleteness. These discrepancies allowed us to determine both $M_\mathrm{F}$ and $\Delta_\mathrm{F}$, which are directly related to the stellar mass limit, $\log_{10} M_\mathcal{C}$, at  redshift $z$ and completeness level by
\begin{equation}
\log_{10} M_\mathcal{C}(z) = M_\mathrm{F}(z) + \Delta_\mathrm{F}(z) \ln\left[ \frac{\mathcal{C}}{1-\mathcal{C}}\right]\ .
\label{eq:completeness_level}
\end{equation}

\begin{figure}
\centering
\resizebox{\hsize}{!}{\includegraphics{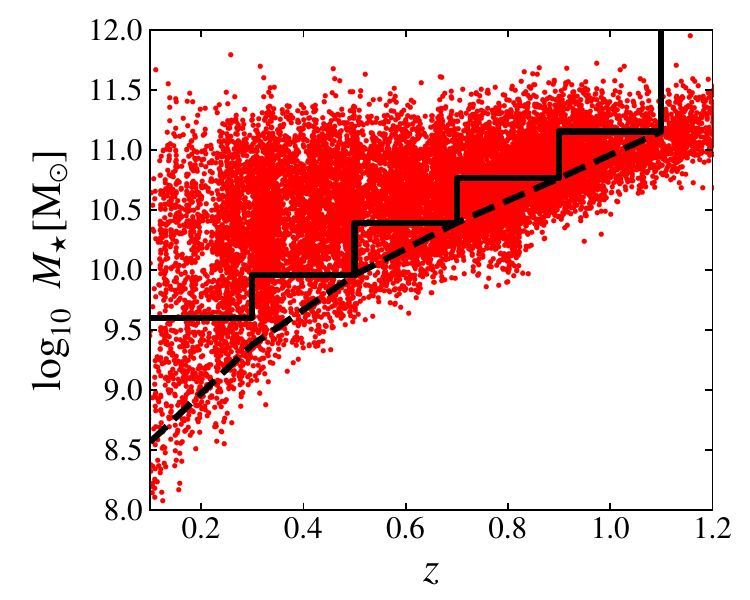}}
\caption{Redshifts ($X$-axis) and stellar masses ($Y$-axis) of ALHAMBRA quiescent galaxies. The dashed line illustrates the $95$~\% stellar mass completeness level of quiescent galaxies from the ALHAMBRA Gold catalogue (complete down to $m_{F814W}=23$). The solid line illustrates the limits assumed in this work to define our sample of galaxies complete in stellar mass at different redshift bins (see Sect.\ref{sec:sample}).}
\label{fig:completeness}
\end{figure}

For this work, we required a conservative stellar mass completeness of at least $\mathcal{C}=0.95$ at any redshift bin (see Fig.~\ref{fig:completeness} and values in Table~\ref{tab:completeness}). From Table~\ref{tab:completeness} we derive that ALHAMBRA is complete up to $\log_{10}M_\star\ge9.4$~dex at $z=0.3$. However, to develop this work, we increased the low-mass limit down to $\log_{10}M_\star\ge9.6$~dex, with the only aim of having a set of equal-sized stellar mass bins of $\sim 0.4$~dex. In Appendix~\ref{sec:appendix_smc}, there is a more detailed and complete explanation of the full process used to determine the ALHAMBRA completeness. When stellar masses are obtained using the MUFFIT and EMILES SSP models, their values differ with respect to those obtained using BC03 SSP models. EMILES stellar masses are systematically higher by about $0.15$ and $0.11$~dex for BaSTI and Padova00 isochrones, respectively. To determine the stellar mass completeness of EMILES predictions, we add $0.15$ (EMILES+BaSTI) and $0.11$~dex (EMILES+Padova00) to the $M_\mathrm{F}$ values provided in Table~\ref{tab:completeness}, whereas $\Delta_\mathrm{F}$ remains unaltered. We checked that this shift analytically reproduces the observed stellar masses properly.

\begin{table}
\centering
\caption{Parameters $M_\mathrm{F}$ and $\Delta_\mathrm{F}$ at different redshifts, and the stellar mass limit at the completeness levels $\mathcal{C}= 0.7$, $0.8$, $0.9$, and $0.95$ for the quiescent sample and BC03 SSP models (details in text).}
\label{tab:completeness}
\centering
\begin{tabular}{lccccc}
\hline\hline
\multirow{2}{*}{Redshift} & \multirow{2}{*}{$0.3$} & \multirow{2}{*}{$0.5$} & \multirow{2}{*}{$0.7$} & \multirow{2}{*}{$0.9$} & \multirow{2}{*}{$1.1$} \\
&&&&& \\
\hline
&&&&& \\
$M_\mathrm{F}$                & $8.99$  &  $9.64$  &  $10.08$ & $10.44$ & $10.84$ \\
$\Delta_\mathrm{F}$         & $0.132$ &  $0.108$ & $0.105$ & $0.110$ & $0.108$ \\
$\log_{10} M_{\mathcal{C}=0.7}$ & $9.10$  & $9.73$   &  $10.17$ & $10.53$ & $10.93$ \\
$\log_{10} M_{\mathcal{C}=0.8}$ & $9.17$  & $9.79$   & $10.23$ & $10.59$ & $10.99$ \\
$\log_{10} M_{\mathcal{C}=0.9}$ & $9.28$  & $9.88$   & $10.31$ & $10.68$ & $11.07$ \\
$\log_{10} M_{\mathcal{C}=0.95}$ & $9.38$  & $9.96$   & $10.39$ & $10.76$ & $11.15$ \\
&&&&& \\
\hline
\end{tabular}
\tablefoot{Parameters for EMILES SSP models are obtained adding $0.15$ and $0.11$~dex to $M_\mathrm{F}$ for BaSTI and Padova00 isochrones, respectively. For EMILES, $\Delta_\mathrm{F}$ is equal to the BC03 values.}
\end{table}

Finally, the total number of galaxies for this study, including the stellar mass completeness constraint, comprises a total of $\sim8\,500$ quiescent galaxies for BC03 SSP models (all the galaxies in our sample are illustrated in Fig.~\ref{fig:completeness}). The number of quiescent galaxies per stellar mass and redshift bin is detailed in Table~\ref{tab:number_bin}. For EMILES SSP models, the final number of quiescent galaxies for this work is $\sim8\,100$.

\begin{table*}
\centering
\caption{Number of quiescent galaxies per stellar mass and redshift bin. The last column gives the total number of galaxies per stellar mass bin. The last row shows the total number of quiescent galaxies per redshift bin. All the cells include redshift and stellar mass bins complete at the level $\mathcal{C}=0.95$, otherwise there is a  dash.}
\label{tab:number_bin}
\centering
\begin{tabular}{rcccccc}
\hline\hline
 & \multirow{2}{*}{$0.1 \le z < 0.3$} & \multirow{2}{*}{$0.3 \le z < 0.5$} & \multirow{2}{*}{$0.5 \le z < 0.7$} & \multirow{2}{*}{$0.7 \le z < 0.9$} & \multirow{2}{*}{$0.9 \le z \le 1.1$} & \multirow{2}{*}{Total} \\ 
&&&&&&\\
\hline
&&&&&&\\
$9.6 \le \log_{10}\ M_\star < 10.0$ & $289$ & -- & -- & -- & -- & $289$ \\ 
$10.0 \le \log_{10}\ M_\star < 10.4$ & $433$ & $996$ & -- & -- & -- & $1429$ \\ 
$10.4 \le \log_{10}\ M_\star < 10.8$ & $403$ & $1035$ & $1088$ & -- & -- & $2526$ \\ 
$10.8 \le \log_{10}\ M_\star < 11.2$ & $238$ & $663$ & $800$ & $1480$ & -- & $3181$ \\ 
 $\log_{10}\ M_\star \ge 11.2$ & $51$ & $140$ & $157$ & $357$ & $417$ & $1122$ \\ 
Total & $1414$ & $2834$ & $2045$ & $1837$ & $417$ & $8547$ \\ 
&&&&&&\\
\hline
\end{tabular}
\end{table*}


\subsection{Photo-$z$ accuracy of ALHAMBRA quiescent galaxies}\label{sec:photoz}

An accurate photo-$z$ determination is essential to properly drive a stellar population study, otherwise any stellar population prediction may be erroneous. As we mention in Sect.~\ref{sec:methods}, we initially used the photo-$z$ constraints provided in the ALHAMBRA Gold catalogue (computed using BPZ2.0) as input for the SED fitting analysis. Therefore, we ran MUFFIT in the photo-$z$ intervals provided in this catalogue, treating the redshift as another free parameter to determine during our stellar population analysis.

In this section we describe how we determined the photo-$z$ precision obtained for the sample of quiescent galaxies studied in this work. To this end, we took advantage of the same subsample of spectroscopic galaxies used by \citet[][priv. comm.]{Molino2014}, which contains publicly available spectroscopic redshifts from surveys that overlap with ALHAMBRA (zCOSMOS, \citealt{Lilly2009}; AEGIS\footnote{All-wavelength Extended Groth strip International Survey}, \citealt{Davis2007}; and GOODS-N\footnote{The Great Observatories Origins Deep Survey-North}, \citealt{Cooper2011}). This spectroscopic subsample comprises galaxies at $0.0 < z < 1.5$ and magnitudes $18 < mF814W < 25$ (mean values of $\langle z \rangle\sim 0.77$ and  $\langle m_{F814W}\rangle \sim22.3$). Because   there is no   standard method for setting numerical values of the photo-$z$ accuracy, we turned to various statistical estimators already used in the literature. The most extended one is probably the normalised median absolute deviation \citep[$\sigma_\mathrm{NMAD}$, ][]{Brammer2008} since this estimator is less affected by catastrophic errors or outliers. Formally,

\begin{equation}
\sigma_\mathrm{NMAD} = 1.48 \times median \left( \frac{|\Delta z -median(\Delta z)|}{1+z_\mathrm{spec}} \right),
\label{eq:nmad}
\end{equation}
where $\Delta z = z_\mathrm{phot} - z_\mathrm{spec}$. Moreover, we propose here another photo-$z$ accuracy estimator: the root mean square (RMS) of the Gaussian distribution built from the values $\Delta z/(1+z_\mathrm{spec}$), which in the following we denote as $\sigma_{\mathrm{photo-}z}$. The number of catastrophic outliers is also an important factor to take into account, so we propose two definitions for it:
\begin{equation}
\eta_1 = \frac{|\Delta z|}{1+z_\mathrm{spec}} > 0.2, {\rm and}
\label{eq:eta1}
\end{equation}
\begin{equation}
\eta_2 = \frac{|\Delta z|}{1+z_\mathrm{spec}} > 5 \times \sigma_\mathrm{NMAD}.
\label{eq:eta2}
\end{equation}

\begin{figure*}
\centering
\resizebox{\hsize}{!}{\includegraphics[trim=0 13 0 5mm,width=0.5\hsize,clip=True]{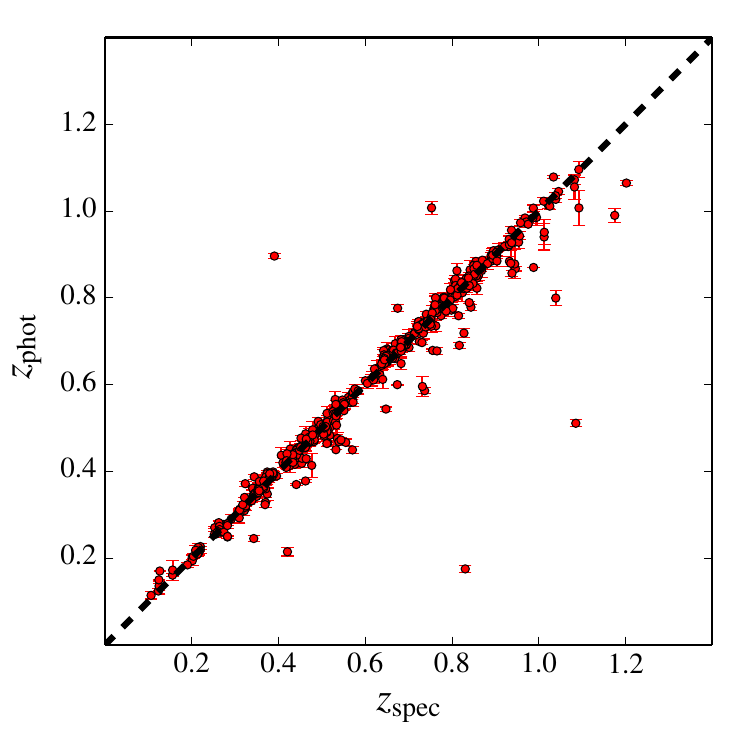}
\includegraphics[trim=0 5 0 0mm,width=0.5\hsize,clip=True]{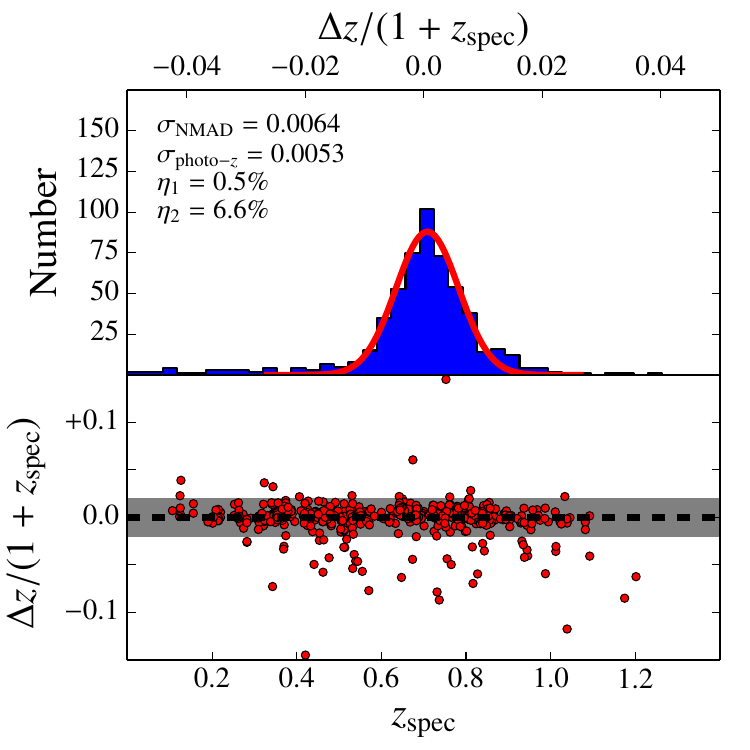}}
\caption{Comparison between the photo-$z$ provided by MUFFIT ($z_\mathrm{phot}$) and their spectroscopic values ($z_\mathrm{spec}$) for $576$ shared quiescent galaxies. Left panel:  One-to-one comparison of redshifts, where dashed black line is the one-to-one relation. Top right: Histogram of values $\Delta z/(1+z_\mathrm{spec})$, along with the Gaussian that best fits this distribution (red solid line) and the photo-$z$ accuracy estimators defined in the text (Eqs.~(\ref{eq:nmad})--(\ref{eq:eta2})). Bottom right: Differences $\Delta z/(1+z_\mathrm{spec})$ as a function of the spectroscopic redshift ($X$-axis). The shaded region illustrates the $3\times\sigma_\mathrm{NMAD}$ uncertainty.}
\label{fig:zphot}
\end{figure*}

Amongst the $\sim8\,500$ quiescent galaxies in our sample complete in stellar mass (see Sect.~\ref{sec:completeness} above), there are $576$ quiescent galaxies with spectroscopic redshifts. Figure~\ref{fig:zphot} illustrates the one-to-one comparison of these $576$ quiescent galaxies, showing the excellent agreement between our photometric predictions and the spectroscopic predictions. For this subsample of $576$ galaxies and according to Eqs.~(\ref{eq:nmad})--(\ref{eq:eta2}), we obtained $\sigma_\mathrm{NMAD}=0.0064$, $\eta_1 = 0.5$~\%, and $\eta_2=6.6$~\%, respectively. Regarding the $\Delta z/(1+z_\mathrm{spec})$ distribution, we fitted this distribution adopting a Gaussian-like function (see Fig.~\ref{fig:zphot}). The Gaussian fit is centred at $0.0006$, which is closely centred to zero or negligible shift, and exhibits an RMS of $\sigma_{\mathrm{photo-}z} = 0.0053$. 

If we compare the photo-$z$ values provided by BPZ2.0 in the ALHAMBRA Gold catalogue with the spectroscopic values, we obtain an accuracy of $\sigma_\mathrm{NMAD}=0.0080$, $\sigma_{\mathrm{photo-}z} = 0.0062$, $\eta_1 = 0.5$~\%, and $\eta_2=7.1$~\%. Therefore, adopting photo-$z$ constraints from external dedicated codes during the MUFFIT SED fitting analysis may improve the photo-$z$ accuracy by  $\sim15-20$~\%. As checked by \citet{DiazGarcia2015}, a photo-$z$ uncertainty at this level ($0.6$~\%) has a minimum effect on the stellar population parameters that are determined via SED fitting in ALHAMBRA: stellar mass, age, metallicity, and extinction.


\section{Star formation rates via SED fitting results}\label{sec:dusty}
Our SED fitting code MUFFIT uses two SSPs for the fitting, which by definition has null SFR since it involves two SSPs of non-zero age \citep[unlike other SED fitting analyses based on $\tau$-models or in more complex SFHs, see e.g.][]{Cid2005,Moustakas2013}. In order to obtain SFRs from the SED fitting results provided by MUFFIT, it is necessary to define a tracer or parameter that allows us to estimate them. Although the SFR of quiescent galaxies is typically low, it can also be used to complement and reinforce the reliability of the results obtained in Sects.~\ref{sec:sample_uvj} and \ref{sec:mcd}. If red galaxies labelled as DSF galaxies after the dust correction (see Eq.~(\ref{eq:quiescent})) also show SFRs proper of star-forming galaxies, we can be more confident that these galaxies were removed from the quiescent sample properly. In this section we propose a methodology for  providing SFR values of galaxies based on the SED fitting results provided by MUFFIT via the $2\,800$~\AA\ luminosity \citep[SFR tracer, see e.g.][]{Kennicutt1998,Madau1998}.

The UV continuum in the range $\lambda\lambda~1\,500$--$2\,800$~\AA\ is a good tracer of SFR in galaxies with ongoing star formation, because this range is mainly dominated by the light emitted by late-O and early-B stars \citep[see e.g.][]{Madau1998}. In particular, we chose the SFR tracers proposed by \citet{Madau1998}, which are based on stellar population models with exponentially declining SFRs, or $\tau$ models, and \citet{Salpeter1955} IMF. Even though the SFR derived from luminosities at $1\,500$~\AA\ and $\tau$ models is slightly less dependent on the duration of the star formation burst, the SFRs derived throughout this section were computed from the rest-frame luminosity at $2\,800$~\AA, $L_{2800\AA}^{z=0}$. Formally,
\begin{equation}\label{eq:sfr2800}
{SFR}_{2800\AA} = 1.27\ 10^{-28} \times L_{2800\AA}^{z=0}\ ,
\end{equation}
where $L_{2800\AA}^{z=0}$ is in units of ergs s$^{-1}$ Hz$^{-1}$, and ${SFR}_{2800\AA}$ in $M_\sun$ yr$^{-1}$. We note that Eq.~(\ref{eq:sfr2800}) does not take dust effects into account. The choice of the SFR tracer at $2\,800$~\AA\ is motivated by the observational wavelength-frame of ALHAMBRA because this starts at $\sim 3\,500$~\AA\ (band $F365W$, full width at half maximum of $\sim 300$~\AA, and effective wavelength $\lambda_\mathrm{eff}=3\,650$~\AA). Therefore, the SFR tracer based on the luminosity at $2\,800$~\AA\ is only directly observed at redshift $z > 0.25$, whereas for $1\,500$~\AA\ at $z>1.3$, which would reduce our sample drastically. Actually, we are able to obtain a prediction of the luminosity at $2\,800$~\AA\ at $z < 0.25$, but this prediction would be an extrapolation of the SED fitting carried out by MUFFIT. For this section, we preferred a more conservative treatment in which the rest-frame flux at $2\,800$~\AA\ must be included in the observational frame ($z \gtrsim 0.25$).

The rest-frame luminosity $L_{2800\AA}^{z=0}$, also free of dust attenuation, was calculated from the SED fitting results provided by MUFFIT. Similarly to the process of removing the dust effects on colours (Sect.~\ref{sec:sample_uvj}), we rebuilt the combination of best-fitting models obtained during the Monte Carlo approach without extinction and for all the galaxies in ALHAMBRA. From this combination of models, we integrated the flux emitted in the rest-frame range $\lambda\lambda~2\,750$--$2\,850$~\AA\ in order to compute $L_{2800\AA}^{z=0}$ and subsequently ${SFR}_{2800\AA}$ via Eq.~(\ref{eq:sfr2800}).

\begin{figure*}
\centering
\resizebox{\hsize}{!}{\centering
\includegraphics[trim={7mm 0 0 0}]{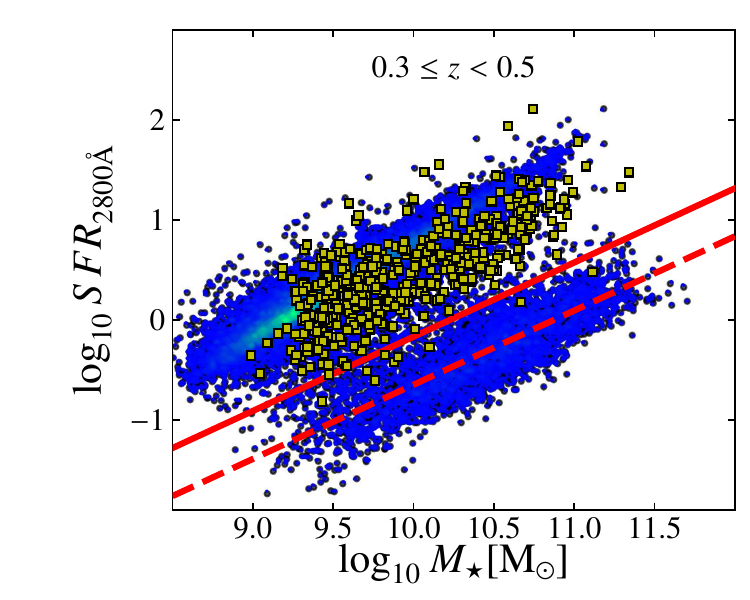}\includegraphics{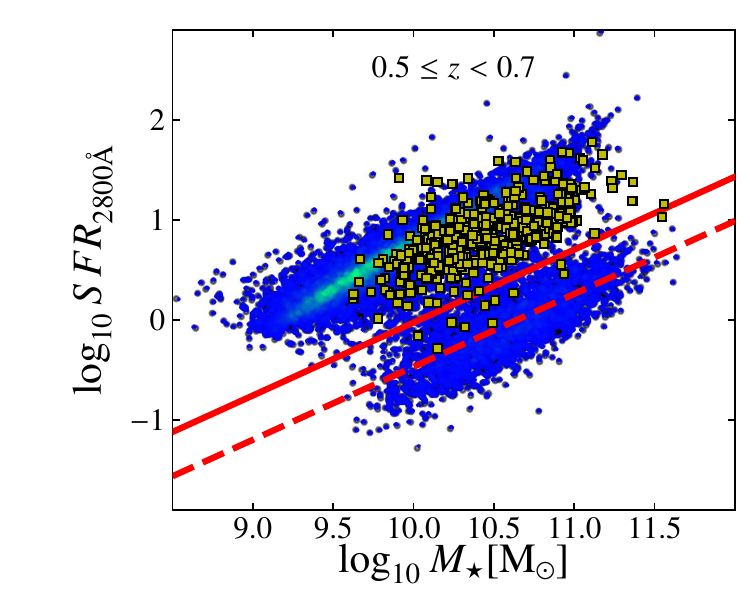}}
\caption{Stellar mass ($X$-axis) vs. star formation rate tracer ${SFR}_{2800\AA}$ ($Y$-axis) defined by \citet[][see Eq.~(\ref{eq:sfr2800})]{Madau1998} of ALHAMBRA galaxies at redshift $0.3 \le z < 0.5$ (left panel) and $0.5 \le z < 0.7$ (right panel). High densities are presented in  green, and  lower densities in blue. The dashed and solid red lines show the fit to the quiescent sequence and the limiting ${SFR}_{2800\AA}$ values for star-forming galaxies, respectively. The yellow markers illustrate galaxies labelled as dusty star-forming galaxies   by the rest-frame stellar mass--colour diagram corrected for extinction (MCDE).}
\label{fig:sfr2800_z}
\end{figure*}

In order to characterise the range of $SFR_{2800\AA}$ values of star-forming galaxies, as well as their dependence on redshift and stellar mass, we studied the distribution of $SFR_{2800\AA}$ as a function of the stellar mass. When viewing the full sample of ALHAMBRA galaxies in the SFR versus stellar-mass plane (see Fig.~\ref{fig:sfr2800_z})  we again differentiate between two groups of galaxies: star-forming and passive galaxies \citep[e.g.][]{Ilbert2010, Moustakas2013, Choi2014}. This figure illustrates a tight correlation between the stellar mass and the SFR of a galaxy: the more massive the galaxy, the higher the SFR of the galaxy, independent of its spectral type \citep{Daddi2007,Elbaz2007,Noeske2007,Moustakas2013}. In order to separate  quiescent and star-forming populations (i.e.~setting limiting $SFR_{2800\AA}$ values for quiescent and star-forming galaxies), we adapted the methodology detailed in Sect.~\ref{sec:mcd}, and repeated it. As a result, we obtain that the representative SFR values of quiescent galaxies, $\log_{10}SFR_{2800\AA}^\mathrm{Q}$, and their limiting values at a $3\sigma$ level, $\log_{10}SFR_{2800\AA}^\mathrm{lim}$, are respectively expressed as

\begin{equation}\label{eq:sfr2800_q}
\log_{10}SFR_{2800\AA}^\mathrm{Q} = 0.75\log_{10}M_\star - 8.00\ ,
\end{equation}
\begin{equation}\label{eq:sfr2800_lim}
\log_{10}SFR_{2800\AA}^\mathrm{lim} = 0.75 (\log_{10}M_\star-10) - 0.18(z-0.1) + 0.11\ .
\end{equation}

Our results point out that the correlation between $\log_{10}SFR_{2800\AA}$ and stellar mass is compatible with no evolution across cosmic time, and a relation of the form $\log_{10}SFR_{2800\AA}\ \propto 0.75\log_{10}M_\star$ can be assumed for $\log_{10}SFR_{2800\AA}^\mathrm{Q}$. As in Sect.~\ref{sec:mcd}, we took advantage of the MLE methodology presented in Appendix~\ref{sec:appendix_mle} to set the $3$$\sigma$ limits of the distributions by the intrinsic scatter of SFR values, $\log_{10}SFR_{2800\AA}^\mathrm{rot}$, defined as

\begin{equation}\label{eq:sfr2800_rot}
\log_{10}SFR_{2800\AA}^\mathrm{rot} = \log_{10}SFR_{2800\AA}-\log_{10}SFR_{2800\AA}^\mathrm{Q}\ .
\end{equation}

There are $828$ DSF galaxies (number defined from the results obtained using BC03 SSP models) that were removed from the quiescent sample (complete in stellar mass and defined by the dust corrected $UVJ$ diagram in Sect.~\ref{sec:sample_uvj}) at redshift $0.3 \le z \le 1.1$. Their $SFR_{2800\AA}$ measurements were compared with Eq.~(\ref{eq:sfr2800_lim}) to determine whether they present proper values of star-forming galaxies. We found that DSF galaxies mainly lie on the lower parts of the stellar mass--SFR plane of star-forming galaxies (see Fig.~\ref{fig:sfr2800_z}), that is, close to the SFR quiescent limit (Eq.~(\ref{eq:sfr2800_lim})). Around $\sim90$~\% of the $828$ DSF galaxies have $SFR_{2800\AA}$ above Eq.~(\ref{eq:sfr2800_lim}), and consequently their SFR values are in the range of star-forming galaxies (see Fig.~\ref{fig:sfr2800_z}). This amount increases  to $\sim97$~\% when we consider SFR errors  ($1\sigma$ uncertainty level), which is almost the total of galaxies labelled as DSF galaxies. Moreover, we found  that  $\sim98.2$~\% of quiescent galaxies (defined by rest-frame colours in Sect.~\ref{sec:sample_uvj}) exhibit a SFR below Eq.~(\ref{eq:sfr2800_lim}), which increases  to $\sim99.8$~\% for a $1\sigma$ uncertainty level. The same conclusion is obtained for EMILES SSP models.

These results additionally support MUFFIT predictions about DSF galaxies and the necessity of removing these galaxies, which comprise $\sim10$~\% of the sample when a $UVJ$ selection is performed without removing the dust effects on colours (see details in Table~\ref{tab:uvj_cont}). It should be noted that $SFR_{2800\AA}$ is also based on the SED fitting predictions carried out by MUFFIT because it is necessary to estimate the luminosity at $\sim2800$~\AA\ after removing dust effects, but it differs with respect to the selection process developed in Sect.~\ref{sec:sample_uvj}, which is based on   colour   and not on a luminosity-based SFR tracer. 


\section{Stellar populations of quiescent galaxies within the stellar mass--colour and $UVJ$ colour--colour diagrams}\label{sec:uvj}

The empirical bimodality of galaxies in the stellar mass--colour and colour--colour diagrams is directly related to their stellar population properties \citep[e.g.][]{Bower1992,Gallazzi2006,Whitaker2010}. This can be interpreted as a hint to explore whether within the quiescent sequence similar stellar population properties are constrained to lie on certain colour ranges. Colour predictions from models predict this, although there are also degeneracies and some stellar population properties are not observed in real galaxies. In this work we are in a privileged position because we are able to estimate the extinction, and thus to determine  the main stellar population parameters that lead to the distribution of colours.

In this section, we explore the distribution of age, metallicity, extinction, and sSFR (Sects.~\ref{sec:av}--\ref{sec:ssfr}, respectively) within the $UVJ$ and stellar mass--colour diagrams (both corrected and non-corrected for extinction). A detailed analysis of the distributions of stellar population properties of quiescent galaxies and their dependence on stellar mass, and their evolution with redshift,  is beyond the scope of this work. We refer readers to the next papers in this series \citep[][]{DiazGarcia2018,DiazGarcia2019} where we will explore this topic in depth.

In all the cases instead of plotting the individual measurements, which include uncertainties, we carried out a bidimensional and locally weighted regression method called LOESS \citep[][]{Cleveland1979,Cleveland1988}. This methodology finds a non-parametric plane to reveal trends on both the stellar mass--colour and $UVJ$ diagrams through the distribution of points (stellar population parameters for this study) minimising the uncertainty effects on the diagrams by a regression technique. We note that the LOESS method does not assume a  model or function to fit the distribution of stellar population properties to a plane, which in practice is similar to estimating average values of parameters across these diagrams \citep[for further details, see][]{Cleveland1979,Cleveland1988}. Specifically, we used the LOESS implementation for the \texttt{Python} language\footnote{http://www-astro.physics.ox.ac.uk/{\textasciitilde}mxc/software/} \citep{Cappellari2013}, where the regularisation factor or smoothness is set to $f=0.6$ and a local linear approximation is assumed; this method
yielded satisfactory results in previous works \cite[e.g.][]{Cappellari2013,Mcdermid2015}. The uncertainties on stellar mass, extinction, age, metallicity, and sSFR of individual galaxies were also taken into account during the regression process.

In addition, we carried out a bi-dimensional fit of the distributions obtained from the LOESS method to quantify the distributions of stellar population properties within the rest-frame $UVJ$ and MCDE diagrams (both corrected for extinction). The distributions of extinction, age, metallicity, and sSFR were quadratically fit to a plane using 
\begin{equation}\label{eq:quadra}
\mathcal{P} = \alpha\cdot\mathcal{X}^2 + \beta\cdot\mathcal{Y}^2 + \gamma\cdot\mathcal{X}\cdot\mathcal{Y} + \delta\cdot\mathcal{X} + \epsilon\cdot\mathcal{Y} + \zeta\ ,
\end{equation}
where $\mathcal{P}$ is the stellar population property (extinction, age, metallicity, and sSFR), $\mathcal{X}$ is either the intrinsic colour $(m_{F551}-J)_\mathrm{int}$ or the stellar mass in dex units (for the dust corrected $UVJ$ or MCDE diagrams, respectively), $\mathcal{Y}$ is the colour $(m_{F365}-m_{F551})_\mathrm{int}$ corrected for extinction, and $\alpha$, $\beta$, $\gamma$, $\delta$, $\epsilon$, and $\zeta$ are the parameters of the quadratically fit. The values of these parameters for the $UVJ$ and MCDE diagrams are provided in Tables~\ref{tab:uvj_corr} and \ref{tab:cmde_corr}, respectively. As a result, average differences between these planes and the distribution of points are  almost  null and the average scatter ($\sigma_\mathrm{fit}$) is below $\sim0.05$~dex in most of the cases (see Tables~\ref{tab:uvj_corr} and \ref{tab:cmde_corr}). Independently of the SSP model set, the stellar population properties exhibiting lower $\sigma_\mathrm{fit}$ values correspond to extinction and sSFR with values of $\sigma_\mathrm{fit} \sim 0.02$~dex, whereas the higher values are related to metallicity. We note that this parametrisation of the distributions may be very useful for future works to estimate these parameters only knowing the positions of quiescent galaxies within these diagrams. In addition, we quantified the correlations between the stellar population properties (extinction, age, metallicity, and sSFR) and both the $(m_{F365}-m_{F551})_\mathrm{int}$ and $(m_{F551}-J)_\mathrm{int}$ colours via a Spearman correlation coefficient ($\rho_\mathrm{UV}$ and $\rho_\mathrm{VJ}$, respectively, see Table~\ref{tab:corr_spear}). This coefficient presents values in the range $-1\le \rho_\mathrm{UV,VJ} \le 1$, where $1$ is total positive correlation, $0$ is no correlation, and $-1$ total anticorrelation.

In Figs.~\ref{fig:sp_UVJ} and \ref{fig:sp_UVJ_av}, we illustrate the distribution of stellar population parameters (non-dust corrected and intrinsic colours, respectively) for quiescent galaxies on the rest-frame $UVJ$-diagram  for BC03 SSP models. To guide the eye, we also plot the typical selection of quiescent galaxies \citep[illustrated in Fig.~\ref{fig:sp_UVJ_av}, see Eq.~(\ref{eq:quiescent}) in this work and][]{Moresco2013}. Interestingly, the only selection criterion $(m_{F365}-m_{F551})_\mathrm{int} \ge 1.5$ (see Fig.~\ref{fig:sp_UVJ}) provides a quiescent sample whose observed colour ranges are well delimited by Eq.~(\ref{eq:quiescent}). In our sample of quiescent galaxies, we only find  $1$~\% of galaxies with $m_{F365}-m_{F551}$ and $m_{F551}-J$ colours outside  the box delimited by Eq.~(\ref{eq:quiescent}).

We illustrate the stellar population parameters of the quiescent galaxies selected by the MCDE (see Eq.~(\ref{eq:CMDE_main}) and Table~\ref{tab:cmde_main}) in Figs.~\ref{fig:CMDE_sp} and \ref{fig:CMDE_sp_av} (corrected and non-corrected for reddening, respectively). The analysis of the distribution of stellar population properties of quiescent galaxies in the stellar mass--colour diagram complements and makes   the interpretation of the analysis within the $UVJ$-diagram easier. We detail the main features of the distribution of stellar population parameters below, which were equally obtained for the three SSP model sets used in this work (BC03 and EMILES, the latter including BaSTI and Padova00 isochrones).


\subsection{Distribution of extinction within the $UVJ$ and stellar mass--colour diagrams}\label{sec:av}

\begin{figure*}
\centering
\includegraphics[trim= 0 25 0 5mm,width=17cm,clip=True]{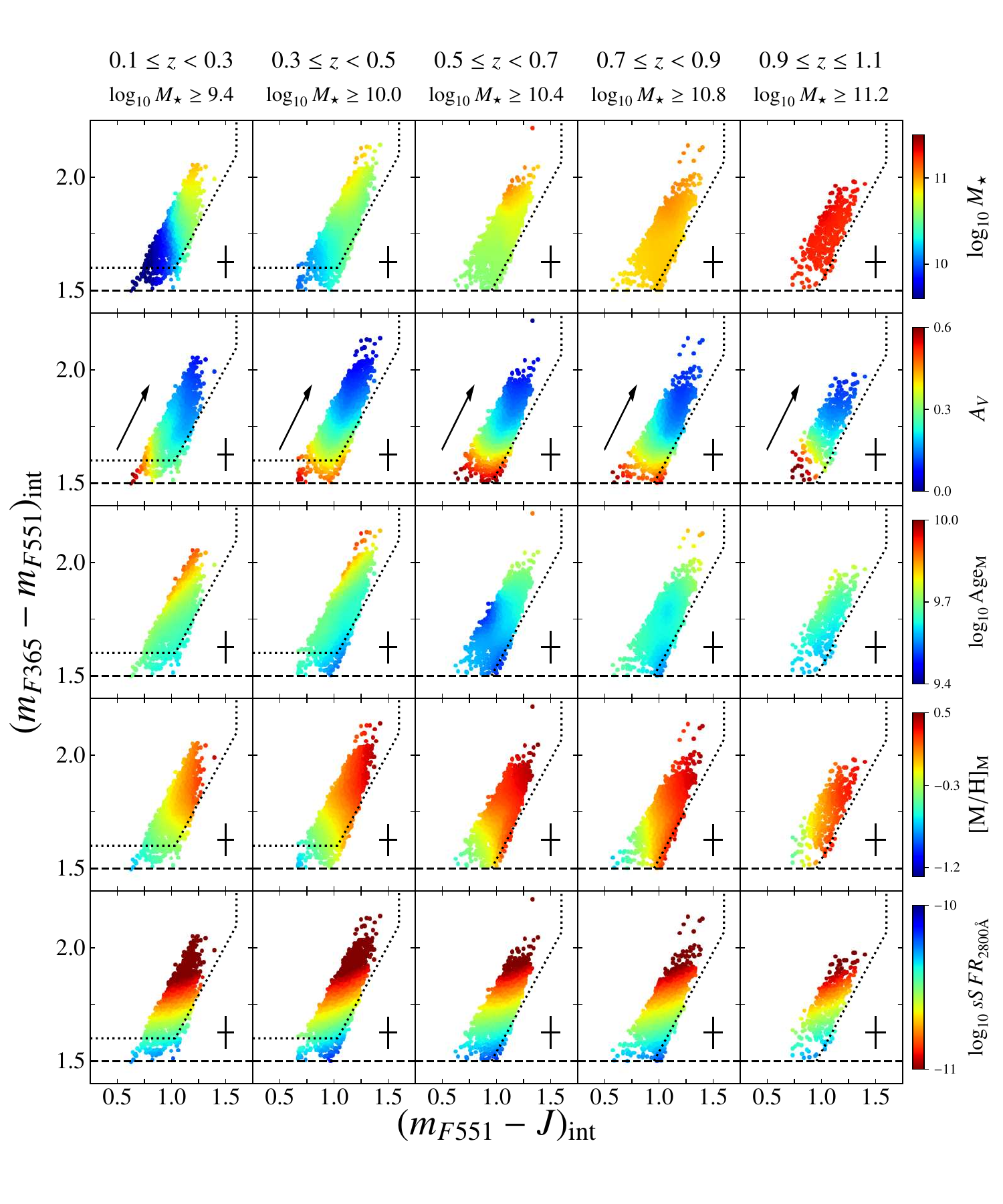}
\caption{Stellar population parameters in the rest-frame $UVJ$ diagram. At different redshift bins, we present the intrinsic colours $(m_{F551}-J)_\mathrm{int}$ ($X$-axis) and $(m_{F365}-m_{F551})_\mathrm{int}$ ($Y$-axis) after correcting for extinction for the mass complete sample of quiescent galaxies (see stellar mass completeness at the top). The different stellar population parameters are colour-coded as a  function of their values and obtained using BC03 SSP models (see the  colour bars to the left of each row). From top to bottom:  stellar mass, extinction, mass-weighted age and metallicity, and specific star formation rate. All the parameters were spatially averaged through a LOESS method. Black crosses illustrate the median uncertainties in both $(m_{F551}-J)_\mathrm{int}$ and $(m_{F365}-m_{F551})_\mathrm{int}$ intrinsic colours. The dotted black line encloses the rest-frame colour ranges assumed for selecting quiescent galaxies in \citet[][see Eq.~(\ref{eq:quiescent})]{Moresco2013}, while the dashed line illustrates our colour limit for selecting quiescent galaxies $(m_{F365}-m_{F551})_\mathrm{int}>1.5$. We illustrate the colour variations owing to a reddening of $A_V=0.5$ (black arrow), assuming the extinction law of \citet{Fitzpatrick1999}.}
\label{fig:sp_UVJ}
\end{figure*}

\begin{figure*}
\centering
\includegraphics[trim= 0 25 0 5mm,width=17cm,clip=True]{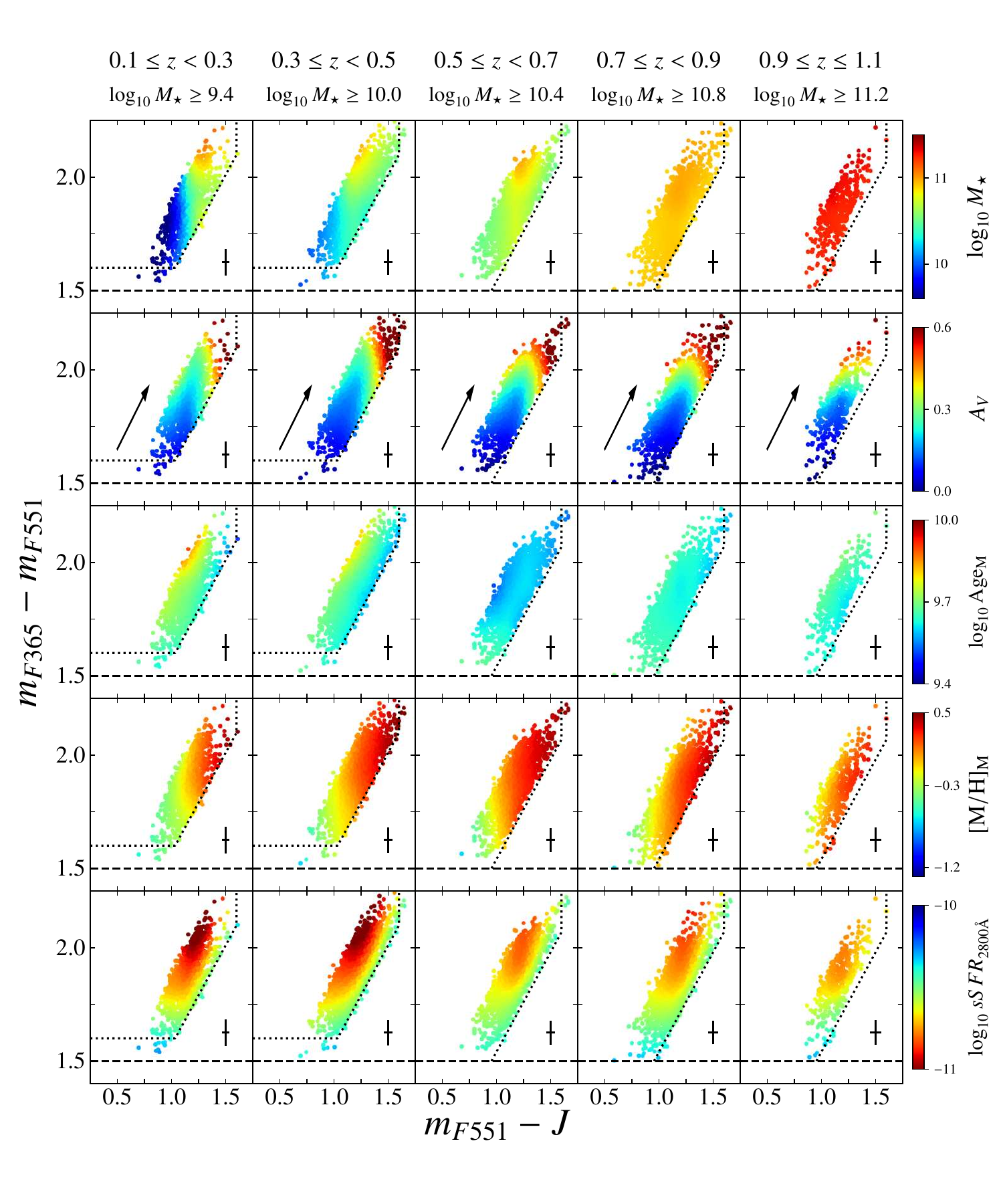}
\caption{Same as Fig.~\ref{fig:sp_UVJ}, but for the rest-frame colours $m_{F551}-J$ ($X$-axis) and $m_{F365}-m_{F551}$ ($Y$-axis), meaning non-dust corrected colours.}
\label{fig:sp_UVJ_av}
\end{figure*}

In Figs.~\ref{fig:sp_UVJ} and \ref{fig:sp_UVJ_av}, we see that fewer massive quiescent galaxies tend to populate the bluest parts of the $UVJ$-diagram, whereas at increasing red colours $(m_{F365}-m_{F551})_\mathrm{int}$ and $(m_{F551}-J)_\mathrm{int}$ quiescent galaxies continuously present larger stellar masses. However, the most massive galaxies are concentrated in the upper part of the $UVJ$-diagram at decreasing redshift, $(m_{F365}-m_{F551})_\mathrm{int} \sim 2.0$. Nevertheless, most massive galaxies are scattered over a wider $(m_{F365}-m_{F551})_\mathrm{int}$ range when we explore the highest redshift panels, $1.5\le (m_{F365}-m_{F551})_\mathrm{int} \le 2.0$. This has been extensively observed in the last years \citep[e.g.][]{Bower1992,Kauffmann2003,Gallazzi2005,Baldry2006,Peng2010}. We note that each panel comprises different stellar mass ranges, as indicated in the upper labels and according to Fig.~\ref{fig:completeness}, hence the less massive galaxies are only present in the local redshift bins.

From our SED fitting analysis, we see in Fig.~\ref{fig:sp_UVJ} that the whole sample shows an expected low dust content ($96$~\% of galaxies present $A_V \le 0.6$), where the more obscured quiescent galaxies lie on the bluer $(m_{F365}-m_{F551})_\mathrm{int}$ and $(m_{F551}-J)_\mathrm{int}$ intrinsic colour regions of the diagram. On the other hand, if dust effects on the colours are not corrected (observational rest-frame colours, Fig.~\ref{fig:sp_UVJ_av}), dusty galaxies populate the red parts of the diagram. Moreover, extinction is anticorrelated with both $(m_{F365}-m_{F551})_\mathrm{int}$ and $(m_{F551}-J)_\mathrm{int}$ dust-corrected colours, with Spearman correlation coefficients of $\rho_\mathrm{UV}\sim-0.92$ and $\rho_\mathrm{VJ}\sim-0.75$, respectively (see Table~\ref{tab:corr_spear}). As a sanity check, the colour changes owing to a dust reddening case with $A_V=0.5$ and $R_V = 3.1$ and the extinction law of \citet[][]{Fitzpatrick1999} are $\Delta(m_{F365}-m_{F551})\sim0.28$ and $\Delta(m_{F551}-J)\sim0.29$ (illustrated in Figs.~\ref{fig:sp_UVJ} and \ref{fig:sp_UVJ_av}). Our results are in good agreement with the colour changes predicted by this extinction law and the observed change in the galaxy distribution between Figs.~\ref{fig:sp_UVJ} and \ref{fig:sp_UVJ_av} at any redshift. We also note that the extinction values provided by MUFFIT are properly decoupled and are not significantly affected by degeneracies with the rest of stellar population parameters (age and metallicity). Otherwise, the extinction distribution in the $UVJ$ diagram would be randomly distributed or it would show other colour dependences.

It is clear from Fig.~\ref{fig:CMDE_sp} that the quiescent galaxies exhibiting the larger extinction are those that also present the bluest intrinsic colours. It is interesting  that there is a clear correlation of extinction with the intrinsic colour $(m_{F365}-m_{F551})_\mathrm{int}$, but also with the stellar mass of the galaxy since $z\sim1$. At fixed intrinsic colour $(m_{F365}-m_{F551})_\mathrm{int}$, the more massive the quiescent galaxy, the greater the reddening by dust. Whilst this correlation exists, discrepancies between extinction values are not very remarkable, since extinction is typically low for the quiescent population.

When we explore the same diagram but without correcting colours for extinction (see Fig.~\ref{fig:CMDE_sp_av}), the reddest galaxies are those with the greatest dust content (for $A_V \gtrsim 0.4$, $m_{F365}-m_{F551} \gtrsim 2$). Thus, at fixed colour $m_{F365}-m_{F551}$, more massive galaxies are less dust-reddened than  less massive ones.

\begin{table*}
\caption{Parameters $\alpha$, $\beta$, $\gamma$, $\delta$, $\epsilon$, and $\zeta$ that more closely fit the distribution of stellar population parameters (extinction, age, metallicity, and sSFR) within the rest-frame $UVJ$ colour--colour diagram corrected for extinction, see Eq.~(\ref{eq:quadra}), for the SSP models of BC03 and EMILES (including BaSTI and Padova00 isochrones). The last column illustrates the average data scatter with respect to the plane fit (Eq.~(\ref{eq:quadra})).}
\label{tab:uvj_corr}
\centering
\begin{tabular}{lcccccccc}
\hline\hline
& & \multirow{2}{*}{$\alpha$} & \multirow{2}{*}{$\beta$} & \multirow{2}{*}{$\gamma$} & \multirow{2}{*}{$\delta$} & \multirow{2}{*}{$\epsilon$} & \multirow{2}{*}{$\zeta$} & \multirow{2}{*}{$\sigma_\mathrm{fit}$} \\
&&&&&&&& \\
\hline
\parbox[t]{2mm}{\multirow{6}{*}{\rotatebox[origin=c]{90}{BC03}}} &&&&&&&& \\
& $A_V$ & $\hfill 1.78 \pm 0.03$ & $\hfill 2.59 \pm 0.06$ & $\hfill -2.70 \pm 0.07$ & $\hfill 0.83 \pm 0.09$ & $\hfill -6.94 \pm 0.17$ & $\hfill 6.55 \pm 0.14$ & $\hfill 0.02$ \\
& $\log_{10}\mathrm{Age}$ & $\hfill 0.78 \pm 0.02$ & $\hfill 1.93 \pm 0.03$ & $\hfill -1.88 \pm 0.04$ & $\hfill 1.49 \pm 0.05$ & $\hfill -4.35 \pm 0.10$ & $\hfill 12.34 \pm 0.08$ & $\hfill 0.04$ \\
& $\mathrm{[M/H]}$ & $\hfill 0.24 \pm 0.07$ & $\hfill -0.76 \pm 0.08$ & $\hfill -1.55 \pm 0.12$ & $\hfill 3.48 \pm 0.15$ & $\hfill 4.64 \pm 0.24$ & $\hfill -6.82 \pm 0.20$ & $\hfill 0.09$ \\
& $\log_{10}sSFR$ & $\hfill 0.42 \pm 0.06$ & $\hfill -1.20 \pm 0.14$ & $\hfill 0.49 \pm 0.15$ & $\hfill -1.30 \pm 0.19$ & $\hfill 1.39 \pm 0.41$ & $\hfill -9.41 \pm 0.30$ & $\hfill 0.04$ \\
&&&&&&&& \\
\hline
\parbox[t]{2mm}{\multirow{6}{*}{\rotatebox[origin=c]{90}{BaSTI}}} &&&&&&&& \\
& $A_V$ & $\hfill 1.25 \pm 0.05$ & $\hfill 0.99 \pm 0.05$ & $\hfill -0.50 \pm 0.07$ & $\hfill -1.77 \pm 0.10$ & $\hfill -3.74 \pm 0.13$ & $\hfill 5.13 \pm 0.11$ & $\hfill 0.02$ \\
& $\log_{10}\mathrm{Age}$ & $\hfill 0.83 \pm 0.03$ & $\hfill 2.23 \pm 0.03$ & $\hfill -2.79 \pm 0.04$ & $\hfill 3.09 \pm 0.06$ & $\hfill -4.71 \pm 0.08$ & $\hfill 12.15 \pm 0.07$ & $\hfill 0.07$ \\
& $\mathrm{[M/H]}$ & $\hfill -0.37 \pm 0.15$ & $\hfill -0.56 \pm 0.08$ & $\hfill 1.01 \pm 0.18$ & $\hfill 2.35 \pm 0.24$ & $\hfill -0.45 \pm 0.21$ & $\hfill -1.34 \pm 0.21$ & $\hfill 0.05$ \\
& $\log_{10}sSFR$ & $\hfill -0.83 \pm 0.07$ & $\hfill -0.79 \pm 0.09$ & $\hfill 1.47 \pm 0.13$ & $\hfill -0.95 \pm 0.16$ & $\hfill -0.34 \pm 0.24$ & $\hfill -8.58 \pm 0.19$ & $\hfill 0.07$ \\
&&&&&&&& \\
\hline
\parbox[t]{2mm}{\multirow{6}{*}{\rotatebox[origin=c]{90}{Padova00}}} &&&&&&&& \\
& $A_V$ & $\hfill 2.55 \pm 0.07$ & $\hfill 2.70 \pm 0.07$ & $\hfill -3.60 \pm 0.11$ & $\hfill 1.65 \pm 0.17$ & $\hfill -7.07 \pm 0.19$ & $\hfill 6.44 \pm 0.16$ & $\hfill 0.03$ \\
& $\log_{10}\mathrm{Age}$ & $\hfill 0.50 \pm 0.04$ & $\hfill 2.05 \pm 0.03$ & $\hfill -2.74 \pm 0.07$ & $\hfill 2.69 \pm 0.09$ & $\hfill -3.46 \pm 0.08$ & $\hfill 11.12 \pm 0.07$ & $\hfill 0.04$ \\
& $\mathrm{[M/H]}$ & $\hfill -6.23 \pm 0.13$ & $\hfill -0.12 \pm 0.07$ & $\hfill 6.15 \pm 0.15$ & $\hfill 4.66 \pm 0.25$ & $\hfill -6.94 \pm 0.22$ & $\hfill 3.10 \pm 0.21$ & $\hfill 0.06$ \\
& $\log_{10}sSFR$ & $\hfill -0.91 \pm 0.12$ & $\hfill -0.97 \pm 0.13$ & $\hfill 2.59 \pm 0.22$ & $\hfill -1.52 \pm 0.28$ & $\hfill -1.94 \pm 0.31$ & $\hfill -6.50 \pm 0.23$ & $\hfill 0.04$ \\
&&&&&&&& \\
\hline
\end{tabular}
\end{table*}

\begin{table*}
\caption{Same as Table~\ref{tab:uvj_corr}, but for the rest-frame stellar mass--colour diagram corrected for extinction (MCDE).}
\label{tab:cmde_corr}
\centering
\begin{tabular}{lcccccccc}
\hline\hline
& & \multirow{2}{*}{$\alpha$} & \multirow{2}{*}{$\beta$} & \multirow{2}{*}{$\gamma$} & \multirow{2}{*}{$\delta$} & \multirow{2}{*}{$\epsilon$} & \multirow{2}{*}{$\zeta$} & \multirow{2}{*}{$\sigma_\mathrm{fit}$}\\
&&&&&&& \\
\hline
\parbox[t]{2mm}{\multirow{6}{*}{\rotatebox[origin=c]{90}{BC03}}} &&&&&&&& \\
& $A_V$ & $\hfill -0.037 \pm 0.003$ & $\hfill 1.16 \pm 0.06$ & $\hfill -0.05 \pm 0.02$ & $\hfill 0.90 \pm 0.06$ & $\hfill -4.44 \pm 0.22$ & $\hfill -0.03 \pm 0.35$ & $\hfill 0.03$ \\
& $\log_{10}\mathrm{Age}$ & $\hfill 0.054 \pm 0.001$ & $\hfill 0.56 \pm 0.02$ & $\hfill 0.06 \pm 0.01$ & $\hfill -1.26 \pm 0.02$ & $\hfill -2.20 \pm 0.08$ & $\hfill 18.07 \pm 0.14$ & $\hfill 0.04$ \\
& $\mathrm{[M/H]}$ & $\hfill -0.058 \pm 0.005$ & $\hfill -0.49 \pm 0.08$ & $\hfill -0.40 \pm 0.04$ & $\hfill 2.08 \pm 0.09$ & $\hfill 7.29 \pm 0.33$ & $\hfill -19.37 \pm 0.59$ & $\hfill 0.07$ \\
& $\log_{10}sSFR$ & $\hfill -0.037 \pm 0.007$ & $\hfill -0.21 \pm 0.11$ & $\hfill 0.20 \pm 0.05$ & $\hfill 0.46 \pm 0.12$ & $\hfill -3.55 \pm 0.40$ & $\hfill -8.28 \pm 0.66$  & $\hfill 0.02$ \\
&&&&&&&& \\
\hline
\parbox[t]{2mm}{\multirow{6}{*}{\rotatebox[origin=c]{90}{BaSTI}}} &&&&&&&& \\
& $A_V$ & $\hfill -0.034 \pm 0.003$ & $\hfill 0.80 \pm 0.04$ & $\hfill 0.05 \pm 0.01$ & $\hfill 0.68 \pm 0.05$ & $\hfill -4.16 \pm 0.18$ & $\hfill 0.84 \pm 0.34$ & $\hfill 0.02$ \\
& $\log_{10}\mathrm{Age}$ & $\hfill 0.012 \pm 0.001$ & $\hfill 0.35 \pm 0.01$ & $\hfill 0.02 \pm 0.01$ & $\hfill -0.35 \pm 0.02$ & $\hfill -1.30 \pm 0.06$ & $\hfill 12.93 \pm 0.10$ & $\hfill 0.05$ \\
& $\mathrm{[M/H]}$ & $\hfill -0.110 \pm 0.002$ & $\hfill 0.71 \pm 0.02$ & $\hfill -0.21 \pm 0.01$ & $\hfill 2.95 \pm 0.04$ & $\hfill 0.05 \pm 0.11$ & $\hfill -17.51 \pm 0.26$ & $\hfill 0.06$ \\
& $\log_{10}sSFR$ & $\hfill -0.021 \pm 0.005$ & $\hfill 0.10 \pm 0.07$ & $\hfill 0.05 \pm 0.03$ & $\hfill 0.37 \pm 0.10$ & $\hfill -2.58 \pm 0.26$ & $\hfill -9.19 \pm 0.54$  & $\hfill 0.03$ \\
&&&&&&&& \\
\hline
\parbox[t]{2mm}{\multirow{6}{*}{\rotatebox[origin=c]{90}{Padova00}}} &&&&&&&& \\
& $A_V$ & $\hfill -0.033 \pm 0.003$ & $\hfill 0.98 \pm 0.06$ & $\hfill 0.05 \pm 0.02$ & $\hfill 0.64 \pm 0.06$ & $\hfill -4.97 \pm 0.21$ & $\hfill 1.82 \pm 0.37$ & $\hfill 0.02$ \\
& $\log_{10}\mathrm{Age}$ & $\hfill 0.037 \pm 0.001$ & $\hfill 0.31 \pm 0.02$ & $\hfill 0.00 \pm 0.01$ & $\hfill -0.85 \pm 0.02$ & $\hfill -0.72 \pm 0.07$ & $\hfill 14.94 \pm 0.13$ & $\hfill 0.03$ \\
& $\mathrm{[M/H]}$ & $\hfill -0.086 \pm 0.003$ & $\hfill 0.93 \pm 0.04$ & $\hfill -0.40 \pm 0.02$ & $\hfill 2.75 \pm 0.05$ & $\hfill 1.57 \pm 0.15$ & $\hfill -18.01 \pm 0.29$ & $\hfill 0.04$ \\
& $\log_{10}sSFR$ & $\hfill -0.041 \pm 0.007$ & $\hfill 0.14 \pm 0.11$ & $\hfill 0.14 \pm 0.05$ & $\hfill 0.68 \pm 0.12$ & $\hfill -4.05 \pm 0.34$ & $\hfill -9.27 \pm 0.66$  & $\hfill 0.02$ \\
&&&&&&&& \\
\hline
\end{tabular}
\end{table*}


\subsection{Ages of quiescent galaxies within the $UVJ$ and stellar mass--colour diagrams}\label{sec:ages}

Older quiescent galaxies in the sample are concentrated in the upper part of the $UVJ$ diagram (details in Figs.~\ref{fig:sp_UVJ} and \ref{fig:sp_UVJ_av}), and they therefore tend to populate the intrinsic redder colours $(m_{F365}-m_{F551})_\mathrm{int}$. Likewise, young quiescent galaxies lie on the bluest colours in concordance with the less massive systems in the sample. In Fig.~\ref{fig:sp_UVJ}, we see that the variety of ages presented by quiescent galaxies is linked to the scatter of $(m_{F365}-m_{F551})_\mathrm{int}$ partly (correlation coefficient of $\rho_\mathrm{UV}\sim0.45$, see Table~\ref{tab:corr_spear}), but not fully linked to this colour. In the rest-frame $UVJ$ diagram, we also find a non-negligible dependence of the age on $(m_{F551}-J)_\mathrm{int}$, although milder than the other intrinsic colour (correlation coefficient of $\rho_\mathrm{VJ}\sim0.15$). For the  EMILES and BaSTI isochrones the age and $(m_{F551}-J)_\mathrm{int}$ colour are not correlated with a value of $\rho_\mathrm{VJ}\sim-0.04$ (see Table~\ref{tab:corr_spear}). Consequently, our results match with previous findings \citep[e.g.][]{Whitaker2010} in which the $(m_{F365}-m_{F551})_\mathrm{int}$ colour is scattered by the ages in the quiescent population. This is not surprising because the $(m_{F365}-m_{F551})_\mathrm{int}$ colour ranges in the $4\,000$~{\AA} break, which is sensitive to age \citep[e.g.][]{Bruzual1983,Balogh1999}, even though this   also degenerates with the metallicity \citep[][]{Worthey1994,Peletier2013}. On the contrary, the non-dust corrected $UVJ$ diagram (see Fig.~\ref{fig:sp_UVJ_av}) shows that the observed colour $m_{F365}-m_{F551}$ is not driven by the age. As illustrated by Fig.~\ref{fig:sp_UVJ_av}, the trends of age with $m_{F365}-m_{F551}$ are not as clear as in the intrinsic colours, or they look inverted. Therefore, extinction   also plays an important role in this aspect by masking and blurring the relation between $(m_{F365}-m_{F551})_\mathrm{int}$ and age.

In Fig.~\ref{fig:CMDE_sp}, we find that the most massive quiescent galaxies are also the oldest and  intrinsically the reddest. This agrees with the `downsizing' scenario   \citep{Cowie1996}, where the more massive galaxies were formed in earlier epochs of the Universe with respect to their less massive counterparts. These very evolved galaxies also present very low extinction.  When studying the distribution of age in the stellar mass--colour diagram without correcting colours for dust (Fig.~\ref{fig:CMDE_sp_av}), the dependence between $m_{F365}-m_{F551}$ and age is softly blurred, although the most massive and reddest ($m_{F365}-m_{F551} \geq 1.8$) are also the oldest.


\subsection{Distributions of metallicity}\label{sec:feh}

Exploring the metallicity distribution in the $UVJ$ diagram (see Fig.~\ref{fig:sp_UVJ}) we find  that there is a tight correlation between metallicity and the intrinsic $(m_{F551}-J)_\mathrm{int}$ colour, where the correlation coefficient is $\rho_\mathrm{VJ}\sim0.85$ (see Table~\ref{tab:corr_spear}). There is a clear trend in which higher metallicity content ($\mathrm{[M/H]_M}>0.1$~dex) is present with  redder $(m_{F551}-J)_\mathrm{int} > 1.1$ colours. Hence, the most metal-rich quiescent galaxies lie on the right-hand side of the $UVJ$ diagram, and this trend is maintained at least up to redshift $z\sim1$. For a fixed $(m_{F551}-J)_\mathrm{int}$ colour, the influence of the metal content on the $(m_{F365}-m_{F551})_\mathrm{int}$ is almost negligible in a wide range of the colour $(m_{F551}-J)_\mathrm{int}$, but it is not null. Whilst the age distribution is distorted by extinction, the metallicity trend with the colour $m_{F551}-J$ is still prominent (see Fig.~\ref{fig:sp_UVJ_av}). The main extinction effect over metallicity distribution is that at the high-redshift panels (or the most massive galaxies, $\log_{10}M_\star > 10.8$~dex), the metallicity exhibits a more remarkable dependence on the $m_{F365}-m_{F551}$ colour with respect to its intrinsic counterpart (see Fig.~\ref{fig:sp_UVJ}) and it is substantially less affected than the age.

In Fig.~\ref{fig:CMDE_sp}, we observe that the stellar mass-metallicity relation is also present, reinforcing the reliability of the stellar population properties obtained by MUFFIT using ALHAMBRA photometry of galaxies. The more massive the galaxy, the more metal rich it is. This is usually referred to as the stellar mass--metallicity relation (MZR), which has been studied previously \citep[][]{Trager2000,Tremonti2004,Gallazzi2005,Panter2008,GonzalezDelgado2014a}. When we focus on the diagram without the dust correction (see Fig.~\ref{fig:CMDE_sp_av}), metallicity values populate well-defined regions of the  stellar mass--colour diagrams.


\subsection{sSFR within the $UVJ$ and stellar mass--colour diagrams}\label{sec:ssfr}

Like the other  stellar population properties explored in this work, lower and higher sSFRs of quiescent galaxies populate different colour ranges in dust-corrected $UVJ$ diagrams (details in Fig.~\ref{fig:sp_UVJ}). Our results indicate that the lowest sSFRs lie on the upper parts of this diagram, independently of the stellar mass and redshift. This is not surprising because at increasing stellar mass, quiescent galaxies exhibit lower sSFR and they are also redder. The correlation between sSFR and the intrinsic colour $(m_{F365}-m_{F551})_\mathrm{int}$ is really remarkable, with a value $\rho_\mathrm{UV}$ close to $-1$ (see Table~\ref{tab:corr_spear}). Furthermore, there is also an anticorrelation with $(m_{F551}-J)_\mathrm{int}$, although milder ($\rho_\mathrm{VJ}\sim -0.65$). It should be noted that sSFRs were obtained from the luminosity at $2\,800$~\AA, Eq.~(\ref{eq:sfr2800}), but we find evidence  to propose the intrinsic colour $(m_{F365}-m_{F551})_\mathrm{int}$ as an alternative sSFR tracer, at least for quiescent galaxies. Nevertheless, this tracer should be treated carefully. When there is no star formation, Eq.~(\ref{eq:sfr2800}) provides non-null SFRs because there is a non-null continuum at $2\,800$~\AA, although these values can be treated as upper sSFR limits. In Fig.~\ref{fig:sp_UVJ_av}, we find  that the correlation between $(m_{F365}-m_{F551})$ and sSFR is not as clear as the correlation with intrinsic colour. However, certain range colours present predominantly lower sSFRs. The lowest sSFR values lie on the upper parts of classic $UVJ$ diagrams (non-dust corrected), whereas the highest values present bluer colours. This   matches with the distribution of stellar mass in these diagrams (see Fig.~\ref{fig:sp_UVJ_av}), which is not surprising. As above, extinction displaces galaxies within these diagrams, which means that dust-obscured quiescent galaxies with higher sSFRs also lie on the upper parts of $UVJ$ diagrams.

\begin{figure*}
\centering
\includegraphics[trim= 0 0 0 0,width=17cm,clip=True]{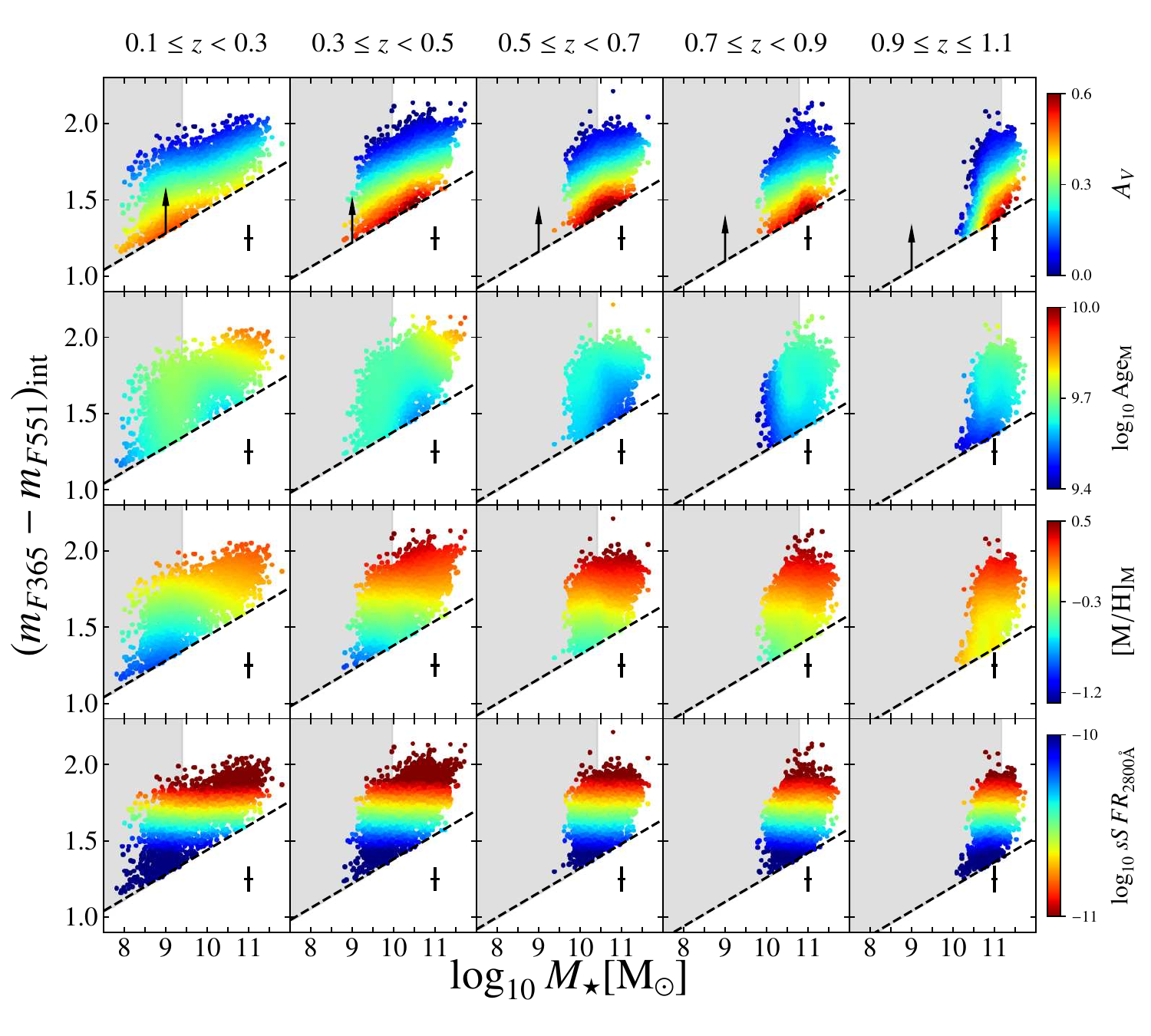}
\caption{Stellar population parameters in the rest-frame stellar mass--colour diagram. At different redshift bins, we present the stellar mass ($X$-axis) and intrinsic colour $(m_{F365}-m_{F551})_\mathrm{int}$ ($Y$-axis) after correcting for extinction. The stellar population  parameters are colour-coded according to their values using BC03 SSP models (see  colour bars). From top to bottom: extinction,  mass-weighted age and metallicity, and specific star formation rate. All the parameters were spatially averaged through a LOESS method. Black crosses illustrate the median uncertainties in both stellar mass and $(m_{F365}-m_{F551})_\mathrm{int}$ intrinsic colour. The dashed line illustrates the colour limit for selecting quiescent galaxies in the MCDE, see Eq.~(\ref{eq:CMDE_main}) and Table~\ref{tab:cmde_main}, for this work. The shaded regions show the stellar mass range in which our quiescent sample is not complete in stellar mass. We illustrate the colour variations owing to a reddening of $A_V=0.5$ (black arrow), assuming the extinction law of \citet{Fitzpatrick1999}.}
\label{fig:CMDE_sp}
\end{figure*}

\begin{figure*}
\centering
\includegraphics[trim= 0 0 0 0,width=17cm,clip=True]{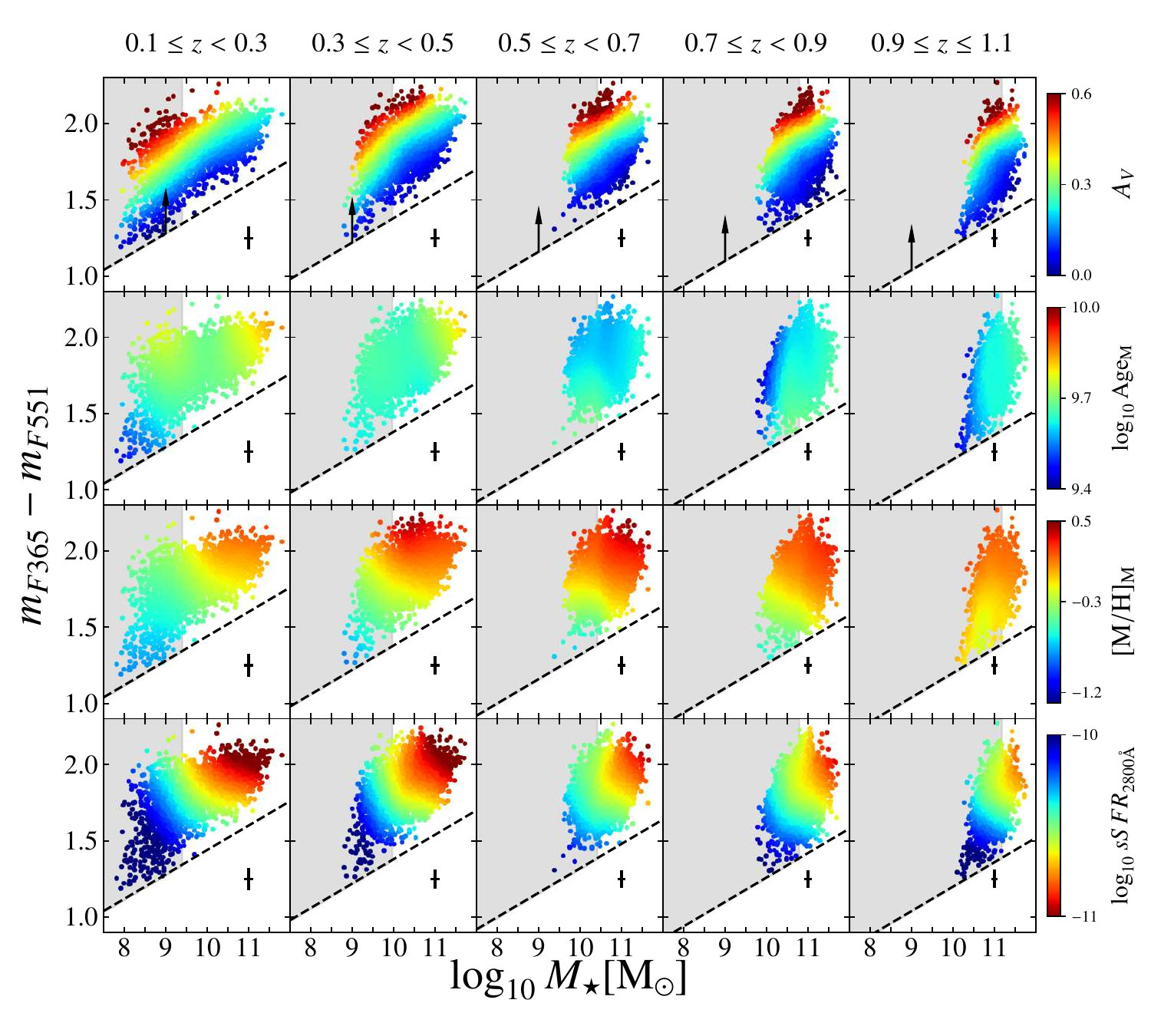}
\caption{Same as Fig.~\ref{fig:CMDE_sp}, but for the rest-frame colour $m_{F365}-m_{F551}$ ($X$-axis, non-dust corrected colour).}
\label{fig:CMDE_sp_av}
\end{figure*}

The sSFR values of quiescent galaxies also present well-defined loci in the MCDE (see lower panels in  Fig.~\ref{fig:CMDE_sp}). At increasing stellar mass, lower sSFR values are obtained. It is interesting that the intrinsic colour $(m_{F365}-m_{F551})_\mathrm{int}$ can be interpreted as an alternative sSFR tracer, also motivated by the high anticorrelation that presents with this colour, $\rho_\mathrm{UV}\sim-1$ (see Table~\ref{tab:corr_spear}). However, to constrain the sSFR of quiescent galaxies using $m_{F365}-m_{F551}$, it is necessary to include the stellar mass. As in previous stellar population properties,  extinction is an important parameter that dilutes the correlations and trends revealed in this work (see Fig.~\ref{fig:CMDE_sp_av}).

\begin{table}
\caption{Spearman correlation coefficients between the dust-corrected colours $(m_{F365}-m_{F551})_\mathrm{int}$ and $(m_{F551}-J)_\mathrm{int}$ and stellar population parameters ($\rho_\mathrm{UV}$ and $\rho_\mathrm{VJ}$, respectively).}
\small
\label{tab:corr_spear}
\centering
\begin{tabular}{ccccccc}
\hline\hline
& \multicolumn{2}{c}{\multirow{2}{*}{BC03}} & \multicolumn{2}{c}{\multirow{2}{*}{BaSTI}} & \multicolumn{2}{c}{\multirow{2}{*}{Padova00} } \\
& \multirow{2}{*}{$\rho_\mathrm{UV}$} & \multirow{2}{*}{$\rho_\mathrm{VJ}$} & \multirow{2}{*}{$\rho_\mathrm{UV}$} & \multirow{2}{*}{$\rho_\mathrm{VJ}$} & \multirow{2}{*}{$\rho_\mathrm{UV}$} & \multirow{2}{*}{$\rho_\mathrm{VJ}$}\\
&&&&&& \\
\hline
&&&&&& \\
 $A_V$ & $\hfill -0.90$ & $\hfill -0.80$ & $\hfill -0.95$ & $\hfill -0.77$ & $\hfill -0.91$ & $\hfill -0.67$ \\
 $\log_{10}\mathrm{Age}$ & $\hfill 0.61$ & $\hfill 0.23$ & $\hfill 0.30$ & $\hfill 0.17$ & $\hfill 0.46$ & $\hfill 0.07$ \\
 $\mathrm{[M/H]}$ & $\hfill 0.68$ & $\hfill 0.84$ & $\hfill 0.24$ & $\hfill 0.75$ & $\hfill 0.46$ & $\hfill 0.80$ \\
 $\log_{10}sSFR$ & $\hfill -0.98$ & $\hfill -0.55$ & $\hfill -0.98$ & $\hfill -0.75$ & $\hfill -0.96$ & $\hfill -0.67$ \\
&&&&&& \\
\hline
\end{tabular}
\end{table}

Although it is well known that the use of different stellar population models yields different quantitative results, we find a great qualitative agreement between all the predictions obtained using the BC03 and EMILES SSP models. All the conclusions drawn in this section can be extrapolated when EMILES SSP models (both BaSTI and Padova00 isochrones) are used.


\section{Discussion}\label{sec:discussion}


\subsection{New insights into the green valley}\label{sec:discussion_green}

Dust corrections play an important role in understanding how quiescent galaxies are distributed across $UVJ$ diagrams as a function of their parameters: stellar mass, age, metallicity, and extinction. One of the most important results in this paper is that the green valley is largely populated by DSF galaxies ($\sim65$~\%), and therefore the number of galaxies in the `real' green valley is much lower than previously claimed. This implies that the transition of galaxies from the blue cloud to the red sequence, and hence the related mechanisms for quenching, should be much more efficient and faster than previously considered. 

\begin{figure}
\centering
\resizebox{\hsize}{!}{\includegraphics{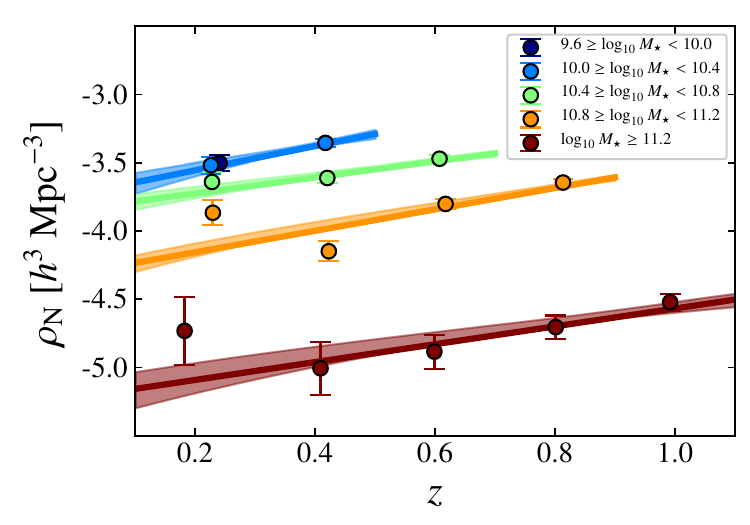}}
\caption{Evolution of co-moving number density, $\rho_\mathrm{N}(z)$, of green valley galaxies with redshift ($X$-axis) at  different stellar mass bins (see inset) using BC03 SSP models. Coloured markers illustrate co-moving number densities at redshift bins and vertical bars are their uncertainties. Solid lines illustrate the best fit to Eq.~(\ref{eq:comoving}), whereas the shaded regions show the 1$\sigma$ uncertainty of the fit.}
\label{fig:comoving_green}
\end{figure}

Any successful mechanisms proposed for shutting down the star formation should account for a less populated green valley and a shorter transition timescale. For instance, the common presence of AGNs in galaxies with intermediate host-galaxy colours at $z \lesssim 1$ \citep{Nandra2007,Bundy2008,Georgakakis2008,Silverman2008,Hickox2009,Schawinski2009,Povic2012} was interpreted as  evidence that AGNs heat up the gas in the host-galaxy \citep{Silk1998,DiMatteo2008} and shut down the star formation rapidly  \citep{Bower2006,Croton2006,Faber2007,Schawinski2007,Schawinski2014}. In light of our results, AGN feedback should be more efficient than previously expected as a quenching mechanism. Although the detailed study of the green valley galaxies and its evolution across cosmic time is beyond the scope of this work, we provide the co-moving number density ($\rho_\mathrm{N}$) of this kind of galaxy up to $z\sim1$ assuming a power-law function of the form
\begin{equation}\label{eq:comoving}
\rho_\mathrm{N}=\rho_0(1+z)^\gamma\ .
\end{equation}
For this work and by the MCDE, we define the green valley as the loci between the limiting values of quiescent galaxies (see Eq.~(\ref{eq:CMDE_main}) and Table~\ref{tab:cmde_main}) and a bluer intrinsic colour of $\Delta(m_{F365}-m_{F551})_\mathrm{int}\sim0.2$ with respect to these relations. It should be noted that this definition strictly involves galaxies transiting from the main to the quiescent sequence, but in a more general case green valley galaxies can be treated as third and overlapping galaxy group between quiescent and star-forming galaxies \citep[see e.g.][]{Ilbert2010}. In Table~\ref{tab:density_par} we list the values of $\rho_0$ and $\gamma$ that best fit Eq.~(\ref{eq:comoving}) at different mass bins and SSP models (see also Fig.~\ref{fig:comoving_green}). To associate uncertainties with the co-moving number densities, we assumed Poisson errors. Independently of the SSP model set used for the analysis, the number density of green valley galaxies decreases at increasing stellar mass. It is also interesting that the number of massive galaxies ($\log_{10}M_\star \gtrsim 10.8$) in the green valley decreases at lower redshifts. However, we note that our results at $\log_{10}M_\star \lesssim 10.8$ only involve a few points ($z \lesssim 0.5$) and cosmic variance is not included in the error budget.

\begin{table*}
\caption{Values $\rho_0$ and $\gamma$ that best fit our co-moving number density calculation (see Eq.~(\ref{eq:comoving})) of green valley galaxies at different stellar mass bins and SSP models.}
\label{tab:density_par}
\centering
\begin{tabular}{rcccccc}
\hline\hline
& \multicolumn{2}{c}{BC03} & \multicolumn{2}{c}{EMILES+BaSTI} & \multicolumn{2}{c}{EMILES+Padova00} \\
& \multirow{2}{*}{$\log_{10}\rho_0$} & \multirow{2}{*}{$\gamma$} & \multirow{2}{*}{$\log_{10}\rho_0$} & \multirow{2}{*}{$\gamma$} & \multirow{2}{*}{$\log_{10}\rho_0$} & \multirow{2}{*}{$\gamma$} \\
&&&&&& \\
\hline
&&&&&& \\
$10.0 \le \log_{10}M_\star < 10.4$ & $-1.42^{+0.29}_{-0.53}$ & $2.64^{+0.76}_{-0.77}$ & $-1.49^{+0.22}_{-0.31}$ & $2.14^{+0.41}_{-0.40}$ & $-1.16^{+0.15}_{-0.21}$ & $2.94^{+0.52}_{-0.49}$ \\
$10.4 \le \log_{10}M_\star < 10.8$ & $-2.09^{+0.34}_{-0.52}$ & $1.85^{+0.40}_{-0.40}$ & $-2.13^{+0.27}_{-0.39}$ & $1.61^{+0.26}_{-0.27}$ & $-1.96^{+0.24}_{-0.34}$ & $1.79^{+0.29}_{-0.29}$ \\
$10.8 \le \log_{10}M_\star < 11.2$ & $-1.64^{+0.15}_{-0.18}$ & $2.64^{+0.31}_{-0.30}$ & $-1.46^{+0.08}_{-0.10}$ & $2.72^{+0.20}_{-0.20}$ & $-1.90^{+0.16}_{-0.19}$ & $2.06^{+0.22}_{-0.21}$ \\
$\log_{10}M_\star \ge 11.2$        & $-2.23^{+0.39}_{-0.64}$ & $2.35^{+0.58}_{-0.58}$ & $-4.18^{+0.96}_{-1.99}$ & $1.09^{+0.35}_{-0.37}$ & $-2.57^{+0.45}_{-0.71}$ & $1.87^{+0.45}_{-0.44}$ \\
&&&&&& \\
\hline
\end{tabular}
\tablefoot{There are no $\rho_0$ and $\gamma$ fitting values for the lowest stellar mass bin, $9.6 \le \log_{10}M_\star < 10$, because this subsample is only available at the lowest redshift bin, $0.1 \le z < 0.3$. All the values were obtained at setting $h=1$.}
\end{table*}

Previous studies \citep{Bell2005,Cowie2008,Brammer2009,Cardamone2010,Mahoro2017} also supported that the population of galaxies in the green valley has bluer intrinsic colours and it is largely dominated by DSF galaxies. Based on the results by \citet{Cardamone2010}, $\sim75$~\% of the galaxies in the green valley have intrinsic blue colours in good agreement with our work ($\sim65$~\%). In the same work and after a dust correction of the $U-V$ colour, most of the galaxies hosting an AGN belong to the red sequence, $(U-V)_\mathrm{int} \gtrsim 1.5$, or the blue cloud, $(U-V)_\mathrm{int} \lesssim 0.8$, with a poor presence of AGNs at intermediate colours, favouring again the idea of either a faster or less frequent quenching mechanism due to AGNs.


\subsection{Lower $U-V$ colour limit of $UVJ$-like diagrams}\label{sec:discussion_diagrams}

Even though the rest-frame $UVJ$ diagram has been extensively used for the selection of quiescent galaxies, the lower limit of colour $U-V$ is still uncertain or fixed arbitrarily. Some authors propose different definitions of the colour limits to minimise the impact of DSF galaxies \citep[see e.g.][]{Moresco2013}. In addition, $UVJ$ diagrams are only defined by two colours, where parameters such as stellar mass are not directly accounted for. The detailed analysis of the distribution of stellar population properties in colour--colour diagrams is key to revealing the likely bias that the different colour selections can introduce. 

\begin{figure*}
\centering
\includegraphics[trim= 0 0 0 0,width=17cm,clip=True]{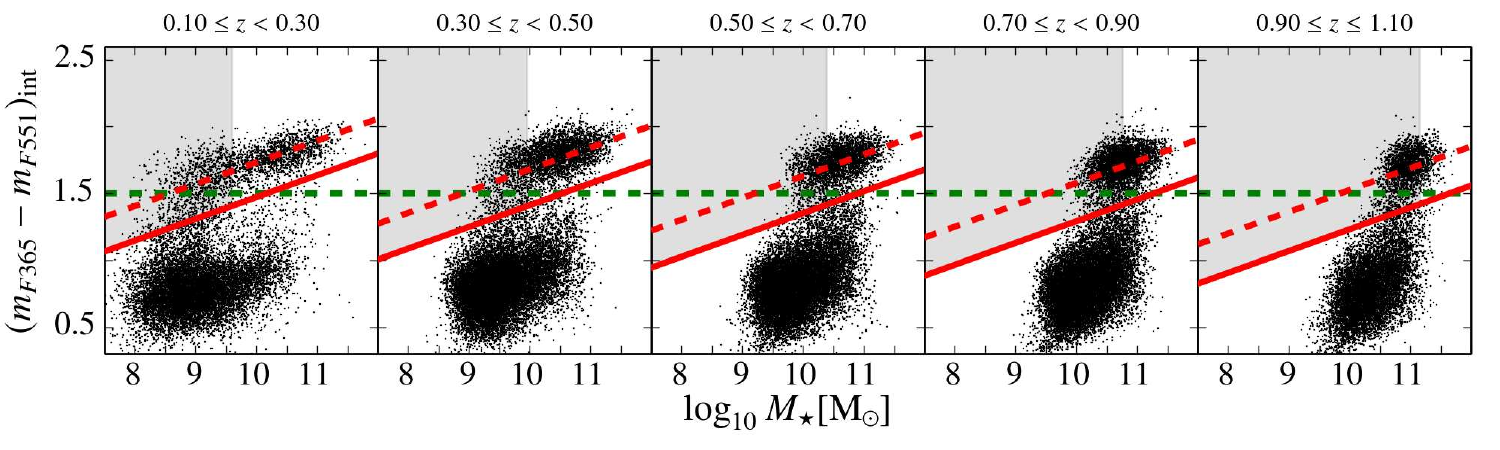}
\caption{Galaxies from the ALHAMBRA survey (black dots) in the rest-frame stellar mass--colour diagram corrected for extinction (MCDE) at different redshift bins. The shaded region illustrates the stellar mass range in which our sample of quiescent galaxies is not complete in stellar mass (see Sect.~\ref{sec:completeness}) at each redshift bin. The dashed green line shows the limiting value $(m_{F365}-m_{F551})_\mathrm{int}=1.5$ used to select quiescent galaxies in Sect.~\ref{sec:sample_uvj}. The dashed red line exhibits the main sequence of quiescent galaxies in the MCDE, see Eq.~(\ref{eq:CMDE_main}) and Table~\ref{tab:cmde_main}, and the solid line is the limiting colour value used to select  quiescent galaxies in this diagram, see Eq.~(\ref{eq:CMDE_main}) and Table~\ref{tab:cmde_main}. The stellar population predictions were obtained through BC03 SSP models.}
\label{fig:CMDE_limit}
\end{figure*}

Our results in Sect.~\ref{sec:uvj} point out that low-mass quiescent galaxies lie at the lower limits of the box classically defined for selecting the quiescent population. Therefore, this kind of colour--colour diagram should be built accordingly with the specific features of each photometric survey. Taking advantage of the MCDE (see Fig.~\ref{fig:CMDE_limit}) we can state that the less massive quiescent galaxies can present $U-V$ colours as blue as the ones of massive star-forming galaxies. These criteria change with redshift because the intrinsic colours of galaxies also vary with cosmic time (see Eq.~(\ref{eq:CMDE_main}) and Table~\ref{tab:cmde_main}). This  is clearly illustrated in Fig.~\ref{fig:CMDE_limit} at $0.1 \le z < 0.3$,  where a bluer colour limit of $(m_{F365}-m_{F551} < 1.5)_\mathrm{int}$ is necessary to not bias the sample in the low-mass regime. However, this limit must be set in conjunction with the stellar mass to avoid the inclusion of massive star-forming galaxies in the quiescent sample.

In this work, we obtain that the quiescent sample from ALHAMBRA (complete in magnitude down to $m_{F814}=23$) defined via the dust-corrected $UVJ$ colour--colour diagram and the MCDE is equivalent in the stellar mass range for which the quiescent sample is complete in stellar mass (see details in Fig.~\ref{fig:CMDE_limit}). For stellar masses below completeness, the only criteria $(m_{F365}-m_{F551})_\mathrm{int} > 1.5$ is not  enough to select quiescent galaxies. Although dust corrections exhibit a clear bimodality (see also Fig.~\ref{fig:ccd_noav}), at $(m_{F365}-m_{F551})_\mathrm{int} \sim 1.5$ there is not a well-defined border limiting the two populations in the $UVJ$ diagrams. This is because these galaxies present similar colours, even after correcting colours for dust-reddening, and the key to split both populations is the inclusion of stellar mass. This would favour the use of the MCDE with respect to the dust corrected $UVJ$ diagrams. Again, we note that in our case the two methods are equivalent to building a quiescent sample complete in stellar mass and luminosity (down to $m_{F814}=23$, see Fig.~\ref{fig:CMDE_limit}).

For deeper surveys, $I > 23$~mag, the selection bias can be even larger if these issues are not taken into account. Less massive quiescent galaxies will be detected, and therefore the $U-V$ lower colour limit should be bluer to not bias them. On the other hand, deeper surveys allow us to explore wider redshift ranges and the use of a unique definition like $(m_{F365}-m_{F551})_\mathrm{int} > 1.5$ is not strictly adequate because galaxy colours evolve with cosmic time. In a general case and taking the explored dust-corrected diagrams in this work into account, a pure non-biased selection of quiescent galaxies via colour diagrams should include dust corrections of the involved colours and galaxy stellar masses. This discussion can be extended to the classical $UVJ$ diagrams in which the colour correction for extinction is not included. Otherwise, samples of red $UVJ$ galaxies at low-mass regimes ($\log_{10}M_\star \sim 9.4$) may include up to   $40$~\% of DSF galaxies. However, the expected contamination for massive quiescent galaxies selected via the non-dust corrected $UVJ$ diagram is less affected by this fact (a $2$--$8$~\% fraction at $\log_{10}M_\star > 11$), especially at lower redshifts.

\subsection{Dusty star-forming galaxies and flux emission at $24$~$\mu\mathrm{m}$}\label{sec:sfr24}

The dust emission at $24$~$\mu\mathrm{m}$ is an additional SFR tracer \citep[see e.g.][]{Kennicutt1998} that can be used to confirm the results provided by MUFFIT about DSF galaxies (see   Sect.~\ref{sec:dusty}). It should be noted that the SFR obtained from the $24$~$\mu\mathrm{m}$ flux is independent of those provided by MUFFIT using ALHAMBRA data.

The ALHAMBRA field number four (hereafter ALHAMBRA-4) partly overlaps with the S-COSMOS field \citep{Sanders2007}, which allowed us to explore the SFRs of both DSF and quiescent galaxies in a shared subsample. In brief, the S-COSMOS is a deep infrared survey acquired with the \textit{Spitzer} Space Telescope at the COSMOS field making use of the  the Infrared Array Camera (IRAC) and  the Multiband Imaging Photometer (MIPS) cameras. We adopted the S-COSMOS MIPS $24$~$\mu\mathrm{m}$ Photometry Catalog October $2008$\footnote{\url{http://irsa.ipac.caltech.edu/data/COSMOS/}} as a reference data set. This $24$~$\mu\mathrm{m}$ catalogue is $\sim90$~\% ($\sim75$~\%) complete above $S_{24\mu\mathrm{m}}\sim80$~$\mu\mathrm{Jy}$ ($S_{24\mu\mathrm{m}}\sim60$~$\mu\mathrm{Jy}$), where the full width at half maximum of the MIPS $24~\mu\mathrm{m}$ PSF is $\sim6\arcsec$ \citep{Lefloch2009}.

To convert the observed flux at $24$~$\mu\mathrm{m}$ to rest-frame luminosities ($L_{24\mu\mathrm{m}}^{z=0}$), we adopted the interstellar medium (ISM) model proposed by \citet{Dale2002} and the photo-$z$ constraints provided by MUFFIT. Briefly, \citet{Dale2002} built a SED model of the ISM from combinations of different dust masses assuming a power-law distribution, where the relative contribution of each component in the model is parametrised by a parameter $\alpha$ \citep[see Eq.~(1) and Table~2 in][]{Dale2002}. Finally, we converted this rest-frame flux to SFRs through the relation provided by \citet{Calzetti2007}:
\begin{equation}\label{eq:sfr24}
{SFR}_{24\mu\mathrm{m}}=1.31\ 10^{-38}  \times {L_{24\mu\mathrm{m}}^{z=0}}^{0.885}\ ,
\end{equation}
where ${SFR}_{24\mu\mathrm{m}}$ is quoted in $M_\sun$~yr$^{-1}$ units.

After cross-correlating the ALHAMBRA and $24~\mu\mathrm{m}$ MIPS catalogues, we found a total of $1261$ common galaxies at $0.2 \le z \le 1.2$, where $58$ galaxies were labelled as DSF galaxies during our SED fitting analysis. It should be  noted that the low number of shared DSF galaxies is mainly due to the reduced overlapping area between the two surveys ($\sim0.25$~deg$^2$). As a result,  $90 \pm 10$~\%  of them ($51$ DSF galaxies) exhibit ${SFR}_{24\mu\mathrm{m}}$ values higher than the lower limit established by Eq.~(\ref{eq:sfr2800_lim}), that is, in the range of star-forming galaxies and reinforcing that these galaxies are reddened by dust. Nevertheless, we find that the subsample of DSF galaxies detected by MIPS is biased to the brightest galaxies, whereas the fainter DSF galaxies may remain under the MIPS detection limit. It is also worth mentioning that discrepancies between the ALHAMBRA and MIPS PSFs ($\sim1\arcsec$ and $6\arcsec$, respectively) may be an extra factor to consider for a fair comparison. This may introduce a detection bias because some DSF galaxies would not be deblended and subsequently included in the MIPS catalogue owing to a lower resolution. In this regard, some of the DSF galaxies may show an enhanced $24\mu\mathrm{m}$ flux due to nearby projected sources. Interestingly,   $91$~\% of the quiescent galaxies in ALHAMBRA-4 were not detected by MIPS. However, this result might be biased owing to the reasons discussed above.


\subsection{Consistency of the results using alternative attenuation laws}\label{sec:discussion_ext}

As our definition of the quiescent sample is based on colours corrected for extinction, the choice of an extinction law may have an impact on our results. To explore whether the choice of another extinction law may vary the main results of this research, we repeated our SED fitting analysis using MUFFIT and the attenuation law of \citet{Calzetti2001} instead of the \citet{Fitzpatrick1999} law.

After running MUFFIT with the attenuation law of \citet{Calzetti2001}, we checked the number of quiescent galaxies that we would obtain using the same criteria as   presented in Sects.~\ref{sec:sample_uvj} and \ref{sec:mcd}. As a result, we find that the number of quiescent galaxies differs by around   $3$~\%. These discrepancies in number are slightly bigger for DSF galaxies. The number of DSF galaxies is  $15$~\%   larger  when using the \citet{Calzetti2001} than the \citet{Fitzpatrick1999} extinction law. Therefore, our results using the \citet{Calzetti2001} law point out that the level of contamination of the quiescent sample owing to the presence of DSF galaxies in $UVJ$ diagrams non-corrected for extinction is around   $25$~\%,  in agreement with the results presented in this work using the \citet{Fitzpatrick1999} extinction law. If we explore the amount of contamination of DSF galaxies in the most and least massive quiescent galaxies of the sample, we obtain again that less massive samples of quiescent galaxies are more affected by this contamination. In particular, quiescent galaxies with stellar masses around $\log_{10}M_\star = 9.6$~dex present a $30$~\% fraction of DSF galaxies when they are selected through a classical $UVJ$ diagram, whereas for $\log_{10} M_\star > 11.2$~dex this number ranges from $5$~\% to $15$~\% at $0.1 < z < 0.3$ and $0.9 < z < 1.1$, respectively.

If we explore the presence of DSF galaxies in the green valley using the \citet{Calzetti2001} attenuation law, we obtain the same result as  using the \citet{Fitzpatrick1999} extinction law. A relevant number of the galaxies that populate the green valley are actually obscured star-forming galaxies. Assuming the same limits for the green valley as in Sects.~\ref{sec:sample_uvj} and \ref{sec:mcd} (i.e.~$0.1$ mag above and below the limiting values expressed by Eq.~(\ref{eq:quiescent}) and Eq.~(\ref{eq:CMDE_main}) for the $UVJ$ and stellar mass--colour diagrams, respectively), the number of DSF galaxies in the green valley is in the range of $25$--$60$~\%.

Finally, we confront one-by-one the stellar population properties obtained  using the \citet{Calzetti2001} and \citet{Fitzpatrick1999} extinction laws for the subsample of quiescent galaxies complete in stellar mass. Our analysis reveals a good agreement between the two predictions and without the presence of systematics. The main consequence of using different attenuation assumptions during the stellar population analysis of quiescent galaxies via SED fitting is the scatter. From this test, we quantify that the scatter is typically below $0.05$~dex for all the parameters explored in this work, except for metallicity that amounts to $0.15$~dex, but without systematics. This result is not surprising because, as revealed in this study, the values of extinction in quiescent galaxies are typically low (average value of $A_V \sim 0.3$). We find that the use of different synthesis models, such as BC03 and EMILES, is a much larger source of systematics   than the use of different extinction or attenuation laws \citep[see also][]{DiazGarcia2018}.


\section{Summary and conclusions}\label{sec:conclusions}

Using the data set provided by the ALHAMBRA multi-filter photometric survey and our optimised SED fitting tool MUFFIT \citep[extensively detailed in][]{DiazGarcia2015} with the BC03 and EMILES SSP models, we explored the stellar content of quiescent galaxies since $z=1.1$, corresponding to the last $8$~Gyr ($60$~\% of the age of the Universe), and their distributions in the dust-corrected stellar mass--colour and the $UVJ$ colour--colour diagrams.

The selection of the quiescent sample was carefully performed to minimise as much as possible the number of contaminants, mainly faint cool stars and dusty star-forming galaxies. In brief, we improved the colour--colour $UVJ$ diagram taking the extinctions provided by MUFFIT into account to generate a dust-corrected $UVJ$ diagram, which minimises the contamination by dusty star-forming galaxies. We also explored the use of  stellar mass--colour diagrams corrected for extinction (MCDE), an alternative diagram for the selection of quiescent galaxies, and compared the results with  to those from the $UVJ$ diagram. Moreover, we carried out a one-by-one visual inspection to remove those sources whose photometry might  be compromised (e.g.~with bad CCD regions or bad photometric apertures) and spurious detections. Based on SED fitting techniques, we removed cool stars from the ALHAMBRA catalogue with magnitudes ranging $22.5 \le m_{F814W} \le 23$. After comparison with the morphological classification of COSMOS for a common sample of sources, the contamination due to cool stars was reduced from $24$~\% to $4$~\%. Finally, we estimated the stellar mass completeness of our sample of quiescent galaxies in ALHAMBRA by applying a novel method for  defining an analytic function for the stellar mass completeness that yields a final sample of $\sim8\,500$ quiescent galaxies at $0.1 \le z \le 1.1$ with a photo-$z$ accuracy of $\sigma_\mathrm{NMAD}=0.006$.

We also developed a reliable methodology for taking advantage of our SED fitting results (based on a mixture of two SSPs) and provide SFR and sSFR values taking the luminosity at $2\,800$~\AA\ as SFR tracer. The exploration of the distribution of SFR and sSFR values in the stellar mass plane reveals a bimodality, where the main sequence  galaxies (star-forming galaxies) populate the upper SFR and sSFR values. In addition, the more massive quiescent galaxies exhibit higher SFRs than their lower mass counterparts. However, the most efficient process of star formation in the quiescent population resides in the low-mass systems, whereas the massive ones present the lowest sSFRs of the quiescent population. 

From the dust-corrected $UVJ$-diagram, we find  a number of striking results:
\begin{itemize}
\item[$\bullet$]
A relevant number of the galaxies that reside in the green valley are actually obscured star-forming galaxies ($\sim30$--$65$~\%). These reveal an intrinsic colour $(m_{F365}-m_{F551})_\mathrm{int}$ characteristic of the star-forming population, although their great dust content significantly reddens their colours. This also implies that the green valley is less populated than expected and the transition of galaxies from the blue cloud to the red sequence, and hence the related mechanisms for quenching, seems to be much more efficient and faster than previously considered. As this result may be very important to derive the quenching timescales of galaxies in future papers, we also provide the co-moving number densities of green valley galaxies after accounting for dust effects.
\item[$\bullet$]
Down to $m_{F814W}=23$, the histogram of the $(m_{F365}-m_{F551})_\mathrm{int}$ colour exhibits a local minimum at $\sim 1.5$ that can be imposed as the bluest colour limit to fairly select the quiescent sample, which remains roughly constant since $z \le 1.1$.
\item[$\bullet$]
Red galaxies that belong to the star-forming sample after the dust correction (intrinsic $m_{F365}-m_{F551}<1.5$) are typically concentrated close to the edges of $UVJ$ diagrams \citep[e.g.][]{Williams2009,Moresco2013}, supporting the reliability of the extinction values provided by MUFFIT.
\item[$\bullet$]
Quiescent galaxies selected through a classical $UVJ$ diagram (not corrected for dust effects) are typically contaminated by a $\sim20$~\% fraction of dusty star-forming galaxies. Our results clearly establish that this contamination is less severe for massive galaxies ($\log_{10}M_\star \ge 11$, $2$--$8$~\% from $z\sim 0.1$ to $z \sim 1.1$) than for the less massive ones ($40$~\%, for $9.2 \le \log_{10}M_\star \le 9.6$ at $0.1 \le z \le 0.3$).
\end{itemize}

The analysis of the distribution of stellar population parameters of quiescent galaxies on a dust-corrected $UVJ$-diagram and MCDE reveals that there is a close correlation between the position of each galaxy in these diagrams and its age, metallicity, extinction, and stellar mass. We conclude the following:

\begin{itemize}
\item[$\bullet$]
The more massive quiescent galaxies lie on the redder parts of the $UVJ$ diagram at greater cosmic times. At higher redshifts, $z\sim 1$, massive quiescent galaxies present a wider spread of $(m_{F365}-m_{F551})_\mathrm{int}$ colours, $1.5\le (m_{F365}-m_{F551})_\mathrm{int} \le 2.0$, than at $z\sim 0.2$, where they lie on the upper and redder parts of the $UVJ$ diagram, $(m_{F365}-m_{F551})_\mathrm{int} \sim 2.0$.
\item[$\bullet$]
The whole quiescent sample shows an expected low dust content ($96$~\% of the galaxies present $A_V \le 0.6$), where quiescent galaxies with bluer intrinsic colours, $(m_{F365}-m_{F551})_\mathrm{int} \lesssim 1.7$ and $(m_{F551}-J)_\mathrm{int} \lesssim 1.2$, are also the galaxies with higher extinction values ($A_V \gtrsim 0.4$). The dust-corrected colours $(m_{F365}-m_{F551})_\mathrm{int}$ and $(m_{F551}-J)_\mathrm{int}$ exhibit a remarkable anticorrelation with extinction with  Spearman correlation coefficients of $\rho_\mathrm{UV}\sim-0.92$ and $\rho_\mathrm{VJ}\sim-0.75$, respectively.
\item[$\bullet$]
We find  a correlation between the colour $(m_{F365}-m_{F551})_\mathrm{int}$ and the age of quiescent galaxies with a value of $\rho_\mathrm{UV}\sim0.45$. The older ages present  redder $(m_{F365}-m_{F551})_\mathrm{int}$ colours, populating the upper parts of the $UVJ$ diagram, $(m_{F365}-m_{F551})_\mathrm{int} > 1.8$. On the other hand, the younger quiescent galaxies present the bluer colours of the diagram. Although there is a correlation with the $(m_{F365}-m_{F551})_\mathrm{int}$ colour, age is also slightly dependent on the $(m_{F551}-J)_\mathrm{int}$ colour ($\rho_\mathrm{VJ}\sim0.15$). The older quiescent galaxies lie on the high-mass end of the MCDE and these  are the reddest ones (both $(m_{F365}-m_{F551})_\mathrm{int}$ and $m_{F365}-m_{F551}$). The less massive quiescent galaxies are younger and bluer, in good agreement with the downsizing scenario.
\item[$\bullet$]
The metallicity distribution in the $UVJ$ diagram is tightly correlated with the colour $(m_{F551}-J)_\mathrm{int}$ and less remarkably with $(m_{F365}-m_{F551})_\mathrm{int}$, with correlation coefficients $\rho_\mathrm{VJ}\sim0.85$ and $\rho_\mathrm{UV}\sim0.45$, respectively. The most metal-rich quiescent galaxies ($\mathrm{[M/H]_M} > 0.1$~dex) also present the reddest colours $(m_{F551}-J)_\mathrm{int} > 1.1$. The more massive the quiescent galaxy, the more metal-rich it is, as revealed by the MCDE. Metallicity values lie on very well-defined positions in the $UVJ$ and the MCDE diagrams.
\item[$\bullet$]
For quiescent galaxies, very low sSFR ($\log_{10}sSFR_{2800\AA}\sim-11$~yr$^{-1}$) lie on the upper parts of rest-frame and dust corrected $UVJ$ diagrams ($(m_{F365}-m_{F551})_\mathrm{int} \gtrsim 1.8$), while values of $\log_{10}sSFR_{2800\AA}\sim-10.5$ are at $(m_{F365}-m_{F551})_\mathrm{int} \lesssim 1.6$. An anticorrelation between sSFR and the intrinsic colour $(m_{F365}-m_{F551})_\mathrm{int}$ is found close to $-1$, which may be used as an sSFR tracer at least for quiescent galaxies. Actually, this tracer should be treated carefully as the sSFRs are computed via a SFR tracer, which is linked to the $2\,800$~\AA\ luminosity, and the continuum yields a non-null sSFR value even when a star formation process is not present.
\end{itemize}

From both stellar mass--colour and $UVJ$ colour--colour diagrams, we conclude that dust corrections play an important role in understanding how quiescent galaxies distribute inside these diagrams as a function of their parameters: mass, age, metallicity, extinction, and sSFR. Without dust corrections, quiescent galaxies with high dust content are in the upper parts of the $UVJ$ diagram and CMD as predicted by extinction laws. The correlation between age and  colour $(m_{F365}-m_{F551})$ is weaker than with $(m_{F365}-m_{F551})_\mathrm{int}$, whereas metallicity still correlates with $(m_{F551}-J),$ but also with $(m_{F365}-m_{F551})$. 

After exploring the MCDE and the rest-frame $UVJ$ colour--colour diagram, we note that to perform a pure non-biased selection of quiescent galaxies from these diagrams, we should include dust corrections of the involved colours and also the galaxy stellar mass. This will allow us to develop future and more reliable stellar population studies without biasing the quiescent galaxy sample, and to reveal the evolution of related properties with cosmic time.

We also quantify the distributions of stellar population properties (extinction, age, metallicity, and sSFR) within the rest-frame $UVJ$ and MCDE diagrams corrected for extinction, which may be very useful for future works to estimate stellar population parameters by the positions of quiescent galaxies within these diagrams.

As this research involves dust-corrected colours, the choice of an extinction law during the SED fitting analysis may have an impact on the sample definition and results obtained in this work. However, we find that the main results in this work are consistent when we use the extinction laws of \citet{Fitzpatrick1999} and \citet{Calzetti2001} separately.


\begin{acknowledgements}

The authors are grateful to the referee for their fruitful comments, which contributed to improving the present research. This work has been supported by the Programa Nacional de Astronom\'ia y Astrof\'isica of the Spanish Ministry of Economy and Competitiveness (MINECO, grants AYA2012-30789 and AYA2015-66211-C2-1-P), by the Ministry of Science and Technology of Taiwan (grant MOST 106-2628-M-001-003-MY3), by the Academia Sinica (grant AS-IA-107-M01), and by the Government of Arag\'on (Research Group E103). L.A.D.G.~acknowledges support from the Caja Rural de Teruel and AYA2015-66211-C2-1-P for partly developing this research. This work has been partly supported by the Programa Nacional de Astronom\'ia y Astrof\'isica of the Spanish Ministry of Economy and Competitiveness (MINECO) under grant AYA2012-30789, and by FEDER funds and the Government of Arag\'on, through the Research Group E103. L.A.D.G.~also thanks    I.F.~for offering the opportunity to develop part of this research at the Mullard Space Science Laboratory (MSSL). We also acknowledge support from the Spanish Ministry for Economy and Competitiveness and FEDER funds through grants AYA2010-15081, AYA2010-15169, AYA2010-22111-C03-01, AYA2010-22111-C03-02, AYA2011-29517-C03-01, AYA2012-39620, AYA2013-40611-P, AYA2013-42227-P, AYA2013-43188-P, AYA2013-48623-C2-1, AYA2013-48623-C2-2, ESP2013-48274, AYA2014-57490-P, AYA2014-58861-C3-1, AYA2016-76682-C3-1-P, AYA2016-77846-P, AYA2016-81065-C2-1, AYA2016-81065-C2-2; Generalitat Valenciana projects Prometeo 2009/064 and PROMETEOII/2014/060; Junta de Andaluc\'{\i}a grants TIC114, JA2828, P10-FQM-6444; and Generalitat de Catalunya project SGR-1398. Throughout this research, we made use of the \texttt{Matplotlib} package \citep{Hunter2007}, a 2D graphics package used for \texttt{Python} that is designed for interactive scripting and quality image generation. This paper is dedicated to Marian Le\'on Canalejo for being there when L.A.D.G.~needed her most and for her patience and continuous encouragement he was finishing his Ph.D.

\end{acknowledgements}


\bibliographystyle{aa}
\bibliography{ms} 

%

\begin{appendix}


\section{Removal of uncertainty effects from distributions of stellar population properties}\label{sec:appendix_mle}

During the last decades, maximum likelihood estimator (MLE) methods have been used for various aims in astronomy \citep[e.g.][]{Naylor2006,Makarov2006,Arzner2007,LopezSanjuan2008,LopezSanjuan2015}. For this work, we performed an MLE to remove uncertainty effects of individual galaxies from distributions of stellar population  properties, that is, a deconvolution to recover the intrinsic distributions of stellar population parameters. In Sects.~\ref{sec:mcd} and \ref{sec:dusty}, the distributions of $(m_{F365}-m_{F551})_\mathrm{int}^\mathrm{rot}$ and $\log_{10}SFR_{2800\AA}^\mathrm{rot}$ values (Eqs.~(\ref{eq:CMDE_rot}) and (\ref{eq:sfr2800_rot}), respectively) presented a challenge to determine the limiting values for quiescent galaxies because   these distributions are affected by the redshift-dependent uncertainties of individual galaxies.

Our starting point is the MLE method detailed by \citet{LopezSanjuan2014}, which was initially developed to disentangle cosmic variance effects. In \citet{LopezSanjuan2014}, the authors assumed that intrinsic distributions are properly described by Gaussian-like probability distributions in the log-space, that is, log-normal distributions in the real space. In addition, for uncertainties in the stellar population parameters of individual galaxies a Gaussian distribution was also assumed. Therefore, uncertainty effects would increase distribution widths with  little impact on the distribution median. Under these assumptions, the likelihood to maximise  is
\begin{equation}\label{eq:mle}
\mathcal{L} = - \frac{1}{2} \sum\limits_{j} \left[ \ln \left( {{p_{\mathrm{e},j}}^2+\sigma^\mathrm{int}_p}^2 \right) + \frac{{\left( \mu_p - p_j \right)}^2}{{{p_{\mathrm{e},j}}^2 + \sigma^\mathrm{int}_p}^2} \right]\ ,
\end{equation}
where $p_j$ is the stellar population parameter (in this work $(m_{F365}-m_{F551})_\mathrm{int}^\mathrm{rot}$ or $\log_{10}SFR_{2800\AA}^\mathrm{rot}$) of the $j$th galaxy in the distribution, $p_{\mathrm{e},j}$ is its $1\sigma$ error, $\mu_p$ the median of the real or intrinsic distribution (without uncertainty effects), and $\sigma^\mathrm{int}_p$ its width or intrinsic dispersion \citep[for details, see][]{LopezSanjuan2014}. For the maximisation of Eq.~(\ref{eq:mle}), we made the most of the \textit{emcee}\footnote{\url{http://dan.iel.fm/emcee}} algorithm \citep{ForemanMackey2013}. This tool is an affine invariant sampling algorithm for a Markov chain Monte Carlo method (MCMC) that can be easily adapted to any function for maximisation. 

\begin{figure}
\centering
\resizebox{\hsize}{!}{\includegraphics{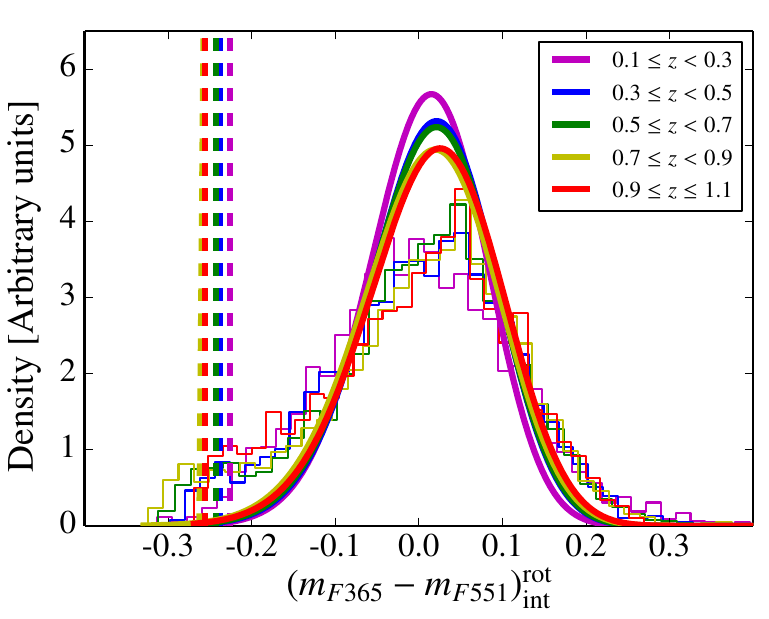}}
\caption{Histogram of the $(m_{F365}-m_{F551})_\mathrm{int}^\mathrm{rot}$ values (see Eq.~(\ref{eq:CMDE_rot}))  for the quiescent sample (intrinsic red galaxies) at different redshifts using the BC03 SSP models. Solid lines illustrate the MLE best fit of the $(m_{F365}-m_{F551})_\mathrm{int}^\mathrm{rot}$ distribution at different redshifts. Dashed coloured lines show the upper $3\sigma$ limit of the distributions provided by the MLE method.}
\label{fig:mle_cmd}
\end{figure}

Figure~\ref{fig:mle_cmd} shows the distribution of $(m_{F365}-m_{F551})_\mathrm{int}^\mathrm{rot}$ values of quiescent galaxies (for further details, see Sect.~\ref{sec:mcd}). It should be noted that Eq.~(\ref{eq:mle}) is defined for Gaussian distributions, and $(m_{F365}-m_{F551})_\mathrm{int}^\mathrm{rot}$ is a log-normal  distribution (see Fig.~\ref{fig:mle_cmd}). Therefore, we maximised Eq.~(\ref{eq:mle}) for the distribution of values in the log-space, that is, $\ln\left[1-(m_{F365}-m_{F551})_\mathrm{int}^\mathrm{rot}\right]$, where a shift in the distribution is necessary to avoid negative values. As expected, medians almost remain unaltered after the MLE deconvolution, whereas the intrinsic distributions are narrower than the observed values (see Fig.~\ref{fig:mle_cmd}). To illustrate, close to the 3$\sigma$ limit there is an overdensity in the number of galaxies, which may be partly interpreted as the transient population of galaxies from the blue cloud to the red sequence (green valley, see Fig.~\ref{fig:mle_cmd}), but also by uncertainties.

\begin{figure}
\centering
\resizebox{\hsize}{!}{\includegraphics{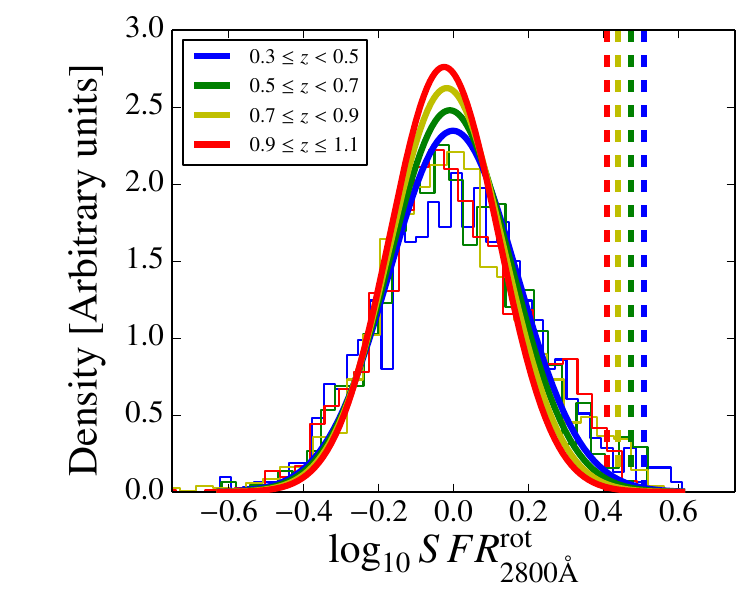}}
\caption{Histogram of the $\log_{10}SFR_{2800\AA}^\mathrm{rot}$ values (see Eq.~(\ref{eq:sfr2800_rot}))   for the quiescent sample defined via star formation rates $SFR_{2800\AA}$ (see Eq.~(\ref{eq:sfr2800})) at different redshifts and using the BC03 SSP models. Solid lines illustrate the  MLE best fit of the $\log_{10}SFR_{2800\AA}^\mathrm{rot}$ distributions. Dashed coloured lines show the upper $3\sigma$ limit of the distributions stated by the MLE method.}
\label{fig:mle_sfr}
\end{figure}

The intrinsic distribution of $\log_{10}SFR_{2800\AA}^\mathrm{rot}$ can be fitted by Gaussian functions at any redshift bin properly (see Fig.~\ref{fig:mle_sfr}). Once again, the most remarkable impact of uncertainties is the widening of the distributions. After removing uncertainties and parametrising the distribution, it is straightforward to set the limiting sSFR values of quiescent galaxies. As mentioned above, we set this limit at a $3\sigma$ level (see Fig.~\ref{fig:mle_sfr}).

\section{Detection and removal of cool stars in the quiescent sample}\label{sec:appendix_faint}

The performance of the MUFFIT module devoted to analysing stars is basically the same as for the galaxy version, but instead of using a mixture of SSP models it takes templates of stars to provide stellar parameters (effective temperature, metallicity, surface gravity, chemical composition, and extinction). 

We ran the stellar version of MUFFIT with all the sources in the ALHAMBRA catalogue that present an apparent magnitude in the range $22.5 \le m_{F814W} < 23$, using as input models the stellar library of \citet{Coelho2005} with the main grid of parameters: effective temperatures in steps of $250$~K, $3500 \le \mathrm{T_{eff}} \le 7000$~K; surfaces gravities with steps of $0.5$, $0.0 \le \log_{10} \le 5.0$; metallicities in the range $\mathrm{[Fe/H]}=-2.5$ to $+0.5$; and chemical compositions $[\alpha/\mathrm{Fe}]=0.0$ and $0.4$. In addition, we added extinction values in the range $A_V=0.0$--$3.1$ to the star models taking $R_V=3.1$ and assuming a \citet{Fitzpatrick1999} extinction law. 

After the SED fitting analysis and thanks to the Monte Carlo process performed by MUFFIT, we obtained two sets of $\chi^2$ values for each source. One set from the SED fitting treating the source as a galaxy and another one as a star. Therefore, the $\chi^2$ distribution with the lower values (the most likely set of templates, galaxy or star) will determine whether the source is a galaxy. Although this method takes advantage of the photometry in all the bands, all the sources show apparent magnitudes in the range $22.5 \le m_{F814W} < 23$, with a reasonably low signal-to-noise ratio that makes the distinction between distributions more difficult in several cases. To solve this drawback, we carried out a Kolmogorov-Smirnov (KS) test that allowed us to determine whether the two $\chi^2$ distributions are different and what  the probability is. We assumed that a source in the range $22.5 \le m_{F814W} < 23$ is a star if the following criterion is satisfied: the median of $\chi^2$ values (from the $\chi^2$ distribution) is lower using star templates than SSP templates, and that the KS test additionally shows that the two distributions of $\chi^2$ values (star and galaxy) are not equivalent at a significance level of $1\sigma$. Under these constraints, we found that in the redshift range $0.1 \le z \le 1.1$ there are $439$ star candidates in the quiescent sample with $22.5 \le m_{F814W} < 23$. Taking into account that there are $2\,284$ quiescent galaxies in our sample at the same magnitude and redshift bin before the faint star/galaxy classification, we   removed   $\sim19$~\% of the sample at $22.5 \le m_{F814W} < 23$. 

To check whether the method developed for ALHAMBRA for removing stars in the range $22.5 \le m_{F814W} < 23$ is reliable, we cross-matched the ALHAMBRA and COSMOS photometric catalogues \citep{Capak2007} to build a subset of shared sources in the two surveys. Making the most of the ACS camera in COSMOS, we compared the stars or point sources detected in COSMOS and those detected by our method to estimate the degree of accuracy. There are $230$ sources in common with our sample of quiescent candidates with apparent magnitudes $22.5 \le m_{F814W} < 23$, in which we found  $50$ star candidates with apparent magnitudes $22.5 \le m_{F814W} < 23$ using the ALHAMBRA photometry and MUFFIT. From the star/galaxy classification of COSMOS \citep{Leauthaud2007} in this subsample of $230$ sources, we checked that $47$ ($94$~\%) of them are classified as point sources in COSMOS. Nevertheless, there are nine stars that were not detected in the common subset with $22.5 \le m_{F814W} < 23$ following the COSMOS classification. This indicates  that  about  $24$~\% of the faint cool stars  should be removed. In the case where we can extrapolate these percentages from the subsample in common with COSMOS to our sample of quiescent galaxies in ALHAMBRA, we would remove $84$~\% of faint cool stars in ALHAMBRA. Furthermore, we would expect a contamination of around $4$~\% of the stars in our sample of quiescent galaxies at $22.5 \le m_{F814W} < 23$. 


\section{Stellar mass completeness determination}\label{sec:appendix_smc}

Our aim is to develop a method for determining and parametrising the stellar mass completeness of the ALHAMBRA galaxies. This method must be  applicable to the sample of quiescent galaxies in this work and also easily calculable for any completeness level $\mathcal{C}$.

Our methodology is based on taking advantage of stellar mass functions, $\Phi(M)$, from deeper surveys. In particular, we took the stellar mass functions from the COSMOS survey for quiescent galaxies \citep{Ilbert2010}. We tried to  measure how the low-mass end of the ALHAMBRA sample differs from the one derived from this much deeper survey, assuming that the differences are dominated by the stellar mass completeness. To include the impact of stellar mass completeness in stellar mass functions, we followed a  process similar to the work described in  \citet[]{Sandage1979}, which relies on a likelihood maximisation. This likelihood, $\mathcal{L}$, encompasses the probability of observing a galaxy accounting for both the selection and observational effects. Formally,
\begin{equation}
\ln \mathcal{L} = \sum\limits_{i=1}^{N_\mathrm{g}} {\frac{\Phi(M_i) f(M_i)}{\int_{M_{i,\mathrm{min}}}^{M_{i,\mathrm{max}}}\! \Phi(M') f(M')\,\mathrm{d}M'}}\ ,
\label{eq:likelihood}
\end{equation}
where $N_\mathrm{g}$ is the number of galaxies in the subsample, $f(M_i)$ is the stellar mass completeness for a galaxy with stellar mass $M_i$ at certain redshift, $M_{i,\mathrm{min}}$ and $M_{i,\mathrm{max}}$ are respectively the minimum and maximum stellar masses at the redshift in which the \textit{i}th galaxy resides, and $\Phi(M)$ the stellar mass function characterised by a Schechter-like function \citep{Schechter1976} through three parameters \citep[$\alpha$, $\mathcal{M}_*$, and the normalisation $\Phi^*$; see][]{Fontana2004,PerezGonzalez2008,Vergani2008,Ilbert2010,Ilbert2013} expressed as

\begin{equation}
\Phi(M)\; \mathrm{d}M = \Phi^* \left(\frac{M}{\mathcal{M}_*}\right)^\alpha \exp{\left(-\frac{M}{\mathcal{M}_*}\right)}\; \mathrm{d}\left(\frac{M}{\mathcal{M}_*}\right). 
\label{eq:smf}
\end{equation}

As in \citet[][]{Sandage1979}, we used a Fermi-Dirac distribution function, $f_\mathrm{FD}$, to parametrise the stellar mass completeness, formally expressed as
\begin{equation}
f_\mathrm{FD}(z,M_\star) = \frac{1}{\exp \left[(M_\mathrm{F}(z)-\log_{10}M_\star)/\Delta_\mathrm{F}(z) \right]+1}\ ,
\label{eq:fermi}
\end{equation}
where $M_\mathrm{F}(z)$ is the stellar mass value (in dex) for which the completeness reaches $50$~\% ($\mathcal{C}=0.5$) and $\Delta_\mathrm{F}(z)$ is related to the decrease rate on the number of galaxies. It should be noted that both $M_\mathrm{F}(z)$ and $\Delta_\mathrm{F}(z)$ are redshift dependent. The stellar mass value limit for a given completeness level and redshift, $M_\mathcal{C}(z)$, can be easily derived from Eq.~(\ref{eq:fermi}) as
\begin{equation}
\log_{10} M_\mathcal{C}(z) = M_\mathrm{F}(z) + \Delta_\mathrm{F}(z) \ln\left[ \frac{\mathcal{C}}{1-\mathcal{C}}\right]\ .
\label{eq:completeness}
\end{equation}

After fixing both $\alpha$ and $\mathcal{M}_*$ to the values obtained by \citet[][the term $\Phi^*$ is not relevant because it is cancelled in Eq.~(\ref{eq:likelihood})]{Ilbert2010}, the maximisation of Eq.~(\ref{eq:likelihood}) provides us $M_\mathrm{F}$ and $\Delta_\mathrm{F}$. For the maximisation of the likelihood, we used again the \textit{emcee} algorithm, for which we set limiting values of $5 \le M_\mathrm{F}(z) \le 14$ and $0.02 \le \Delta_\mathrm{F}(z) < 2$.

As a result, we show in Table~\ref{tab:completeness} the values $M_\mathrm{F}$ and $\Delta_\mathrm{F}$ that maximise Eq.~(\ref{eq:likelihood}),  and also show the stellar mass limits for different completeness levels for the quiescent galaxies in ALHAMBRA down to $m_{F814W}=23$. These values were obtained for BC03 SSP models, that is, the same model set used by \citet{Ilbert2010} for the COSMOS survey. We checked to see if there were systematic discrepancies between the stellar masses obtained using EMILES and those obtained using BC03 SSP models. Systematically, stellar masses for quiescent galaxies are about $0.15$~dex ($0.11$~dex) more massive using the EMILES SSP models with BaSTI (Padova00) isochrones in comparison with the BC03 ones. For EMILES, the completeness is  reproduced well when we add this systematic difference ($0.15$ and $0.11$~dex) in $M_\mathrm{F}$ without altering $\Delta_\mathrm{F}$. As a sanity check, we compared our predictions of mass limits at $\mathcal{C}=0.95$ with those obtained following the methodology proposed by \citet{Pozzetti2010}. We found good agreement between the two methods, with mean differences of $\sim0.03$~dex at $0.1 < z < 1.1$.


\section{Stellar population properties of quiescent galaxies using EMILES SSP models}\label{sec:appendix_sup}

\begin{figure*}
\centering
\includegraphics[trim= 0 3mm 0 1mm,width=17cm,clip=True]{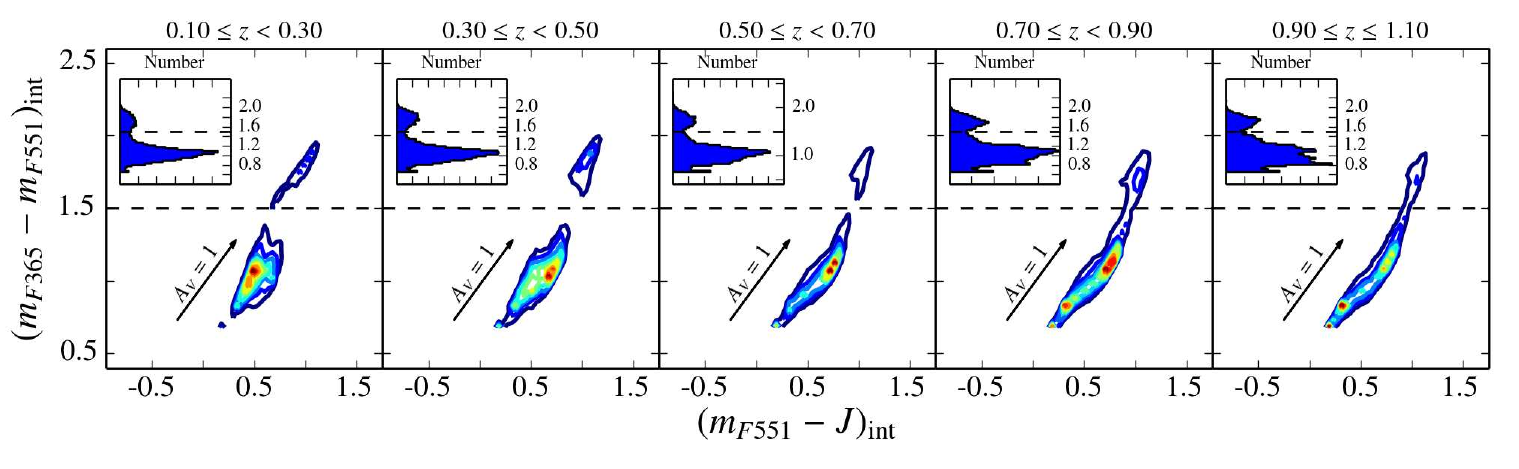}
\includegraphics[trim= 0 3mm 0 1mm,width=17cm,clip=True]{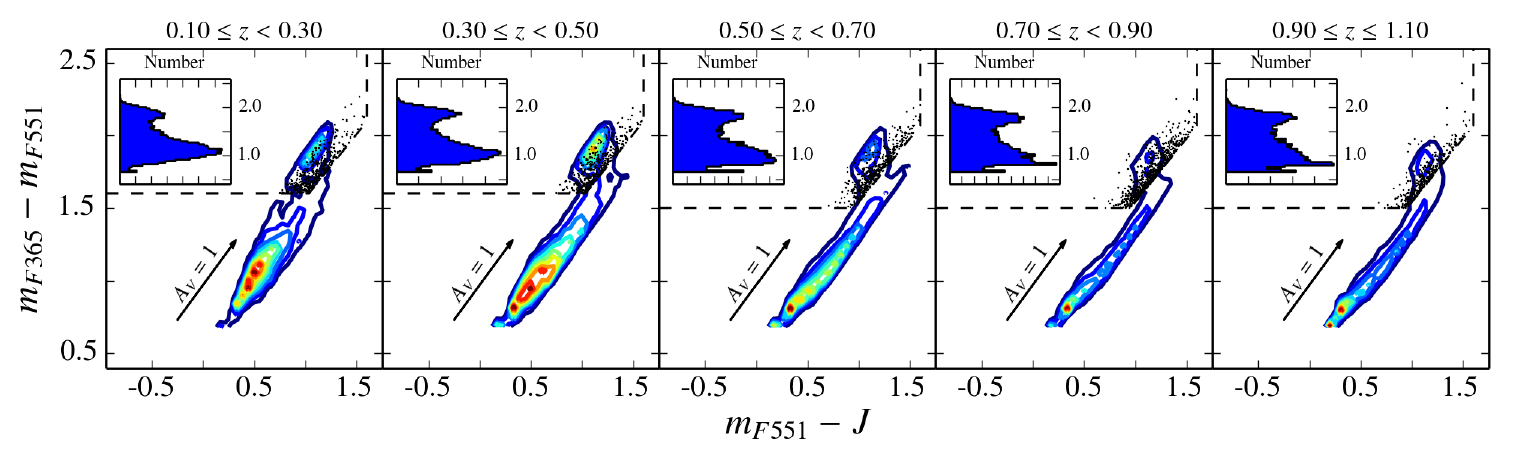}
\caption{Density surface and distribution of rest-frame colours for galaxies from the ALHAMBRA survey obtained using EMILES+BaSTI SSP models. Top: Rest-frame intrinsic colours $(m_{F551}-J)_\mathrm{int}$ ($X$-axis) and $(m_{F365}-m_{F551})_\mathrm{int}$ ($Y$-axis) after correcting for extinction at different redshifts. Bottom: Rest-frame colours without removing dust effects. Redder (bluer) density-curve colours are related to higher (lower) densities. Inner panels, histograms of the intrinsic (top) and observed (bottom) rest-frame colour $m_{F365}-m_{F551}$. Dashed lines in the top panels illustrate our limiting value $(m_{F365}-m_{F551})_\mathrm{int}=1.5$ for quiescent galaxies, and in the bottom panels the quiescent $UVJ$ sample defined by \citet[][Eq.~(\ref{eq:quiescent})]{Moresco2013}. Black dots are galaxies labelled as quiescent by the $UVJ$ criteria of \citet{Moresco2013} that lie in the star-forming region after removing extinction effects. We illustrate the colour variations due to a reddening of $A_V=1$ (black arrow), assuming the extinction law of \citet{Fitzpatrick1999}.}
\label{fig:ccd_noav_basti}
\end{figure*}

\begin{figure*}
\centering
\includegraphics[trim= 0 17.1mm 0 0,width=17cm,clip=True]{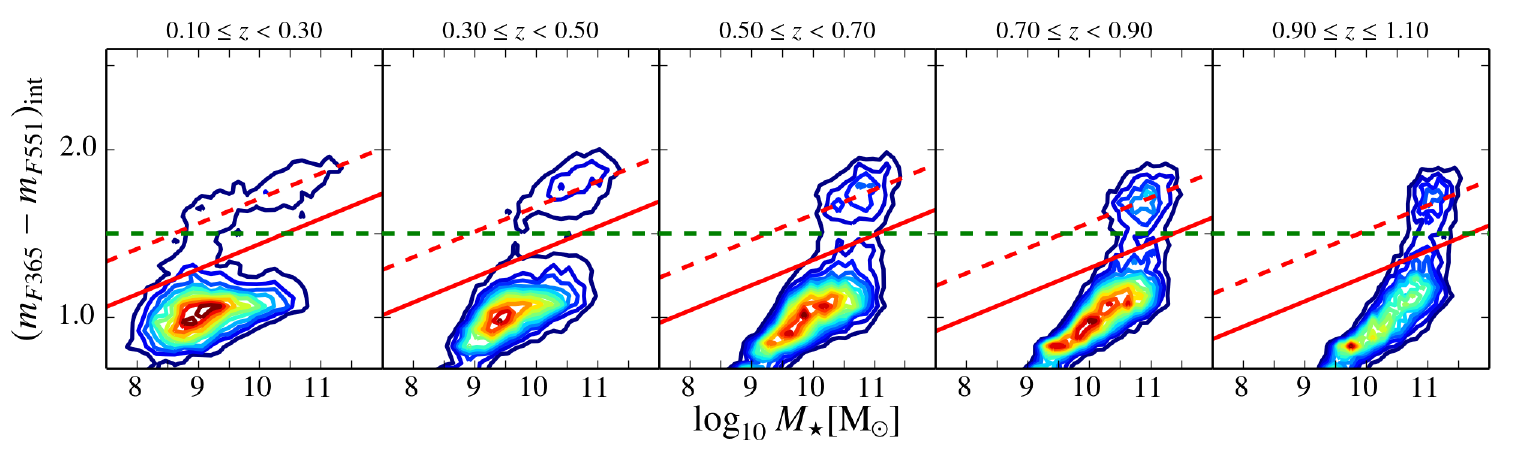}
\includegraphics[trim= 0 0 0 8mm,width=17cm,clip=True]{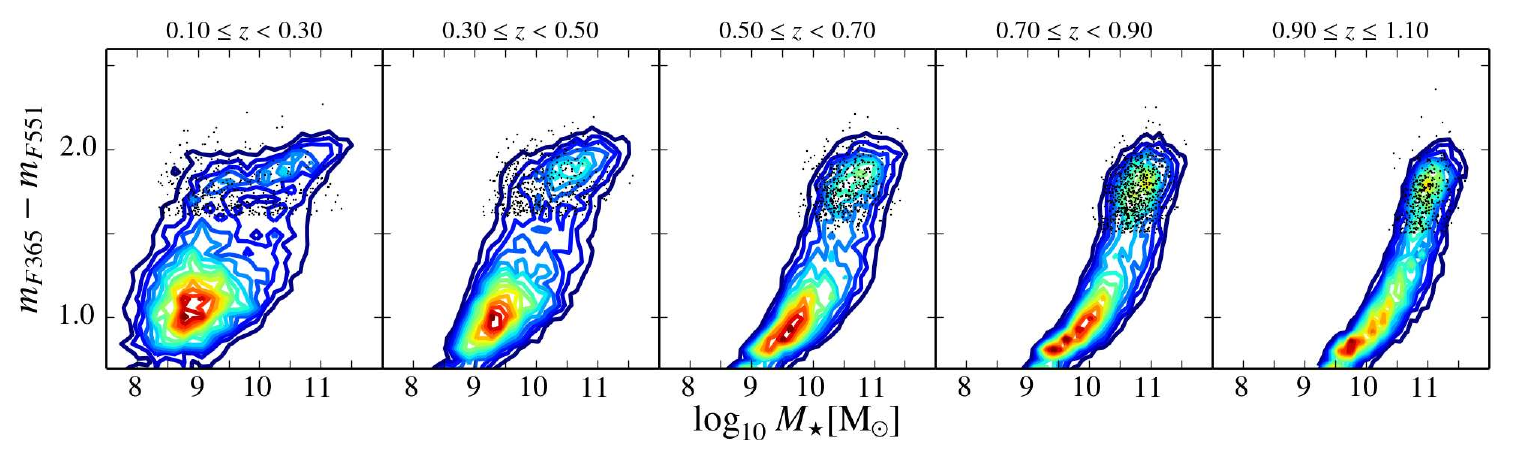}
\caption{Density surface and distribution of stellar mass vs. rest-frame colour $(m_{F365}-m_{F551})$ for galaxies from the ALHAMBRA survey using EMILES+BaSTI SSP models. Top: Rest-frame intrinsic colour $(m_{F365}-m_{F551})_\mathrm{int}$ ($Y$-axis) after correcting for extinction at different redshift. Bottom: Rest-frame colours without removing dust effects. Redder (bluer) density-curve colours are related to high (low) densities. Dashed green lines in the top panels illustrate the limiting value $(m_{F365}-m_{F551})_\mathrm{int}=1.5$ used for selecting quiescent galaxies in Sect.~\ref{sec:sample_uvj}. Dashed and solid red lines show the fit to the quiescent sequence, $(m_{F365}-m_{F551})_\mathrm{int}^\mathrm{Q}$, and the limiting intrinsic colours, $(m_{F365}-m_{F551})_\mathrm{int}^\mathrm{lim}$, values of quiescent galaxies respectively (see details in Sect.~\ref{sec:mcd}). Black dots in the bottom panels are galaxies labelled as quiescent with the $UVJ$ criteria of \citet{Moresco2013} that lie in the star-forming region after removing extinction effects.}
\label{fig:CMDE_basti}
\end{figure*}

\begin{figure*}
\centering
\includegraphics[trim= 0 25 0 5mm,width=17cm,clip=True]{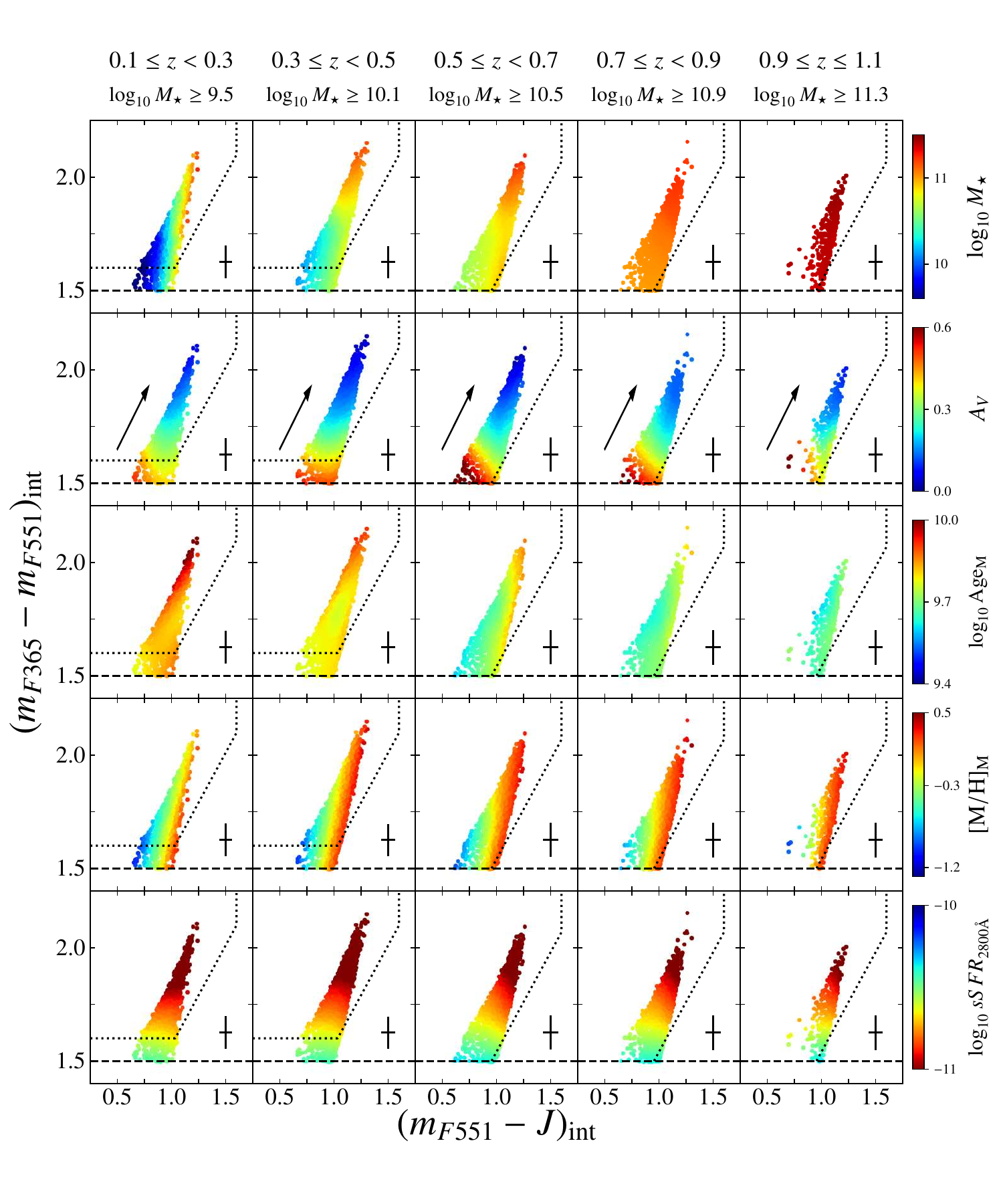}
\caption{Stellar population parameters in the rest-frame $UVJ$ diagram. At different redshift bins, we present the intrinsic colours $(m_{F551}-J)_\mathrm{int}$ ($X$-axis) and $(m_{F365}-m_{F551})_\mathrm{int}$ ($Y$-axis) after correcting for extinction for the mass complete sample of quiescent galaxies (see stellar mass completeness at the top). The different stellar population parameters are colour-coded as a  function of their values and obtained using EMILES+BaSTI SSP models (see the  colour bars to the left of each row). From top to bottom:  stellar mass, extinction, mass-weighted age and metallicity, and specific star formation rate. All the parameters were spatially averaged through a LOESS method. Black crosses illustrate the median uncertainties in both $(m_{F551}-J)_\mathrm{int}$ and $(m_{F365}-m_{F551})_\mathrm{int}$ intrinsic colours. The dotted black line encloses the rest-frame colour ranges assumed for selecting quiescent galaxies in \citet[][see Eq.~(\ref{eq:quiescent})]{Moresco2013}, while the dashed line illustrates our colour limit for selecting quiescent galaxies $(m_{F365}-m_{F551})_\mathrm{int}>1.5$. We illustrate the colour variations owing to a reddening of $A_V=0.5$ (black arrow), assuming the extinction law of \citet{Fitzpatrick1999}.}
\label{fig:sp_UVJ_basti}
\end{figure*}

\begin{figure*}
\centering
\includegraphics[trim= 0 25 0 5mm,width=17cm,clip=True]{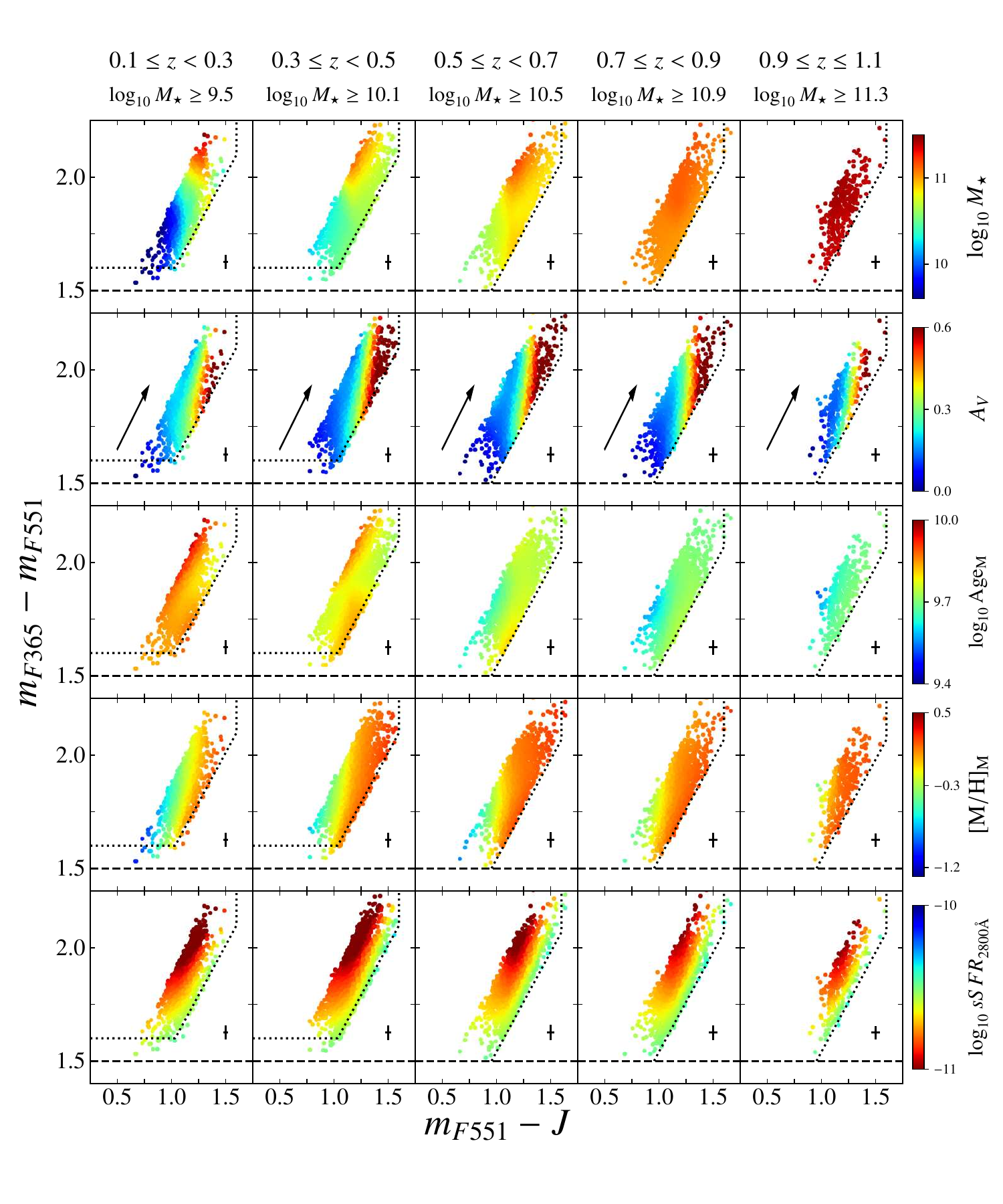}
\caption{Same as Fig.~\ref{fig:sp_UVJ_basti}, but for the rest-frame colours $m_{F551}-J$ ($X$-axis) and $m_{F365}-m_{F551}$ ($Y$-axis), meaning non-dust corrected colours.}
\label{fig:sp_UVJ_av_basti}
\end{figure*}

\begin{figure*}
\centering
\includegraphics[trim= 0 0 0 0,width=17cm,clip=True]{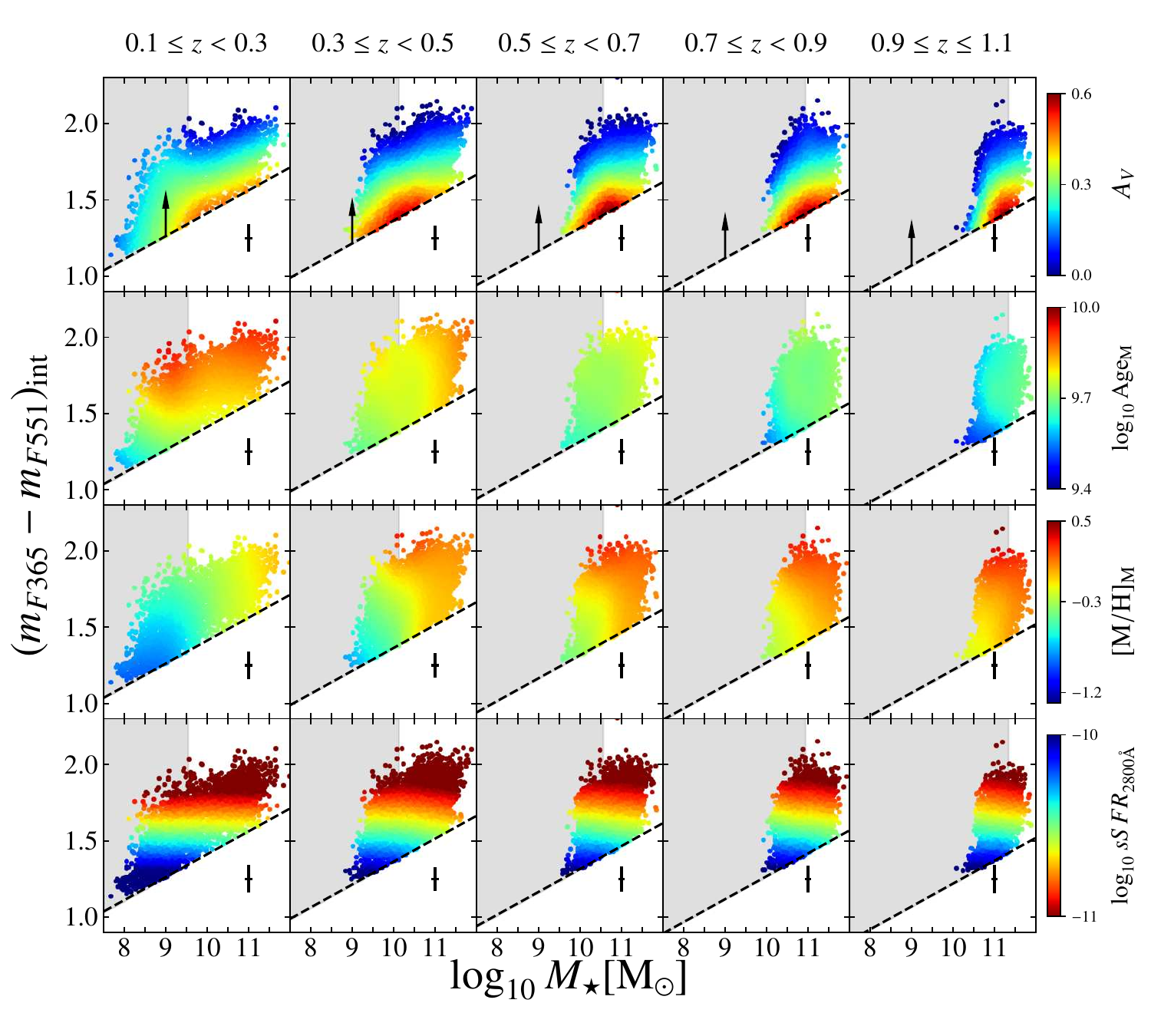}
\caption{Stellar population parameters in the rest-frame stellar mass--colour diagram. At different redshift bins, we present the stellar mass ($X$-axis) and intrinsic colour $(m_{F365}-m_{F551})_\mathrm{int}$ ($Y$-axis) after correcting for extinction. The stellar population  parameters are colour-coded according to their values using EMILES+BaSTI SSP models (see  colour bars). From top to bottom: extinction,  mass-weighted age and metallicity, and specific star formation rate. All the parameters were spatially averaged through a LOESS method. Black crosses illustrate the median uncertainties in both stellar mass and $(m_{F365}-m_{F551})_\mathrm{int}$ intrinsic colour. The dashed line illustrates the colour limit for selecting quiescent galaxies in the MCDE, see Eq.~(\ref{eq:CMDE_main}) and Table~\ref{tab:cmde_main}, for this work. The shaded regions show the stellar mass range in which our quiescent sample is not complete in stellar mass. We illustrate the colour variations owing to a reddening of $A_V=0.5$ (black arrow), assuming the extinction law of \citet{Fitzpatrick1999}.}
\label{fig:CMDE_sp_basti}
\end{figure*}

\begin{figure*}
\centering
\includegraphics[trim= 0 0 0 0,width=17cm,clip=True]{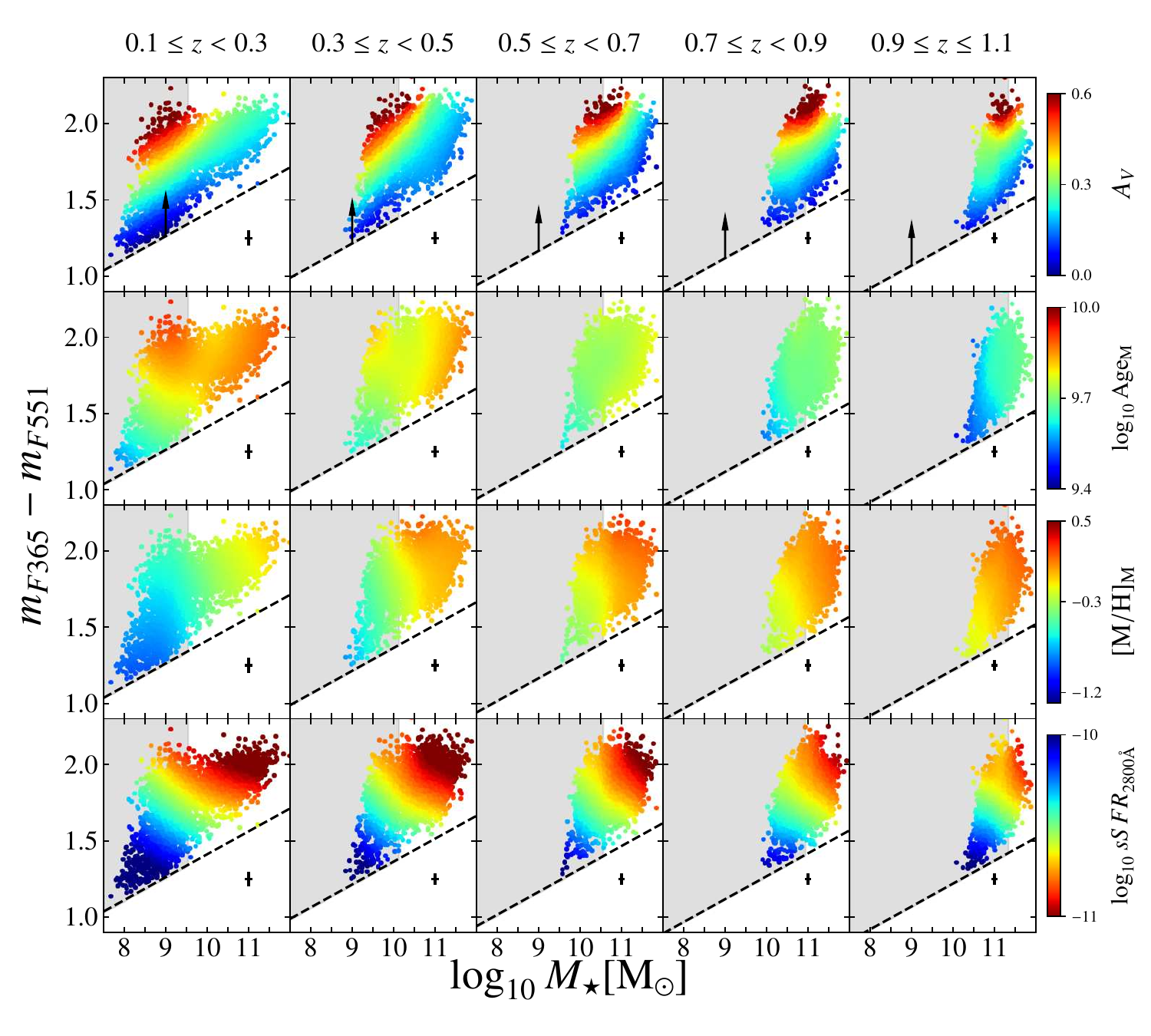}
\caption{Same as Fig.~\ref{fig:CMDE_sp_basti}, but for the rest-frame colour $m_{F365}-m_{F551}$ ($X$-axis, non-dust corrected colour).}
\label{fig:CMDE_sp_av_basti}
\end{figure*}

\begin{figure*}
\centering
\includegraphics[trim= 0 0 0 0,width=17cm,clip=True]{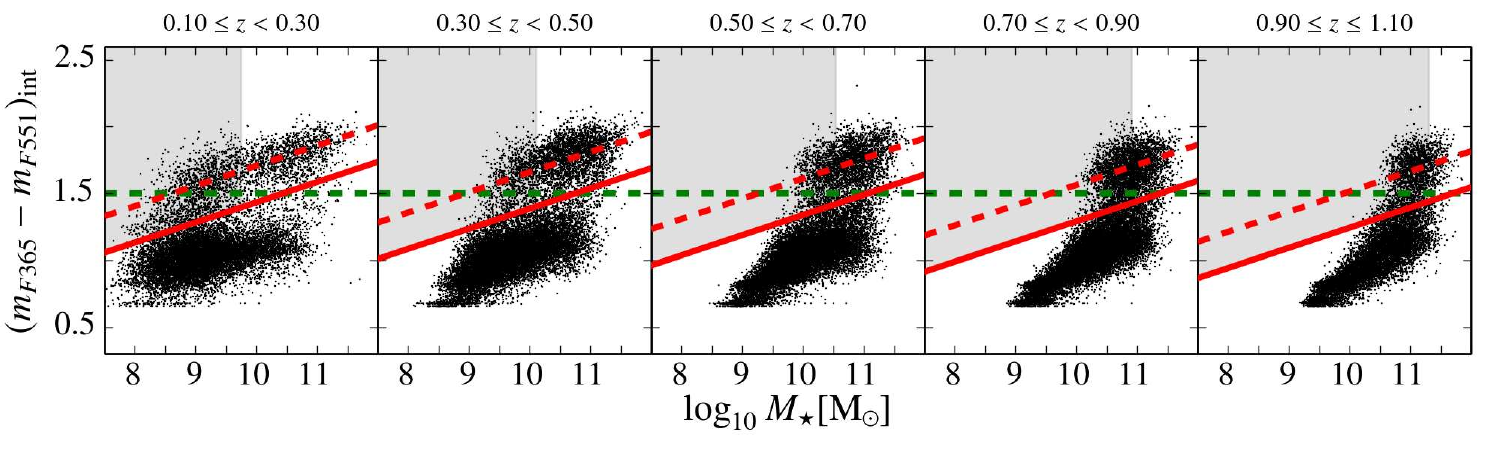}
\caption{Galaxies from the ALHAMBRA survey (black dots) in the rest-frame stellar mass--colour diagram corrected for extinction (MCDE) at different redshift bins. The shaded region illustrates the stellar mass range in which our sample of quiescent galaxies is not complete in stellar mass (see Sect.~\ref{sec:completeness}) at each redshift bin. The dashed green line shows the limiting value $(m_{F365}-m_{F551})_\mathrm{int}=1.5$ used to select quiescent galaxies in Sect.~\ref{sec:sample_uvj}. The dashed red line exhibits the main sequence of quiescent galaxies in the MCDE, see Eq.~(\ref{eq:CMDE_main}) and Table~\ref{tab:cmde_main}, and the solid line is the limiting colour value used to select  quiescent galaxies in this diagram, see Eq.~(\ref{eq:CMDE_main}) and Table~\ref{tab:cmde_main}. The stellar population predictions were obtained through EMILES+BaSTI SSP models.}
\label{fig:CMDE_limit_basti}
\end{figure*}

\begin{figure*}
\centering
\includegraphics[trim= 0 3mm 0 1mm,width=17cm,clip=True]{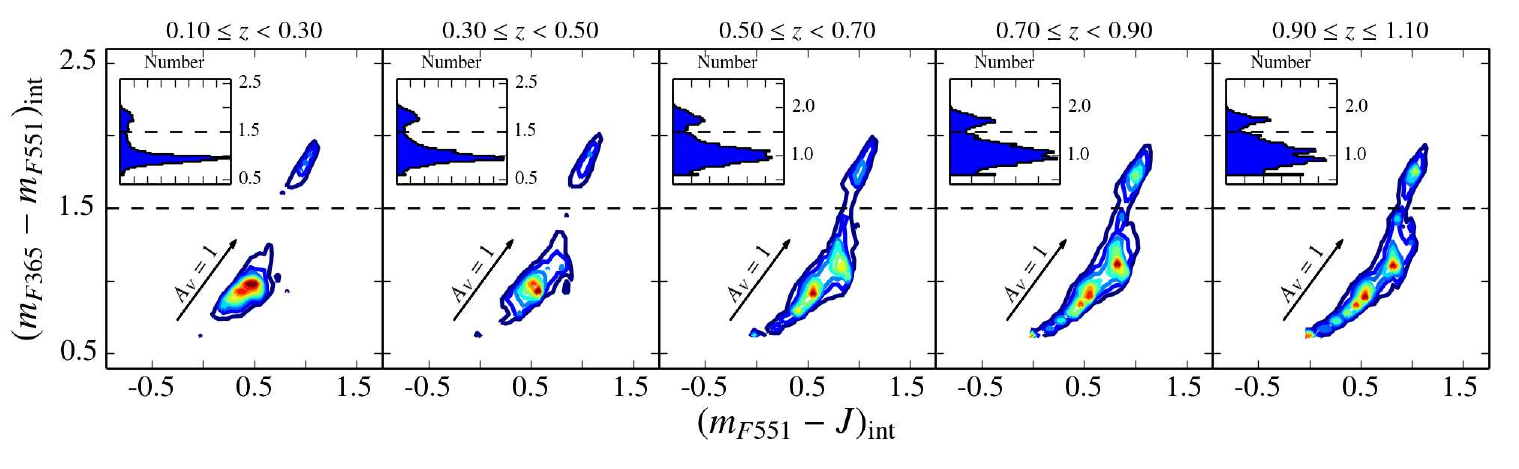}
\includegraphics[trim= 0 3mm 0 1mm,width=17cm,clip=True]{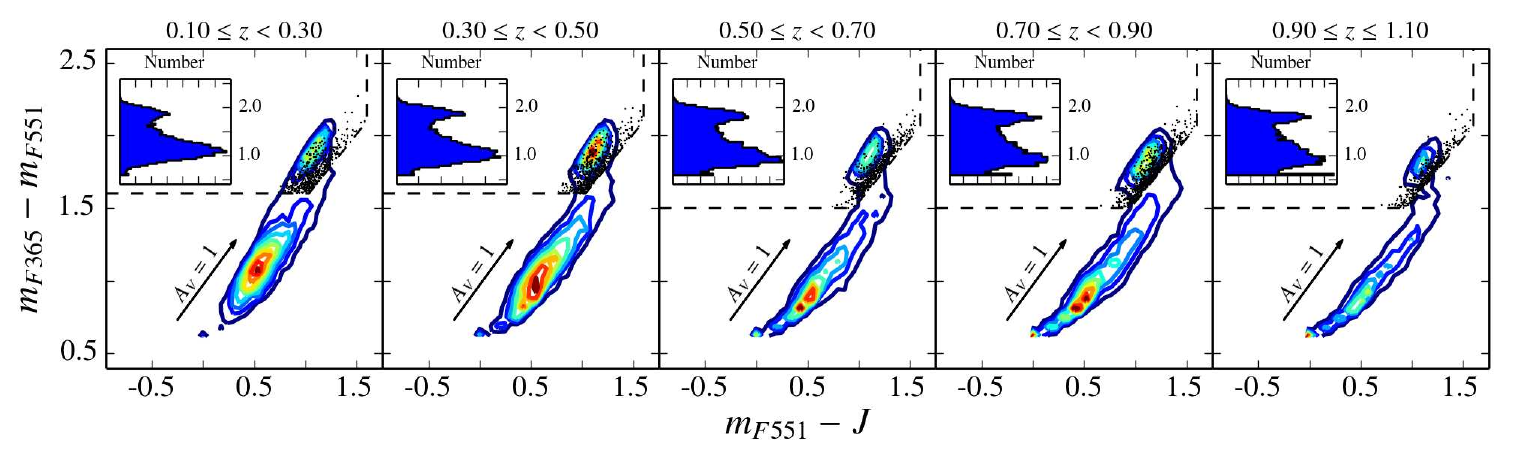}
\caption{Same as Fig.~\ref{fig:ccd_noav_bast
i}, but for Padova00 isochrones.}
\label{fig:ccd_noav_padova}
\end{figure*}

\begin{figure*}
\centering
\includegraphics[trim= 0 17.1mm 0 0,width=17cm,clip=True]{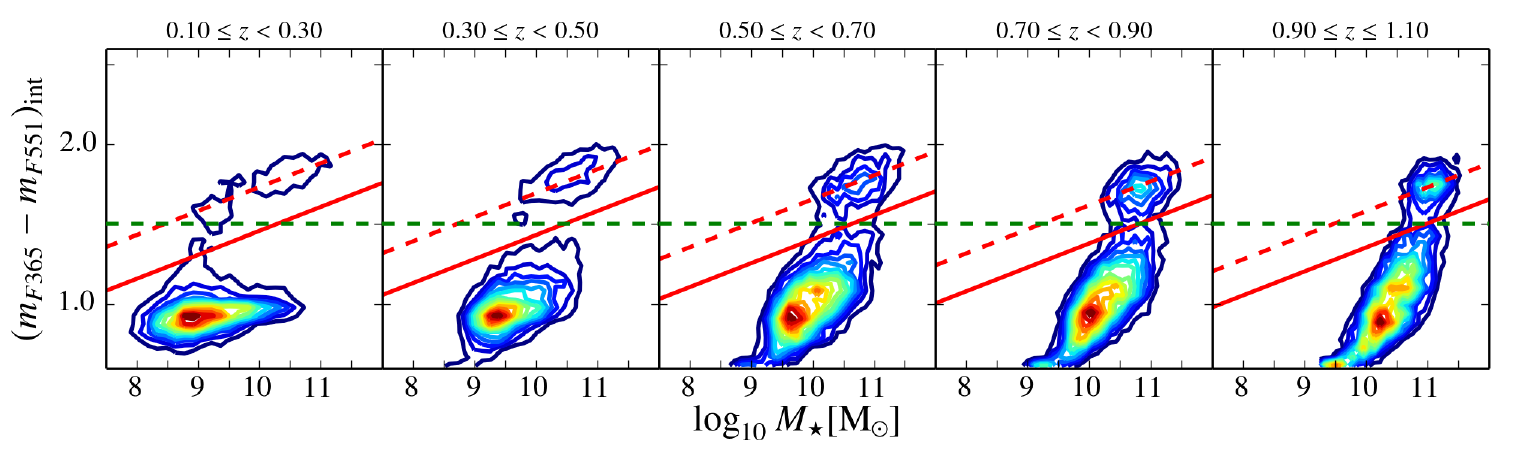}
\includegraphics[trim= 0 0 0 8mm,width=17cm,clip=True]{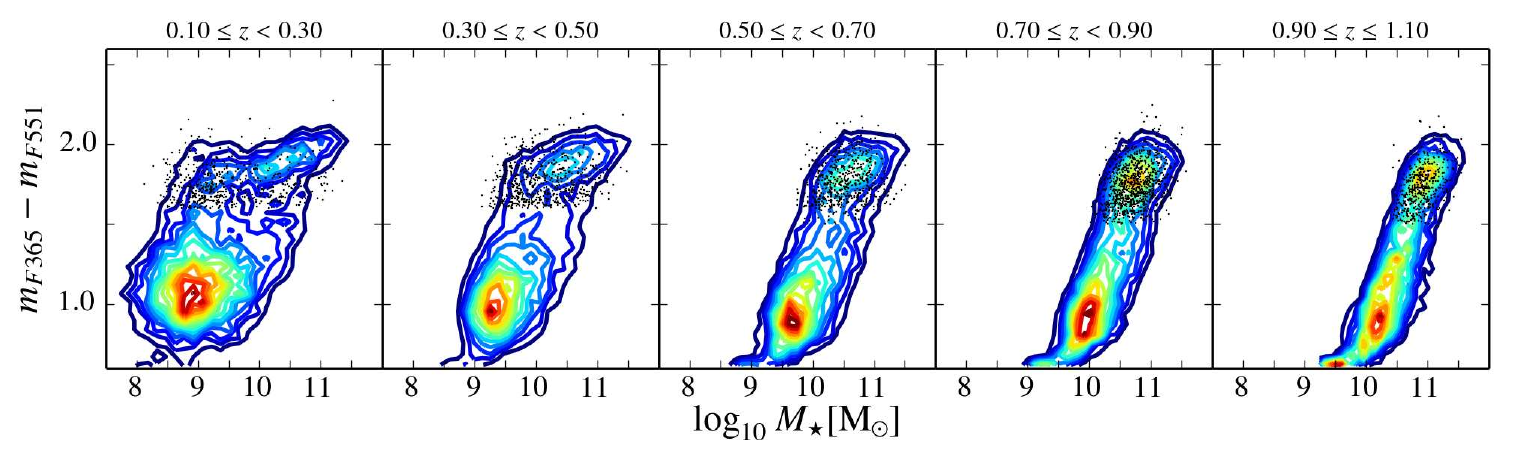}
\caption{Same as Fig.~\ref{fig:CMDE_basti}, but using Padova00 isochrones.}
\label{fig:CMDE_padova}
\end{figure*}

\begin{figure*}
\centering
\includegraphics[trim= 0 25 0 5mm,width=17cm,clip=True]{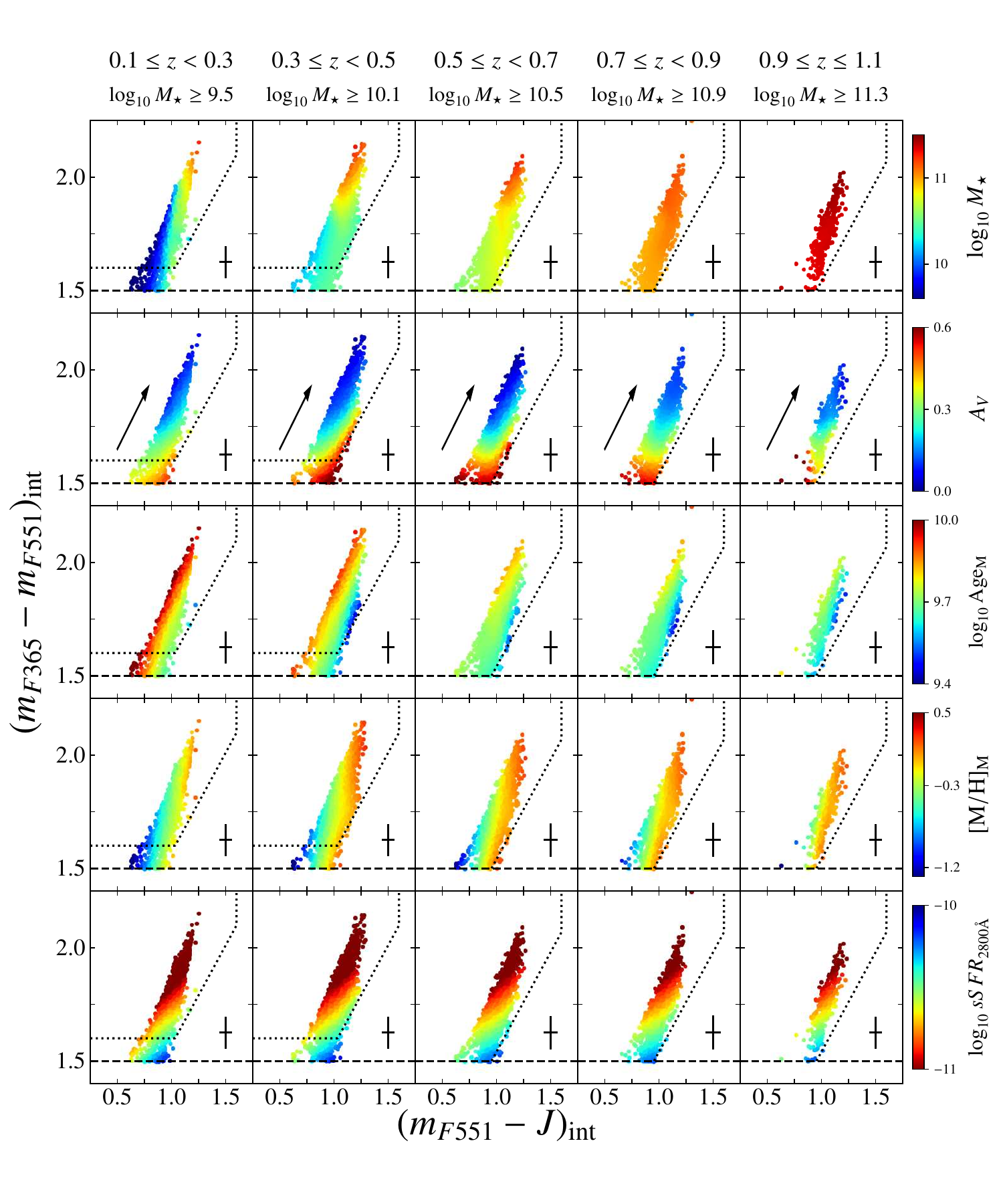}
\caption{Same as Fig.~\ref{fig:sp_UVJ_basti}, but using EMILES+Padova00 SSP models.}
\label{fig:sp_UVJ_padova}
\end{figure*}

\begin{figure*}
\centering
\includegraphics[trim= 0 25 0 5mm,width=17cm,clip=True]{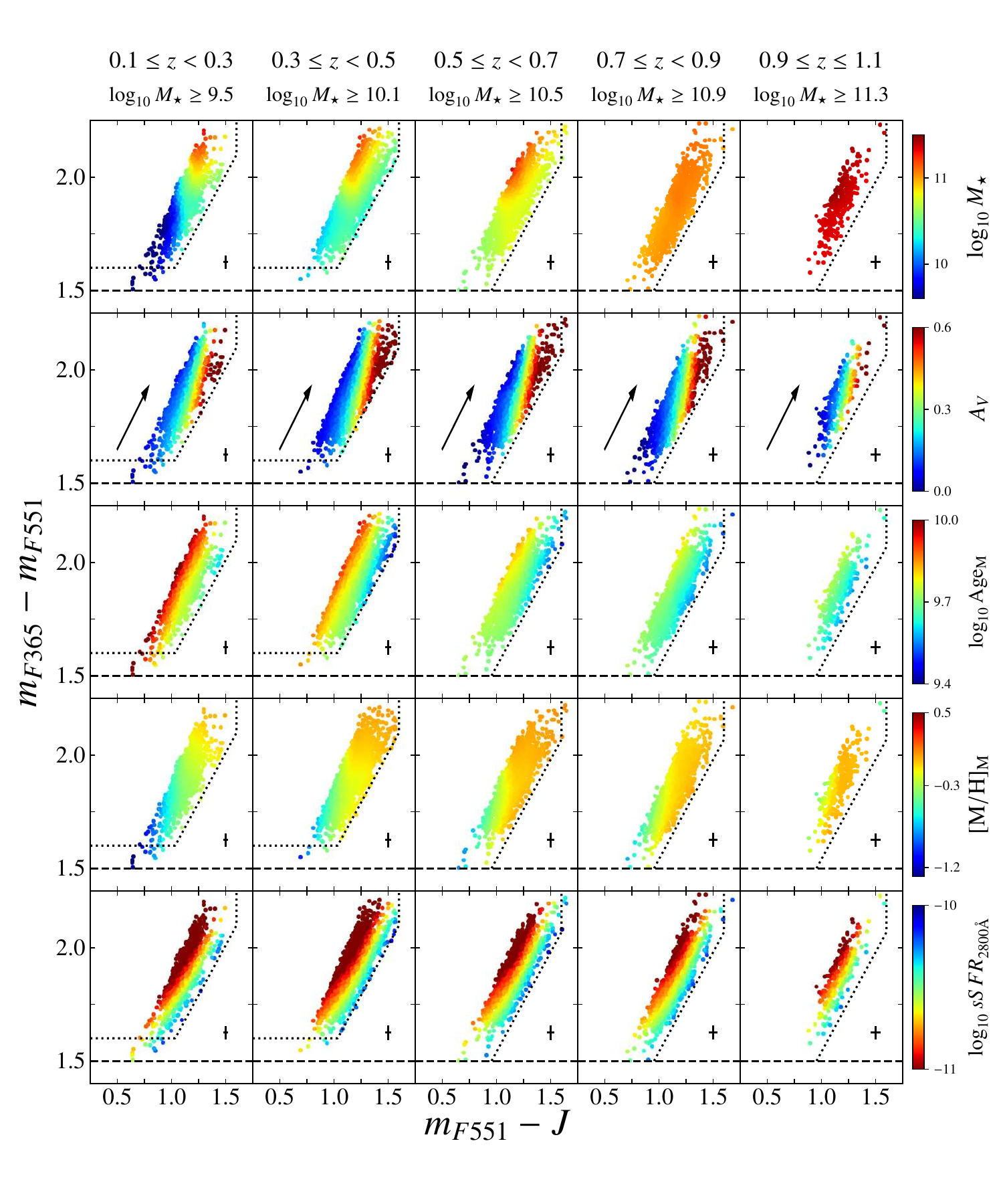}
\caption{Same as Fig.~\ref{fig:sp_UVJ_av_basti}, but using EMILES+Padova00 SSP models.}
\label{fig:sp_UVJ_av_padova}
\end{figure*}

\begin{figure*}
\centering
\includegraphics[trim= 0 0 0 0,width=17cm,clip=True]{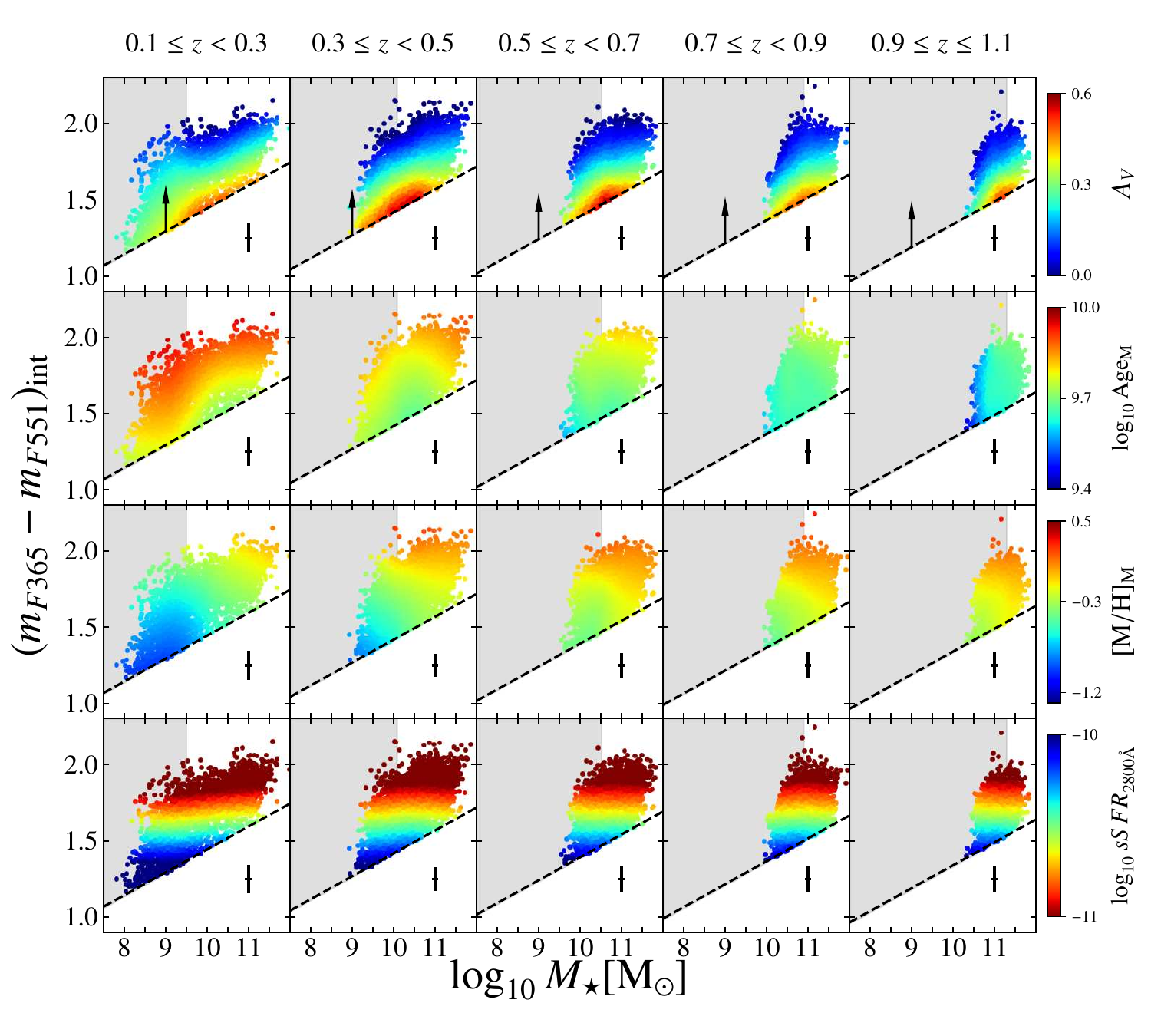}
\caption{Same as Fig.~\ref{fig:CMDE_sp_basti}, but using EMILES+Padova00 SSP models.}
\label{fig:CMDE_sp_padova}
\end{figure*}

\begin{figure*}
\centering
\includegraphics[trim= 0 0 0 0,width=17cm,clip=True]{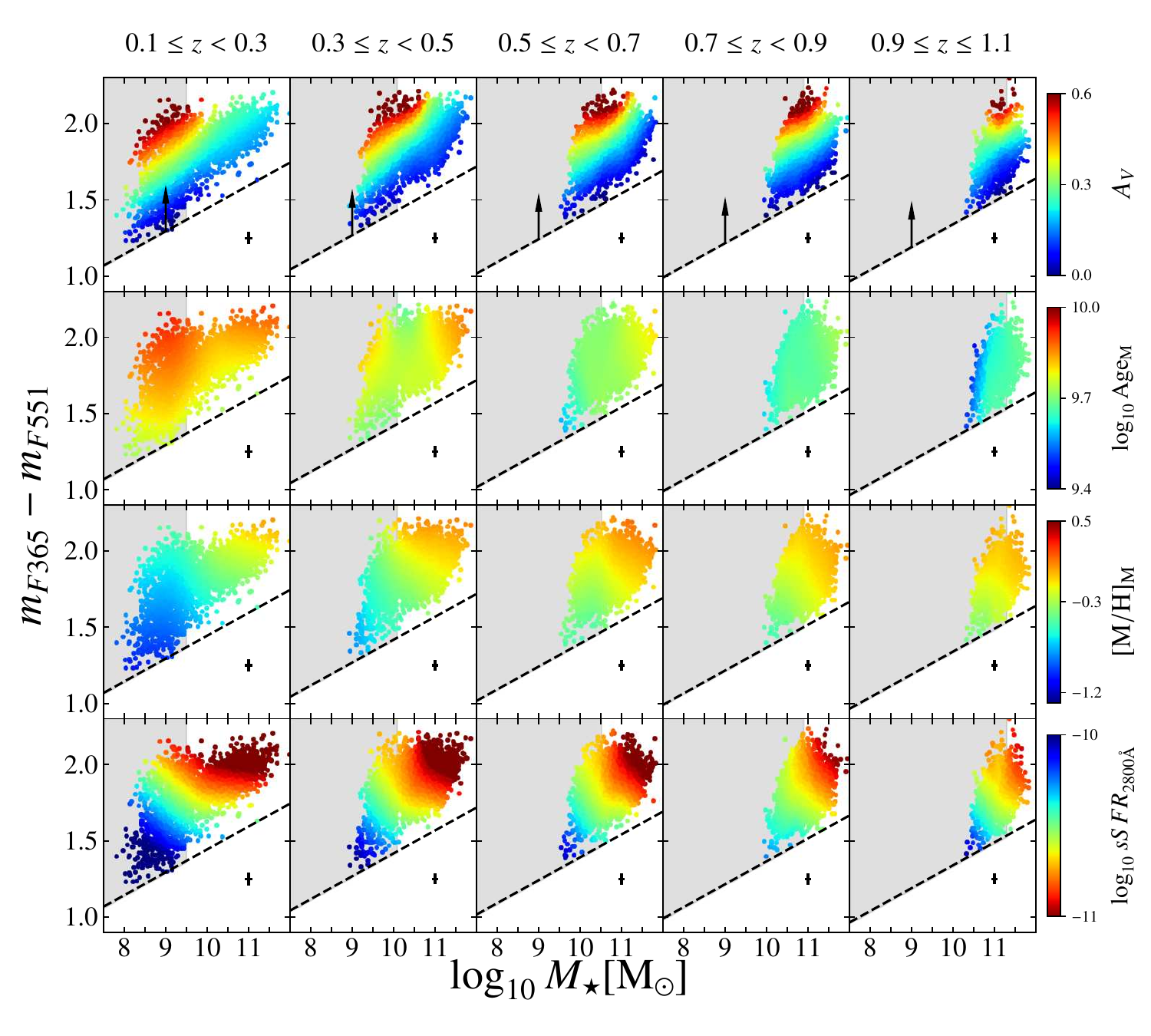}
\caption{Same as Fig.~\ref{fig:CMDE_sp_av_basti}, but using EMILES+Padova00 SSP models.}
\label{fig:CMDE_sp_av_padova}
\end{figure*}

\begin{figure*}
\centering
\includegraphics[trim= 0 0 0 0,width=17cm,clip=True]{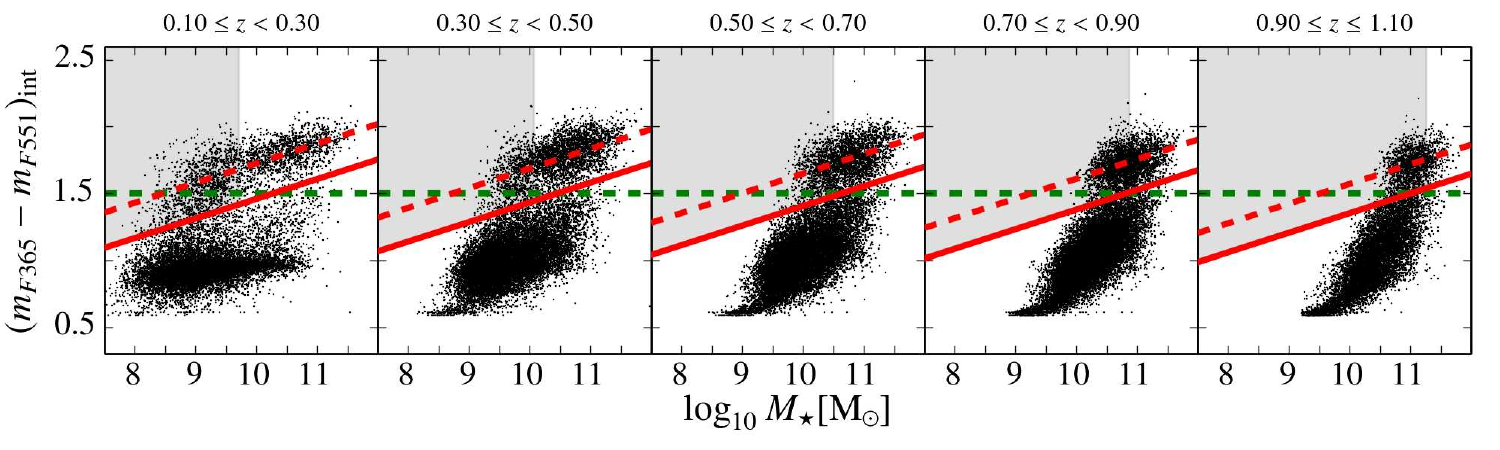}
\caption{Same as Fig.~\ref{fig:CMDE_limit_basti}, but using EMILES+Padova00 SSP models.}
\label{fig:CMDE_limit_padova}
\end{figure*}

As detailed in Sect.~\ref{sec:methods}, we assessed potential systematics from the use of different population synthesis models by building sets of composite stellar population models through two separate sets of SSP models. Below, we illustrate some of the results presented and discussed in Sects.~\ref{sec:sample}--\ref{sec:discussion}, but using the EMILES SSP models for BaSTI (Figs.~\ref{fig:ccd_noav_basti}--\ref{fig:CMDE_limit_basti}) and Padova00 isochrones (Figs.~\ref{fig:ccd_noav_padova}--\ref{fig:CMDE_limit_padova}).

\begin{figure}
\centering
\resizebox{\hsize}{!}{\includegraphics{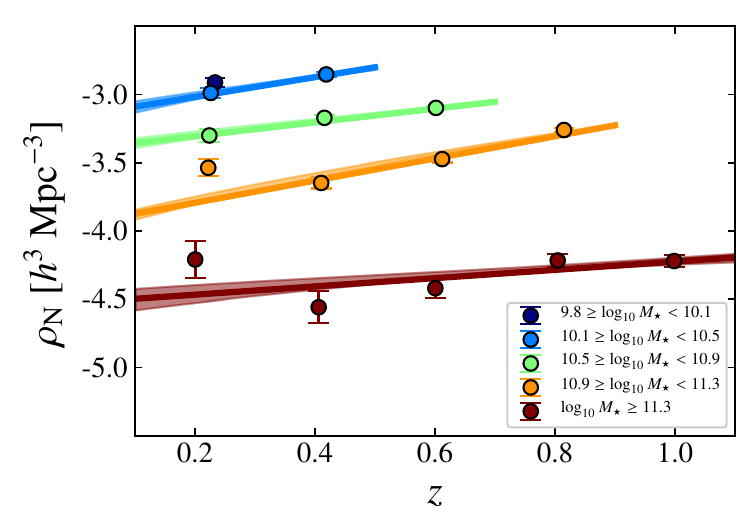}}
\caption{Evolution of co-moving number density, $\rho_\mathrm{N}(z)$, of green valley galaxies with redshift ($X$-axis) at  different stellar mass bins (see inset) using EMILES+BaSTI SSP models. Coloured markers illustrate co-moving number densities at redshift bins and vertical bars are their uncertainties. Solid lines illustrate the best fit to Eq.~(\ref{eq:comoving}), whereas the shaded regions show the 1$\sigma$ uncertainty of the fit.}
\label{fig:comoving_green_basti}
\end{figure}

\begin{figure}
\centering
\resizebox{\hsize}{!}{\includegraphics{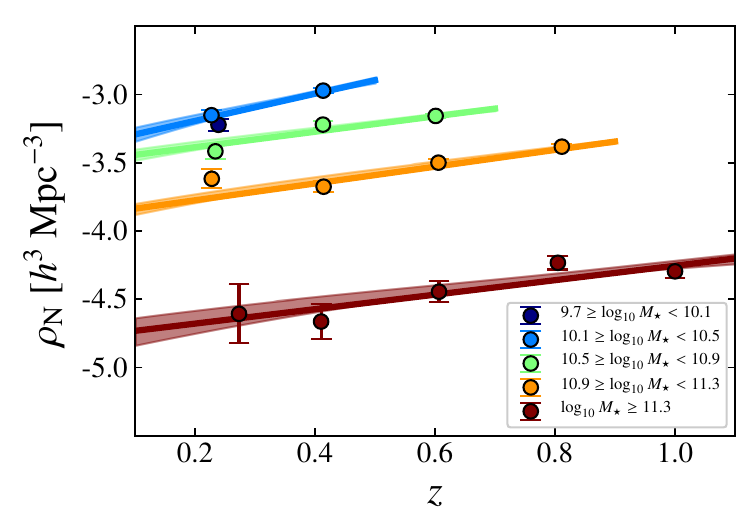}}
\caption{Same as Fig.~\ref{fig:comoving_green_basti}, but using EMILES+Padova00 SSP models.}
\label{fig:comoving_green_padova}
\end{figure}


\end{appendix}

\end{document}